\documentclass[onecolumn,amsmath,amssymb,12pt,superscriptaddress,nofootinbib]{revtex4-2}
\usepackage[T1]{fontenc}
\usepackage{lmodern}
\usepackage[english]{babel}
\usepackage{amsthm}
\usepackage{amssymb}
\usepackage{amsmath}
\usepackage{amsthm}
\usepackage[]{graphicx}
\usepackage{tensor}
\usepackage{framed}
\usepackage{framed}
\usepackage[bookmarks,linktocpage, colorlinks=true, plainpages = false, citecolor = blue,  linkcolor=blue, urlcolor = blue, filecolor = blue]{hyperref} 
\usepackage{natbib}
\usepackage{dcolumn}
\usepackage{bm}
\usepackage{float}
\usepackage{tikz}
\usetikzlibrary{decorations.pathmorphing,patterns,shapes}
\usepackage[utf8]{inputenc}
\usepackage[T1]{fontenc}
\usepackage{xcolor}
\usepackage{physics} 
\usepackage{slashed}
\usepackage{dsfont}
\usepackage{float}
\usepackage{simplewick}
\usepackage{microtype}
\usepackage{graphicx}
\usepackage{vwcol} 
\usepackage{url}
\usepackage{hyperref}
\usepackage{indentfirst}
\usepackage{caption}
\usepackage{subcaption}
\usepackage{setspace}
\def\af{a_{f}}
\def\phif{\phi_{f}}
\def\xf{x_{f}}
\def\yf{y_{f}}

\begin{document}
	
	\allowdisplaybreaks
	\begin{titlepage}
		
		\title{
			Revisiting the no-boundary proposal with a scalar field
		}
		\author{Caroline Jonas}
		\email{caroline.jonas@aei.mpg.de}
		\affiliation{Max--Planck--Institute for Gravitational Physics (Albert--Einstein--Institute), 14476 Potsdam, Germany}
		\author{Jean-Luc Lehners}
		\email{jlehners@aei.mpg.de}
		\affiliation{Max--Planck--Institute for Gravitational Physics (Albert--Einstein--Institute), 14476 Potsdam, Germany}
		\author{Vincent Meyer}
		\email{vmeyer@aei.mpg.de}
		\affiliation{Humboldt-Universit\"{a}t zu Berlin, Zum Grossen Windkanal 2, \\ 12489 Berlin, Germany}
		
		\begin{abstract}
			\vspace{.5cm}
			\noindent Recent works have suggested that the no-boundary proposal should be defined as a sum over regular, not necessarily compact, metrics. We show that such a prescription can be implemented in the presence of a scalar field. For concreteness, we consider the model of Garay et al., in which the potential is a sum of exponentials, and which lends itself to an analytical treatment. Compared to the earlier implementation, we find that saddle points with unstable fluctuations can be eliminated by imposition of an appropriate regularity condition. This leads to the appearance of additional saddle points, corresponding to unclosed geometries. We argue that such saddles will occur generically, though we also find in our example that they are subdominant to the closed, Hartle-Hawking, saddle points. When the potential is positive, classical spacetime is only predicted for inflationary histories. When the potential is negative, we recover the AdS gravitational path integral, with a stable scalar field included. One puzzle that we find is that in general the path integral must be restricted to sum only over specific, discrete and late time dependent initial values of the scalar field. Only when the scalar is required to take real values is this puzzle eliminated, a situation that moreover leads to advantageous phenomenological characteristics. 
		\end{abstract}
		\maketitle
	\end{titlepage}
	\tableofcontents

	%%%%%%%%%%%%%%%%%%%%%%%%%
	%%%%%%%%%%%%%%%%%%%%%%%%%
	%%%%%%%%%%%%%%%%%%%%%%%%%
	%%%%%%%%%%%%%%%%%%%%%%%%%
	%%%%%%%%%%%%%%%%%%%%%%%%%
	%%%%%%%%%%%%%%%%%%%%%%%%%
	%%%%%%%%%%%%%%%%%%%%%%%%%
	%%%%%%%%%%%%%%%%%%%%%%%%%
	%%%%%%%%%%%%%%%%%%%%%%%%%
	%%%%%%%%%%%%%%%%%%%%%%%%%

	\section{Introduction} 
	
	The no-boundary proposal is an attempt to describe the quantum wave function of the universe. It was first introduced in the 1980s by Hartle and Hawking \cite{Hawking:1981gb,Hartle:1983ai}, and has recently been the focus of renewed interest, see e.g.~\cite{Feldbrugge:2017kzv,Halliwell:2018ejl,Janssen:2019sex,DiTucci:2019bui,Matsui:2020tyd}. This is because, on the one hand, it has proven remarkably difficult to construct competing theories of initial conditions and, on the other hand, there is a certain naturalness to the no-boundary proposal that makes it compelling. Recent works have reinforced this last point: for instance, if one studies gravitational path integrals in the presence of a negative cosmological constant, one is led to the equivalent of no-boundary conditions, but in spacetimes that are asymptotically Anti-de Sitter \cite{DiTucci:2020weq}. What is remarkable about this setting is that via AdS/CFT it is known what result to expect for the wave function \cite{Caputa:2018asc} and thus we can be rather confident that, in this setting, the no-boundary prescription is correct. Then, via analytic continuation in the cosmological constant, one recovers the standard cosmological no-boundary proposal \cite{Lehners:2021jmv}. It has further been shown that the no-boundary proposal is robust to the inclusion of higher order terms in the Riemann curvature, which are expected to arise via quantum gravitational corrections \cite{Jonas:2020pos,Cano:2020oaa}. Finally, the case has been made that, possibly together with loitering/emergent scenarios, the no-boundary proposal constitutes the only currently known way of rendering cosmological amplitudes finite and well-defined \cite{Jonas:2021xkx}. Despite these appealing features we should point out that a crucial issue, namely whether or not the no-boundary proposal is in agreement with observations, is still open, for a recent discussion see \cite{Matsui:2020tyd}. This is partly due to the fact that the predictions of the no-boundary proposal depend on the dynamical theory under consideration, and partly because the no-boundary proposal explains aspects of early universe theories that are otherwise already assumed, such as the classicality of spacetime and the ground state of fluctuations \cite{Halliwell:1984eu,Hartle:2008ng}, so that it may be difficult to cristallise out truly new predictions. Still, we will have a little more to say about observational predictions in the discussion section.
	
	Traditionally, the no-boundary proposal was defined as a path integral summing over geometries that are both compact and regular \cite{Hartle:1983ai}. However, this definition is in tension with the uncertainty principle, as compactness is a condition on the scale factor and regularity a condition on its conjugate momentum (i.e.~the expansion rate of the universe). If one imposes compactness alone, then in general one finds that the path integral contains two types of saddle points: those with stable fluctuations and their complex conjugate saddle points with unstable fluctuations. This hinders a clear-cut definition of the no-boundary proposal, at least in certain simple minisuperspace models \cite{Feldbrugge:2017fcc,Feldbrugge:2017mbc}. By contrast, if one imposes regularity alone, then one may eliminate unstable saddle points \cite{DiTucci:2019dji,DiTucci:2019bui} (early suggestions to use a momentum condition, though with somewhat different motivations, include \cite{Louko:1988bk,Bousso:1998na}, and momentum conditions were also used in \cite{Janssen:2019sex}). The price to pay is that it is not guaranteed that the saddle points will also be compact, which is certainly a condition for interpreting the no-boundary wave function as describing the creation of the universe. For pure gravity and isotropic metrics, this has been shown to work, and the saddle points indeed end up being both regular and compact \cite{DiTucci:2019dji,DiTucci:2019bui}.
	
	However, the link to observations is done using a scalar field, either with an inflationary \cite{Baumann:2009ds} or an ekpyrotic \cite{Lehners:2008vx} potential. In both cases, it has been shown that the no-boundary proposal can explain the nucleation of a classical spacetime from nothing, as both mechanisms drive the wave function to a semi-classical WKB form \cite{Hartle:2008ng,Lehners:2015sia}. For ekpyrosis, one faces the challenge of understanding the bounce to the expanding phase. In phenomenological models, such a bounce can be included \cite{Lehners:2015efa}. But on a fundamental level bounces remain ill understood, as it remains unclear whether quantum gravity can allow for effective null energy violations on large scales \cite{Brandenberger:2016vhg}. Hence we will not consider ekpyrotic models further in this work.
	
	We will rather focus on inflationary potentials here, revisiting a minisuperspace model considered previously by Garay et al.~\cite{Garay:1990re} and that has the property that it can be treated analytically to a large extent. Garay et al. defined the no-boundary wave function by a sum over compact metrics. Since they did not discuss cosmological fluctuations in their work, they were not concerned by the fact that with this definition, stable and unstable saddle points are typically linked by steepest descent contours, making it impossible to avoid the unstable saddles \cite{Feldbrugge:2017mbc}. Our goal is to see whether the proposal to define the no-boundary wave function by a sum over regular metrics is viable in this case. 
	
	We find that, just as for the pure gravity case, imposing regularity is equivalent to requiring the absence of momentum flow at the creation of the universe \cite{Lehners:2021jmv}. However, in contrast to the pure gravity case where this condition was unique, here it leads to a family of possible initial conditions, with a parameter $\gamma$ that may loosely speaking be thought of as the initial value of the scalar field (this interpretation is not always valid, as we will discuss). By construction, the saddle points are always regular, but only for certain values of $\gamma$ can the saddle points also be compact. Even then, we find that closed saddle point geometries appear alongside unclosed geometries, which we argue to be generically present. Still, an important result is that the unclosed geometries are always subdominant to the closed ones, so that the no-boundary wave function ends up being dominated by closed, regular geometries, as desired. As previously noted by Garay et al.~\cite{Garay:1990re}, the dominant saddle points do not lead to a WKB wave function in all regions of phase space, and in fact a classical spacetime is only predicted in inflationary regions of the potential. We also extend our analysis to include a negative potential, in which case we recover geometries that are asymptotically AdS, but with a dynamical scalar field.  
	
	An important question highlighted by our work is how to define the no-boundary wave function in general. As we will see, the regularity condition (which may also be seen as the absence of momentum flow) does not in and of itself uniquely fix the initial conditions. Further input is required, such as the condition that compact saddle points exist. This issue, and the resulting puzzles that it implies, are discussed in more detail in the discussion section. However, let us point out already that recent works proposing to define quantum field theories on complex spacetimes using specific convergence conditions offer a clue as to how the puzzle may ultimately be resolved.

		%%%%%%%%%%%%%%%%%%%%%%%%%%%%%%%%
	%%%%%%%%%%%%%%%%%%%%%%%%%%%%%%%%%%%%%%%%%%%%%%%%%%%%%%%%%%%%%%%%%%%%%%%%%%%%%%%%%%%%%%%%%%%%

	\section{Model and path integral}
	
	We are interested in the no-boundary wave function of the universe $\Psi(h_{ij},\Phi_f),$ a functional of a 3-dimensional hypersurface $h_{ij}$ on which a scalar field configuration $\Phi_f$ resides. The arguments may be thought of as a possible current state of the universe, with the wave function providing probabilities for inequivalent histories of the universe \cite{Hartle:2008ng}. The wave function is formally defined as a path integral
	\begin{align}
	\Psi(h_{ij},\Phi_f) = \int Dg_{\mu\nu}D\Phi \, e^{\frac{i}{\hbar}S}\,,
	\end{align}
	where the integration runs over a restricted class of 4-manifolds with metrics $g_{\mu\nu}$ and scalar field configurations. A precise definition of the above path integral remains an important goal of quantum cosmology. In the present paper, we will investigate how to define this integral in a minisuperspace setting, where we restrict the metric to be of FLRW form and the scalar to be homogeneous.
	
	The action is taken to be that for general relativity minimally coupled to a scalar field. A Gibbons-Hawking-York (GHY) term is added on the final boundary, but not on the initial boundary, since we would like to impose a regularity (Neumann) condition there, rather than fixing the metric (which would have required a Dirichlet condition),
	\begin{align}
		S=&\ \frac{1}{16\pi G}\int_{M}\dd^4x\sqrt{-g}\left[R-8\pi G\left(g^{\mu\nu}\partial_\mu\Phi\partial_\nu\Phi+\tilde{V}(\Phi)\right)\right]-\frac{1}{8\pi G}\int_{\partial M_f}\dd^3x\sqrt{h}K\,.
	\end{align}
	The scalar potential $\tilde{V}(\Phi)$ will be specified shortly. We work in minisuperspace with an FLRW ansatz for the metric. More specifically, we consider the universe to be homogeneous and isotropic, with scale factor $a$:
	\begin{align}
		\dd s^2&=\frac{2G}{3\pi}\Big[-\frac{N^2(\tau)}{a^2(\tau)}\dd \tau^2+a^2(\tau)\dd\Omega_3^2\Big]\,.\label{eq:metricintau}
	\end{align}
	The time coordinate has been defined in a somewhat unusual way for later convenience \cite{Garay:1990re}. The scalar field is also assumed to be homogeneous, $\Phi=\Phi(\tau)$. One can rescale the scalar field and the potential to remove numerical factors from the action, to find
	\begin{align}
		&\Phi(\tau)=\sqrt{\frac{3}{4\pi G}}\,\phi(\tau)\quad;\quad\tilde{V}(\Phi)=\left(\frac{9}{8G^2}\right)V(\phi)\,;\label{eq:rescaling}\\
		\Rightarrow\ &S=\ \frac{1}{2}\int\dd \tau\Big[N-\frac{a^2\dot{a}^2}{N}+\frac{a^4\dot{\phi}^2}{N}-a^2NV(\phi)\Big]-\frac{1}{2}\left[\frac{a^3\dot{a}}{N}\right]^{0}\,.\label{eq:actionaandphi}
	\end{align}
	The last term is a surface term at $\tau=0.$ The surface term at $\tau=1$ has disappeared after integration by parts.
	
	As in \cite{Garay:1990re}, we use the following scalar potential
	\begin{equation}
		V(\phi)=\alpha\cosh(2\phi)+\beta\sinh(2\phi)\,,
	\end{equation}
	which renders the system analytically solvable for arbitrary real parameters $\alpha$ and $\beta$. This becomes obvious by applying the following change of variables:
	\begin{equation}
		x=a^2\cosh(2\phi)\quad;\quad y=a^2\sinh(2\phi)\,;\label{eq:defxandy}
	\end{equation}
	which transforms the action into
	\begin{equation}
		2S[x,y]=\int\dd\tau 	N\left[\frac{\dot{y}^2-\dot{x}^2}{4N^2}+1-\alpha x-\beta y\right]-\left[\frac{x\dot{x}-y\dot{y}}{2N}\right]^0\,.\label{eq:actionxandy}
	\end{equation}
	We define the momenta conjugate to $x,y$ by
	\begin{align}
	\Pi_{x}(\tau)=-\frac{\dot{x}}{2N}\,, \qquad \Pi_{y}(\tau)=\frac{\dot{y}}{2N}\,,
	\end{align}
	The equations of motion and constraint deriving from the action are very simple,
	\begin{align}
	\ddot{x}=2N^2 \alpha\,, \qquad \ddot{y}=-2N^2\beta\,,  \\
	\frac{1}{4N^2}(\dot{x}^2-\dot{y}^2)+1-\alpha x-\beta y=0\,. \label{constraint}
	\end{align}
	
	Before proceeding, we must first discuss the boundary conditions that should be imposed on the fields. There are two complementary ways of doing this, and fortunately both agree. The first is to continue with the action integral, and look at the variational problem that it implies. Varying w.r.t. the fields $x$ and $y$ we obtain the surface terms 
	\begin{align}
	\delta (2S) \supset \left[ -x \, \delta\left( \frac{\dot{x}}{2N}\right) + y \, \delta\left( \frac{\dot{y}}{2N}\right)\right]^0 + \left[ \frac{1}{2N} \left( -\dot{x} \, \delta x + \dot{y} \, \delta y\right)\right]^1\,.
	\end{align}
	Thus at $\tau=0$ we may impose a condition on the momenta $\Pi_x \equiv -\Pi_{x}(\tau=0), \Pi_y \equiv \Pi_{y}(\tau=0)$. (Note that, for later calculational convenience, we define $\Pi_x$ with an additional minus sign.) The appropriate regularity condition is obtained by looking at the constraint \eqref{constraint} when $x,y \to 0,$ implying that we should impose 
	\begin{align}
	\Pi_x^2 - \Pi_y^2 = -1 \qquad \textrm{(regularity condition)}\,.
	\end{align}
	At $\tau=1$ we can simply fix the field values $x(1)\equiv x_f, \, y(1) \equiv y_f.$
	
	A second approach is to look at the Wheeler-DeWitt (WdW) equation, which is the quantum version of the classical constraint \eqref{constraint}. As discussed in \cite{Lehners:2021jmv}, the no-boundary proposal can be seen as the requirement that the universe is entirely self-contained, and that consequently no momentum flows into it from ``outside'' at the nucleation of the universe. In momentum space, the WdW equation is obtained as the operator version of the constraint \eqref{constraint}, replacing $x \to -i \frac{d}{d\Pi_x}, y \to i \frac{d}{d\Pi_y}$,
	\begin{align}
	\left( \Pi_x^2-\Pi_y^2+1+i\alpha \frac{d}{d\Pi_x} - i \beta \frac{d}{d\Pi_y}\right) \Psi = 0\,. \label{WdWmomentum}
	\end{align}
	The absence of momentum flow, $\frac{d\Psi}{d\Pi_{x,y}}=0,$ thus yields the exact same condition, $\Pi_x^2-\Pi_y^2=-1.$ 
	
	It will be useful to parameterise the initial momenta as
	\begin{align}
	\Pi_x = i \cosh(2\gamma)\,,\ \Pi_y = i \sinh(2\gamma)\,,
	\end{align}
	This means that we still have a free parameter $\gamma$ specifying the initial conditions. Determining and interpreting $\gamma$ will be one of the main tasks of the present work. Before continuing, let us remark that one condition on $\gamma$ is familiar from earlier works \cite{DiTucci:2019dji,DiTucci:2019bui,DiTucci:2020weq}, namely that the sign of the imaginary part of $\Pi_x$ must be chosen such that perturbations of the geometry and matter are stable, rather than unstable\footnote{This condition may also be seen as choosing the Hartle-Hawking wave function \cite{Hartle:1983ai}, rather than Vilenkin's tunnelling wave function \cite{Vilenkin:1982de}.}. The appropriate choice is
	\begin{align}
	\Im\left[ \Pi_x \right] = - \Im \left[\Pi_{x(0)}\right] > 0 \qquad \textrm{(stability condition)}\,.
	\end{align}
	
	With the boundary conditions specified so far (fixing the momenta at $\tau=0$ and the field values at $\tau=1$), it is straightforward to find solutions to the equations of motion for $x$ and $y,$
		\begin{align}
			&\bar{x}(\tau)=\alpha N^2\tau^2+2N\Pi_x \tau + \xf-2N\Pi_x-\alpha N^2\,,\label{eq:solx} \\
			&\bar{y}(\tau)=-\beta N^2 \tau^2 +2N\Pi_y \tau + \yf-2N\Pi_y+\beta N^2\,.\label{eq:soly}
		\end{align}
	We may use these solutions to perform the path integrals over $x,y$ by writing a generic field evolution as a sum of such a solution plus a variation $X,Y:$
	\begin{equation}
		x(\tau)=\bar{x}(\tau)+X(\tau)\,,\quad y(\tau)=\bar{y}(\tau)+Y(\tau)\,.
	\end{equation}
	It is important that the variations are completely arbitrary, except that the total fields must obey the boundary conditions, implying that we must fix
	\begin{equation}
		\dot{X}(0)=0\,,\ X(1)=0\,,\ \dot{Y}(0)=0\,,\ Y(1)=0\,. \label{bcfluct}
	\end{equation}
	The action can then be rewritten as
	\begin{align}
		S&=S_\text{on-shell}+\frac{1}{2}\int\dd\tau N\left(\frac{\dot{Y}^2-\dot{X}^2}{4N^2}\right)+\left[\frac{\dot{\bar{y}}Y-\dot{\bar{x}}X}{4N}\right]^1-\left[\frac{x\dot{X}-y\dot{Y}}{4N}\right]^0\,.\label{eq:fullactionfluct}
	\end{align}
	Due to \eqref{bcfluct} the surface terms vanish, and because the original action was quadratic we now obtain Gaussian integrals over $X$ and $Y,$
	\begin{equation}
		\Psi=\int\dd N\,e^{iS_\text{on-shell}/\hbar}\cdot\int_{\dot{X}(0)=0}^{X(1)=0}\mathcal{D}X\int_{\dot{Y}(0)=0}^{Y(1)=0}\mathcal{D}Y\,\exp(\frac{i}{2\hbar}\int\dd\tau\frac{\dot{Y}^2-\dot{X}^2}{4N})\,.
	\end{equation}
	The calculation of the fluctuation integrals with the required mixed Neumann-Dirichlet boundary conditions can be done explicitly, see the appendix of \cite{DiTucci:2020weq} for details, and the result is simply a numerical prefactor
	\begin{equation}
		\Psi[\xf,\yf,\gamma]=\frac{8\hbar}{\pi}\int\dd N \exp(\frac{i}{\hbar}S_\text{on-shell}[N,\xf,\yf,\gamma])\,.
	\end{equation} 
	In appendix \ref{sec:fluct} we provide a quick, indirect way of verifying this result. We are thus left with an ordinary integral over the lapse $N,$ with the integrand arising from the explicit integration over the solutions \eqref{eq:solx}, \eqref{eq:soly},
	\begin{align}
		S_\text{on-shell}=\frac{N^3(\alpha^2-\beta^2)}{6}+\frac{N^2(\alpha\Pi_x+\beta\Pi_y)}{2}-\frac{N(\alpha\xf+\beta\yf)}{2}+\frac{\yf\Pi_y-\xf\Pi_x}{2}\,.\label{eq:actiononshellN}
	\end{align}
	The saddle points of the integral are located at
	\begin{equation}
		\left\lbrace
		\begin{aligned}
			&N_{\text{saddle}\,\pm}^{\alpha\neq\beta}=\frac{-(\alpha\Pi_x+\beta\Pi_y)\pm\sqrt{(\alpha\Pi_x+\beta\Pi_y)^2+(\alpha^2-\beta^2)(\alpha \xf+\beta \yf)}}{\alpha^2-\beta^2}\,;\\
			&N_\text{saddle}^{\alpha=\beta}=\frac{1}{2}\frac{\xf+\yf}{\Pi_x+\Pi_y}\,.\label{eq:saddle}
		\end{aligned}
		\right.
	\end{equation}
	Thus, for fixed boundary conditions, in particular fixed $\gamma,$ we obtain two saddle points in a general potential, and a single saddle point for the exponential potential ($\alpha = \beta$). In terms of the original variables, the saddle points may be rewritten as
		\begin{equation}
		\left\lbrace
		\begin{aligned}
			&N_{\text{saddle}\,\pm}^{\alpha\neq\beta}=\frac{1}{\alpha^2-\beta^2}\left(-iV(\gamma)\pm\sqrt{(\alpha^2-\beta^2)a_f^2\cosh(2\phi_f)-V(\gamma)^2} \right)\,;\\
			&N_\text{saddle}^{\alpha=\beta}=-\frac{i}{2}a_f^2\exp\left( 2\phi_f-2\gamma\right)\,.\label{eq:saddleorigvar}
		\end{aligned}
		\right.
	\end{equation}
	
	As is evident, the saddle points (and thus also the semi-classical wave function) depend on the initial conditions via the parameter $\gamma.$ Since $\gamma$ fixes the initial momenta, and not the field values, it is not guaranteed that the saddle point geometries will be compact, i.e.~it is not guaranteed that they close at $\tau=0.$ In fact, this will only occur for specific values of $\gamma.$ We will adopt the prescription that we will simply sum over all values of $\gamma$ that lead to at least one closed saddle points, i.e.~we will take the total wave function to be given by
	\begin{align}
	\Psi[x_f,y_f] = \sum_{\gamma \, \textrm{s.t.} \, \exists \, \bar{a}(0)=0} \Psi[x_f,y_f,\gamma]\,.
	\end{align}
	We should point out that this corresponds to a generalisation of the path integral, where we now also sum over specific boundary conditions. As often, it is illuminating to study a few concrete examples to better see what consequences this definition entails. This will be the subject of the next section.

	\section{Examples}
	
\begin{figure}
		\includegraphics[width=9cm]{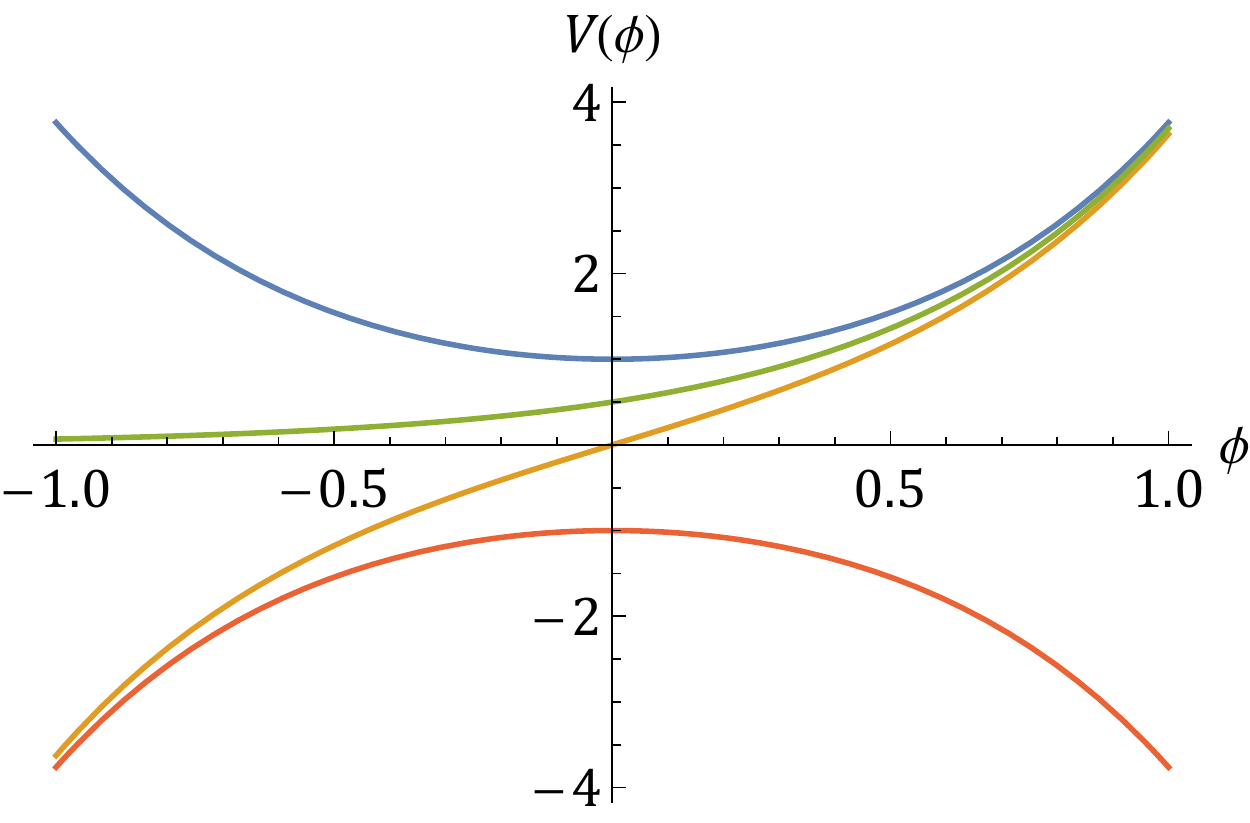}\caption{Different scalar field potentials: $(\alpha,\beta)=(1,0)$ hyperbolic cosine in blue, $(\alpha,\beta)=(0,1)$ hyperbolic sine in orange, $(\alpha,\beta)=(1/2,1/2)$ exponential in green and $(\alpha,\beta)=(-1,0)$ negative hyperbolic cosine in red.}\label{fig:potentials}
	\end{figure}

	Despite its simplicity, the form of the potential $V(\phi)=\alpha \cosh(2\phi) + \beta \sinh(2\phi)$ that we are considering allows for a rich variety of qualitatively different behaviours, see Fig. \ref{fig:potentials}. We will look at the case where the potential is positive, inflationary, and has a minimum ($\beta=0,\alpha>0$). A second case is the opposite potential, with an AdS maximum at negative values of the potential ($\beta=0, \alpha <0$). Further, we can consider an exponential without minimum ($\alpha=\beta$) and a potential that is unbounded both above and below ($\alpha=0$). We will examine these cases in turn. 
	\subsection{$\alpha=1\,,\ \beta=0$: $V(\phi)=\cosh(2\phi)$.}\label{subsubsec:alpha1beta0}
	
	The $\cosh$ potential is the most relevant example to study: it is inflationary, with a slow-roll region for very small values of $|\phi|.$ Also, when the scalar field vanishes, the model reduces to gravity with a cosmological constant (plus scalar fluctuations), and hence we expect to recover the known no-boundary wave function in this limit. As we will see, this potential is the only one where a classical spacetime is predicted, which is a non-trivial result (a related discussion appeared in \cite{Garay:1990re}). The model was also previously used to study large homogeneous quantum fluctuations in inflation \cite{Bramberger:2019zks}.
		
\begin{figure}[h!]
		\centering
		\includegraphics[width=10cm]{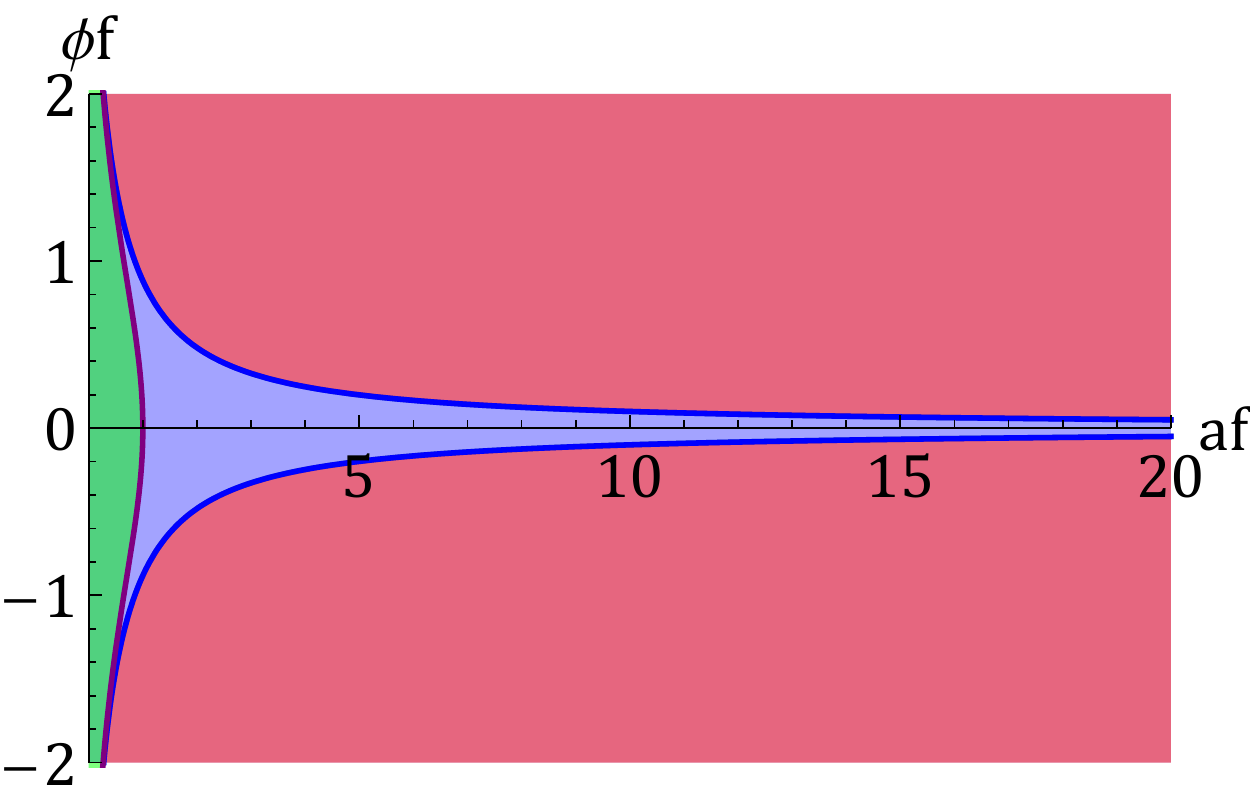}{\caption{\small Phase space regions \eqref{eq:regionsalpha1beta0}. Region I is in blue, region II is in green and region III is in red. The purple curve is the condition $\af^2\cosh[2](\phif)=1$, and the blue curve is the condition $\af^2\sinh[2](\phif)=1$. } \label{fig:phasespacealpha1beta0}}
	\end{figure}

	Let us start by recalling the locations of the saddle points \eqref{eq:saddle}:
	\begin{equation}
		N_{\text{saddle}}^{\pm}=-i\cosh(2\gamma)\pm\sqrt{\af^2\cosh(2\phif)-\cosh[2](2\gamma)}\,.\label{eq:Nsaddlealpha1}
	\end{equation}
	The constraint equation \eqref{constraint}, which is satisfied at the saddle points, directly implies that at the initial time $\tau=0$ we have $x|_0=0$.
	Therefore, given that $a^4=x^2-y^2,$ the initial geometry closes if and only if $y|_0=0$. From \eqref{eq:soly} we see that this is satisfied for
	\begin{equation}
		N_\text{closed}=\frac{\yf}{2\Pi_y}=\frac{a^2_\text{f}\sinh(2\phif)}{2i\sinh(2\gamma)}\,.\label{eq:Nclosed}
	\end{equation}
	Equating the saddle point value \eqref{eq:Nsaddlealpha1} with \eqref{eq:Nclosed}, we find that closed saddle point geometries must obey the equation
	\begin{equation}
		\af^2\sinh[2](2\phif)+4\sinh(2\gamma)\sinh(2\gamma-2\phif)=0\,.\label{eq:closed_saddle}
	\end{equation}
	We solve this equation for $\gamma\in\mathbb{C}$, and find three types of solutions valid in three different regions of the $(\af,\phif)$ phase space:
	\begin{equation}
		\left\lbrace
		\begin{aligned}
			\gamma_{\mp}^{\text{I}}=&\,\frac{\phif}{2}\mp\frac{i}{2}\arcsin(\abs{\sinh(\phif)}\sqrt{\af^2\cosh[2](\phif)-1})\,;\\
			\gamma_{\mp}^{\text{II}}=&\,\frac{1}{2}\cosh^{-1}\bigg(\sqrt{1+\frac{\sinh[2](2\phif)}{4}\Big(2-\af^2\cosh(2\phif)\mp\sqrt{\left(2-\af^2\cosh(2\phif)\right)^2-\af^4}\Big)}\bigg)\,;\\
			\gamma_{\mp\,(n)}^{\text{III}}&=\mp\frac{1}{2}\cosh[-1](\frac{1}{2}\sinh(2\phif)\sqrt{\af^2\cosh(2\phif)-2\mp\sqrt{\left(\af^2\cosh(2\phif)-2\right)^2-\af^4}})\\
			&\,\quad+i\left(\frac{\pi}{4}+\frac{n\cdot\pi}{2}\right),\quad n\in\lbrace0,1\rbrace\,.
		\end{aligned}
		\right.\label{eq:gammasolalpha1}
	\end{equation}
	The above expressions are derived in appendix \ref{appendix:gamma}, and the regions of validity, represented in Fig.~\ref{fig:phasespacealpha1beta0}, are found to be:
	\begin{equation}
		\left\lbrace
		\begin{aligned}
			&\text{region I:}\qquad &\af^2\cosh[2](\phif)>1\ \text{and}\ \af^2\sinh[2](\phif)<1\,;&\qquad\\
			&\text{region II:}\qquad &\af^2\cosh[2](\phif)<1\,;&\qquad\\
			&\text{region III:}\quad &\af^2\sinh[2](\phif)>1\,.&\qquad
		\end{aligned}
		\right.\label{eq:regionsalpha1beta0}
	\end{equation}
	
	We are now ready to  compute the lapse integral. This is done by applying Picard-Lefschetz theory, as in \cite{Feldbrugge:2017kzv}. For this, we must study the steepest descent contours (of the lapse integrand) emanating from the saddle points. On these ``thimbles'' the lapse integral may then be defined as a sum of absolutely convergent integrals, which we can approximate to leading order in $\hbar$ by their saddle point values. 
	
\vspace{0.5 cm}	
{\noindent\it{Region I: complex saddle points.}}

For both $\gamma^{\text{I}}_\mp$ values, we find two saddle point values for the lapse $N$ given by \eqref{eq:Nsaddlealpha1}. All four saddle points are complex, but two of them are closed and the other two unclosed. This is because, for a given $\gamma^{\text{I}}_\mp$ only one choice of sign in \eqref{eq:Nsaddlealpha1} may be made to match with the required value \eqref{eq:Nclosed} for a closed geometry. But for each value of $\gamma$ there are two saddle points, hence the second one must be unclosed. That the regularity condition may nevertheless be satisfied is due to the fact that the potential itself may vanish at $\tau=0.$ We explain this in somewhat more detail in appendix \ref{appendix:picardthm}. Basically, this result follows from the fact that we have extended the fields to the complex plane, so that the potential is now a holomorphic (in fact entire) function of the fields. Picard's little theorem implies that such a function will generically vanish for some field value(s). Thus, the appearance of unclosed saddle points was in fact to be expected.

\begin{figure}
	\centering
	\begin{subfigure}{0.49\linewidth}
		\includegraphics[width=0.49\textwidth]{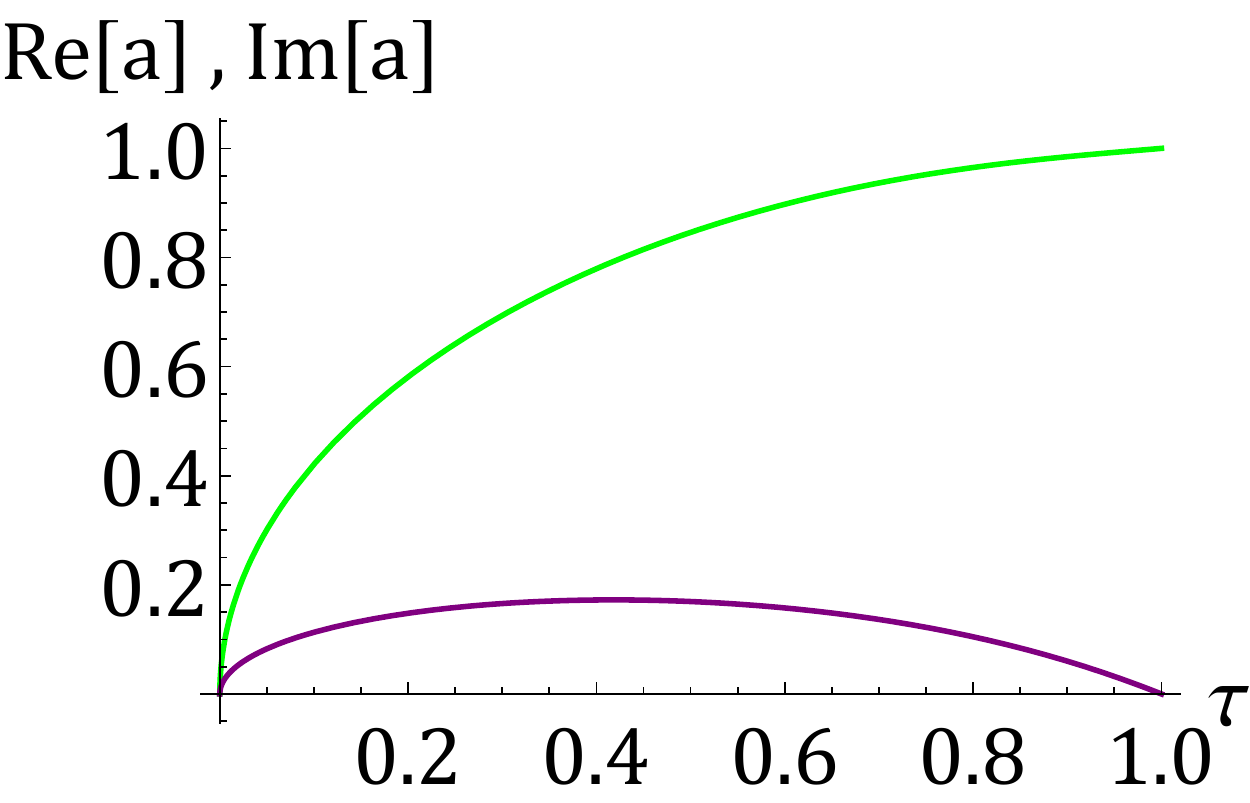}%\caption{Complex fields at $N_+[\gamma_-]$.}\label{fig:geom_a_complexN1phi1}
		\includegraphics[width=0.49\textwidth]{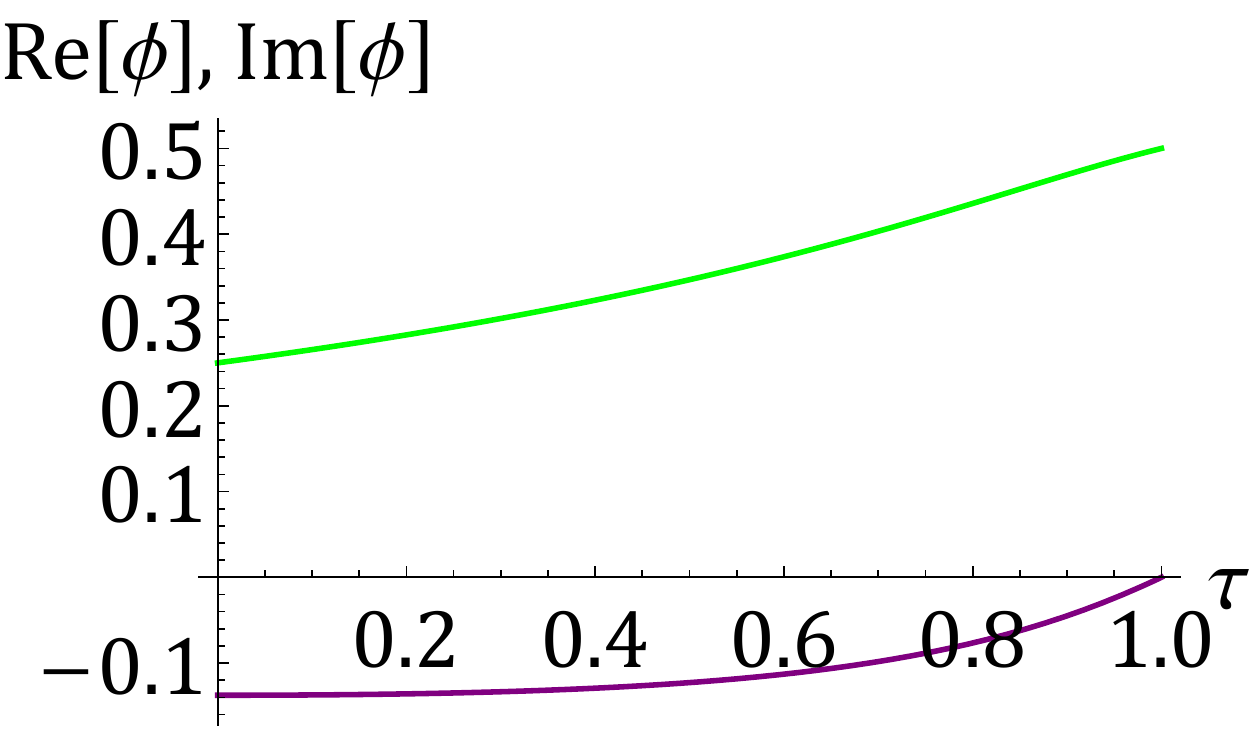}\caption{Complex fields at $N_+[\gamma_-]$.}\label{fig:geom_a_complexN1phi1}
	\end{subfigure}
	\begin{subfigure}{0.49\linewidth}
		\includegraphics[width=0.49\textwidth]{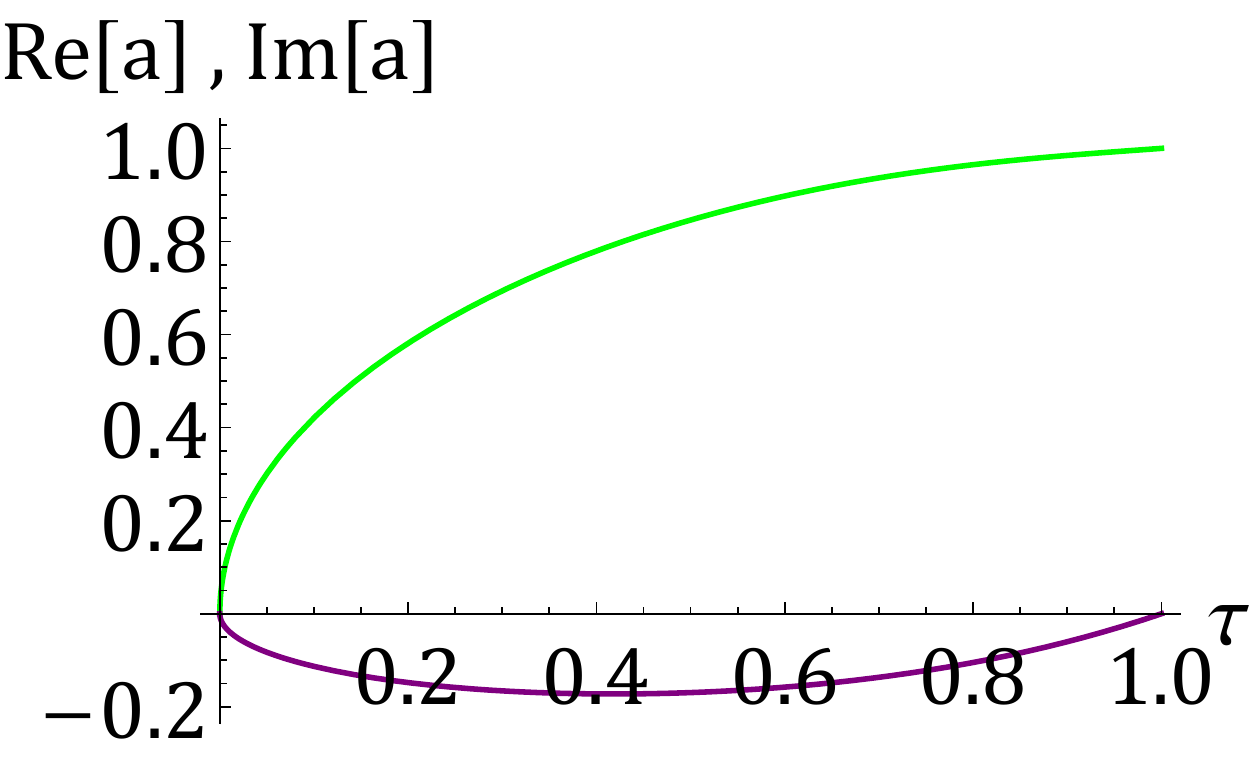}
		\includegraphics[width=0.49\textwidth]{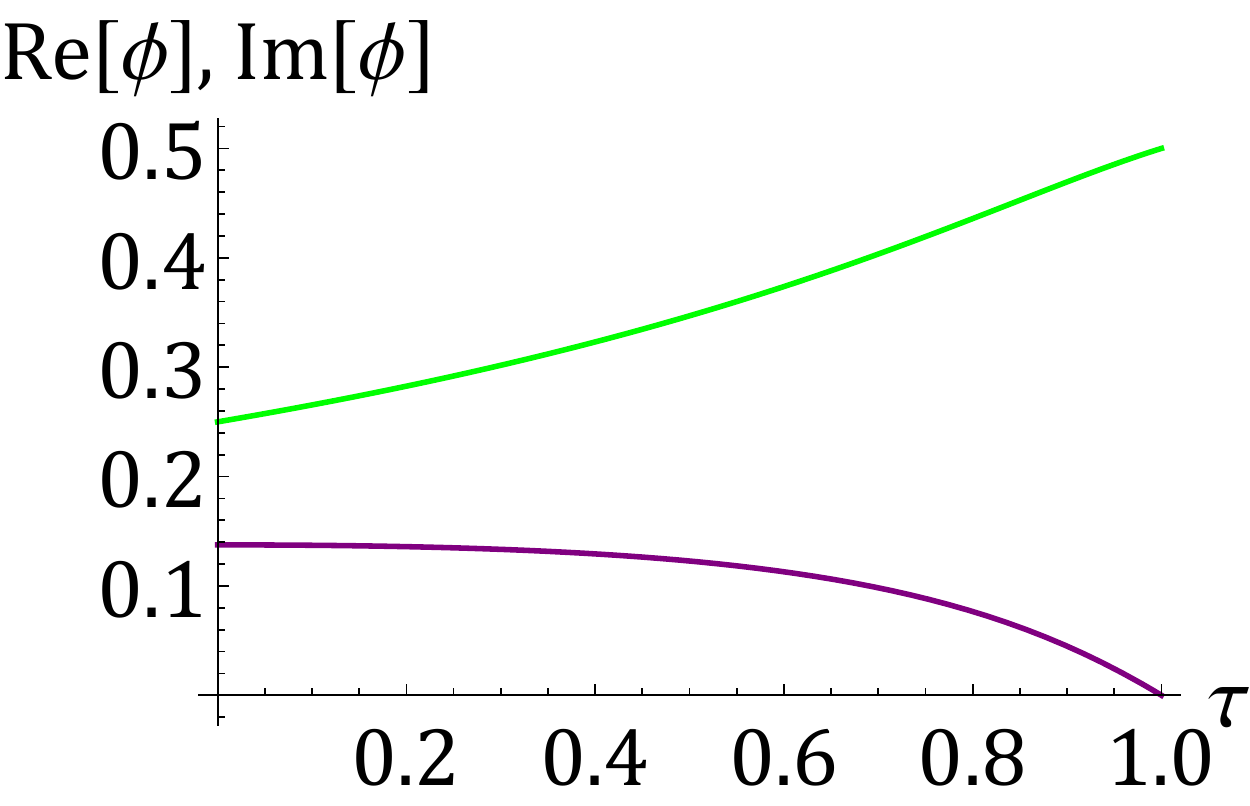}\caption{Complex fields at $N_-[\gamma_+]$.}\label{fig:geom_a_complexN2phi2}
	\end{subfigure}
	\begin{subfigure}{0.49\linewidth}
		\includegraphics[width=0.49\textwidth]{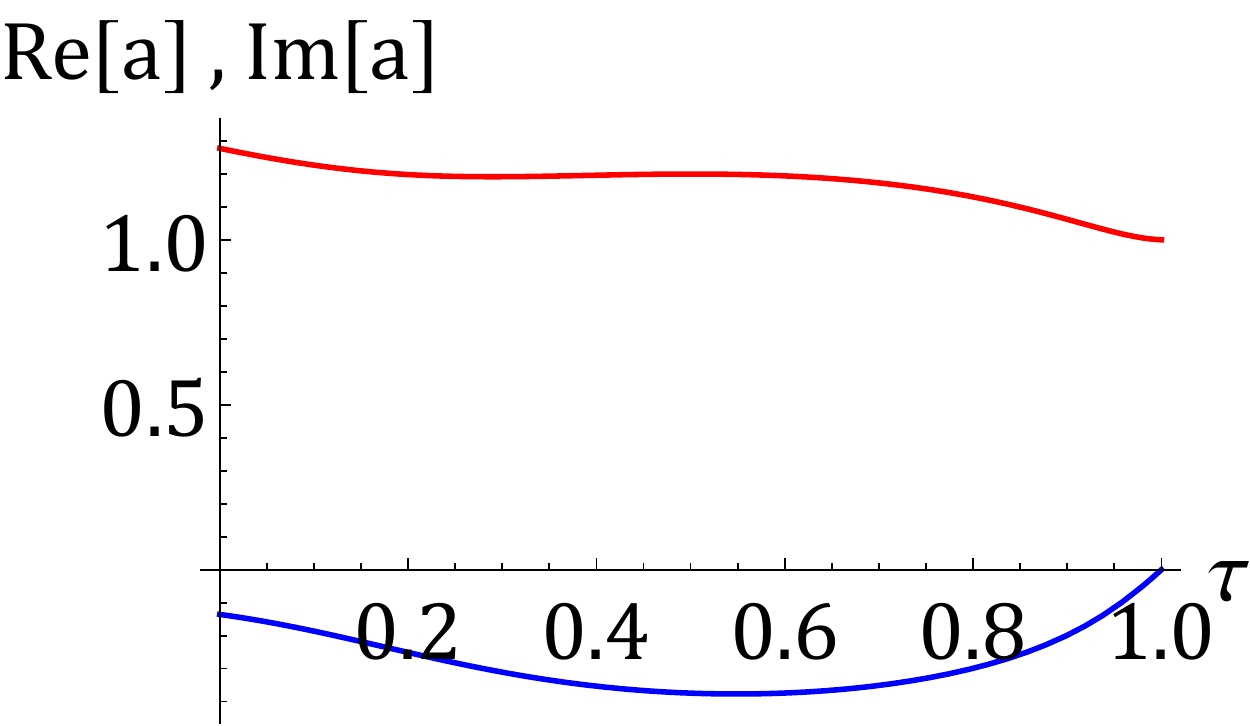}
		\includegraphics[width=0.49\textwidth]{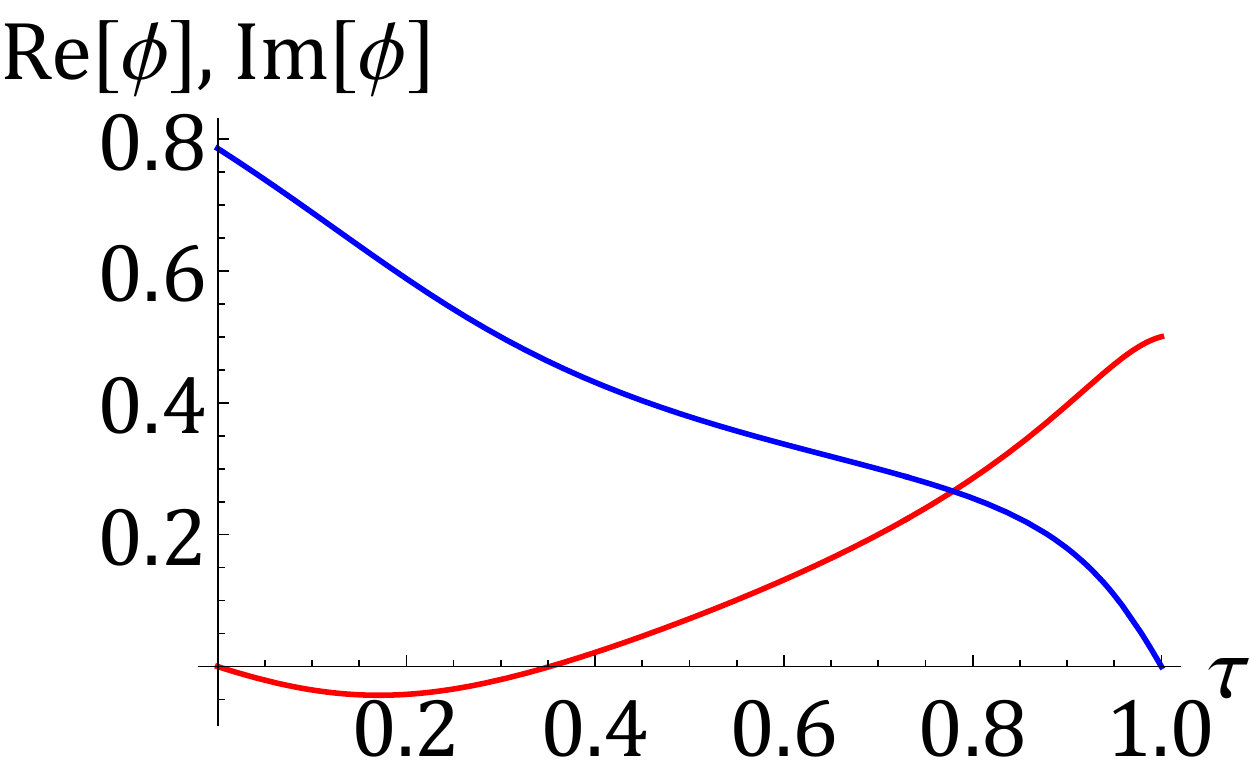}\caption{Complex fields at  $N_-[\gamma_-]$.}\label{fig:geom_a_complexN2phi1}
	\end{subfigure}
	\begin{subfigure}{0.49\linewidth}
		\includegraphics[width=0.49\textwidth]{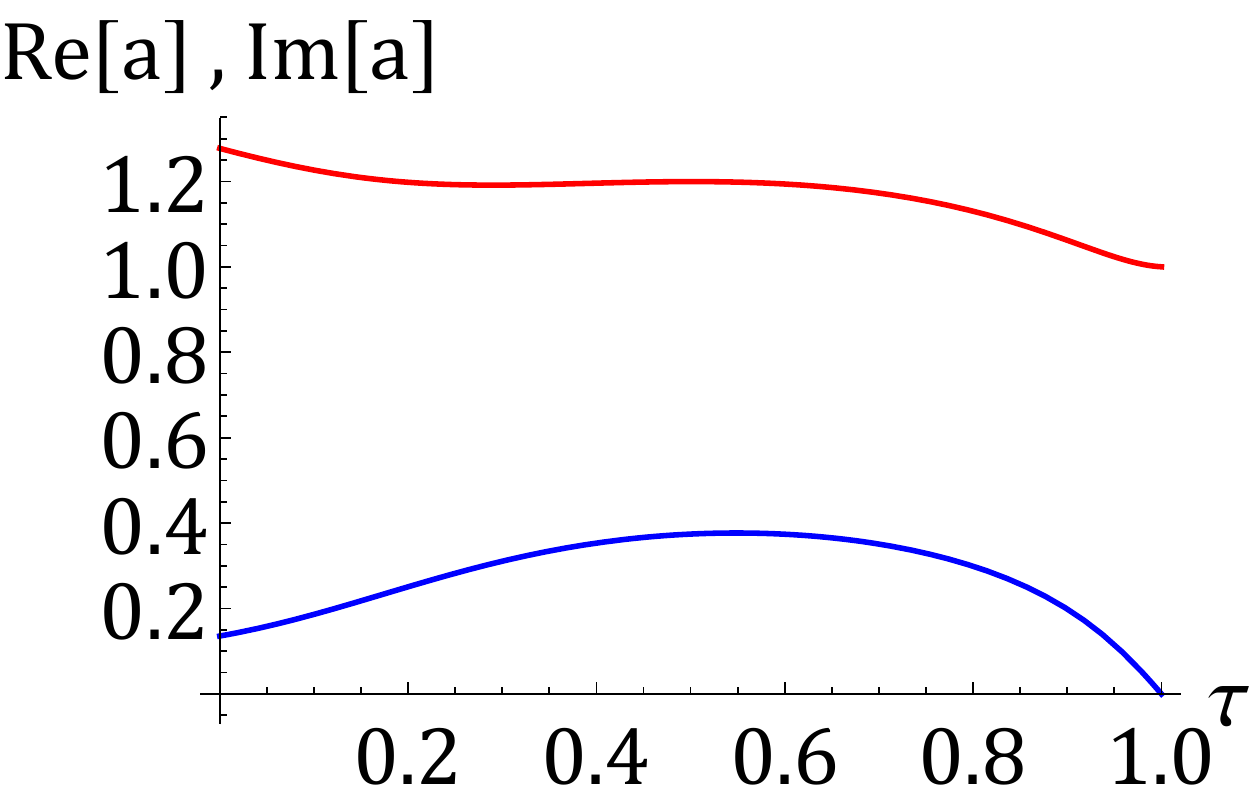}
		\includegraphics[width=0.49\textwidth]{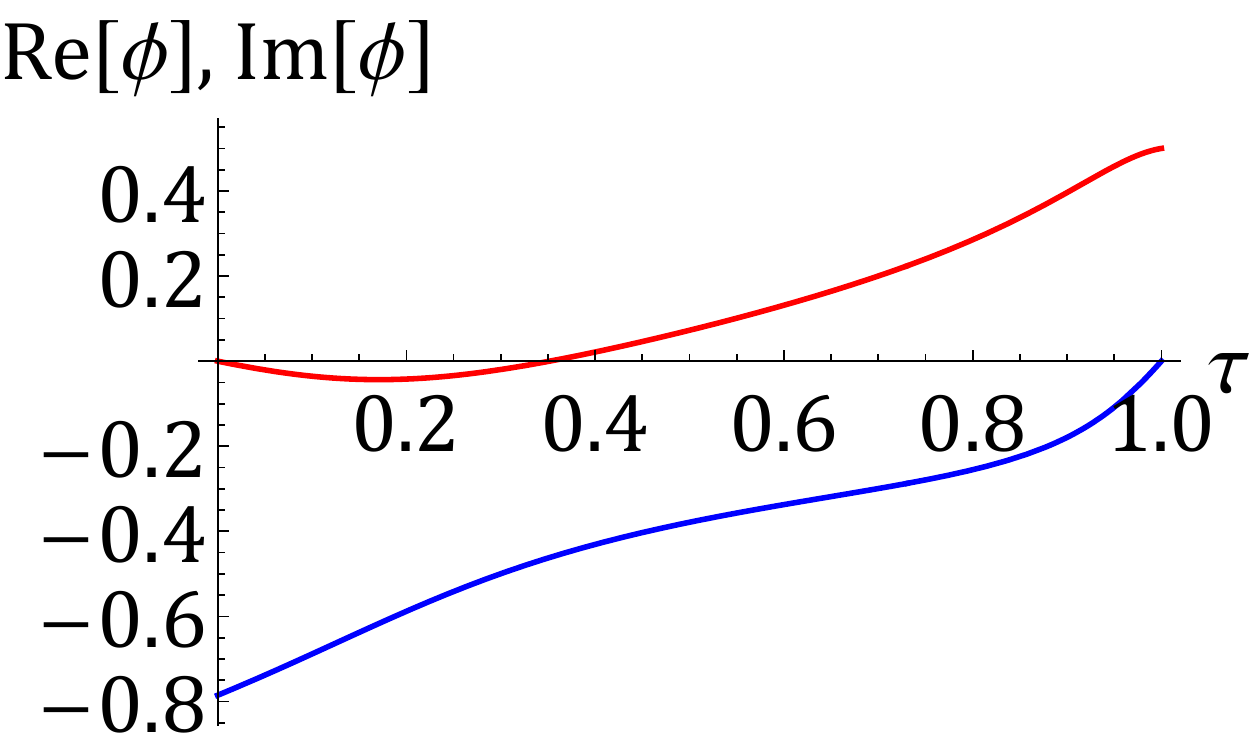}\caption{Complex fields at $N_+[\gamma_+]$.}\label{fig:geom_a_complexN1phi2}
	\end{subfigure}
	\caption{\small Case $\alpha=1,\ \beta=0$, with $N$ complex: $\af=1$, $\phif=0.5$. In this case $\gamma_\mp=0.25\mp0.137 i$ and the saddle points are $N_+[\gamma_-]=0.521 -0.853 i$, $N_-[\gamma_+]=-0.521-0.853i$, $N_-[\gamma_-]=-0.804-1.317 i$, $N_+[\gamma_+]=0.804 -1.317 i$. $N_+[\gamma_-]$ and $N_-[\gamma_+]$ are the closed saddle points. The graphs show the evolution of the scale factor and scalar field at the different saddle points. Real part in green for closed and in red for unclosed saddle points. Imaginary part in purple for closed and in blue for unclosed saddles.}\label{fig:complexsaddle1}
\end{figure}

We illustrate the field evolutions at the saddle points with the numerical examples provided in Fig.~\ref{fig:complexsaddle1}. The saddle points appear in pairs, with opposite real parts of the lapse. This corresponds to geometries and scalar field evolutions that are complex conjugates of each other. As one can see very clearly in the figure, two of the saddle point geometries start out at zero size, while two others have an initial hypersurface with a non-zero, complex scale factor. In all cases, the scalar starts from a complex field value at $\tau=0,$ for the unclosed saddles this is $\phi=\pm i\pi/4,$ where the potential vanishes. For the closed saddle points, we have that
\begin{align}
\phi(\tau=0) = \lim\limits_{\tau \to 0} \frac{1}{2} \textrm{arctanh} \left(\frac{\bar{y}(\tau)}{\bar{x}(\tau)}\right) = \frac{1}{2} \textrm{arctanh} \left(\frac{\Pi_y}{\Pi_x}\right) = \gamma\,.
\end{align}
Thus, for closed saddles, we have the simple interpretation that $\gamma$ corresponds to the initial scalar field value. At $\tau=1,$ all field values are real, as required.

\begin{figure}
	\begin{subfigure}{0.45\linewidth}
		\includegraphics[width=7cm]{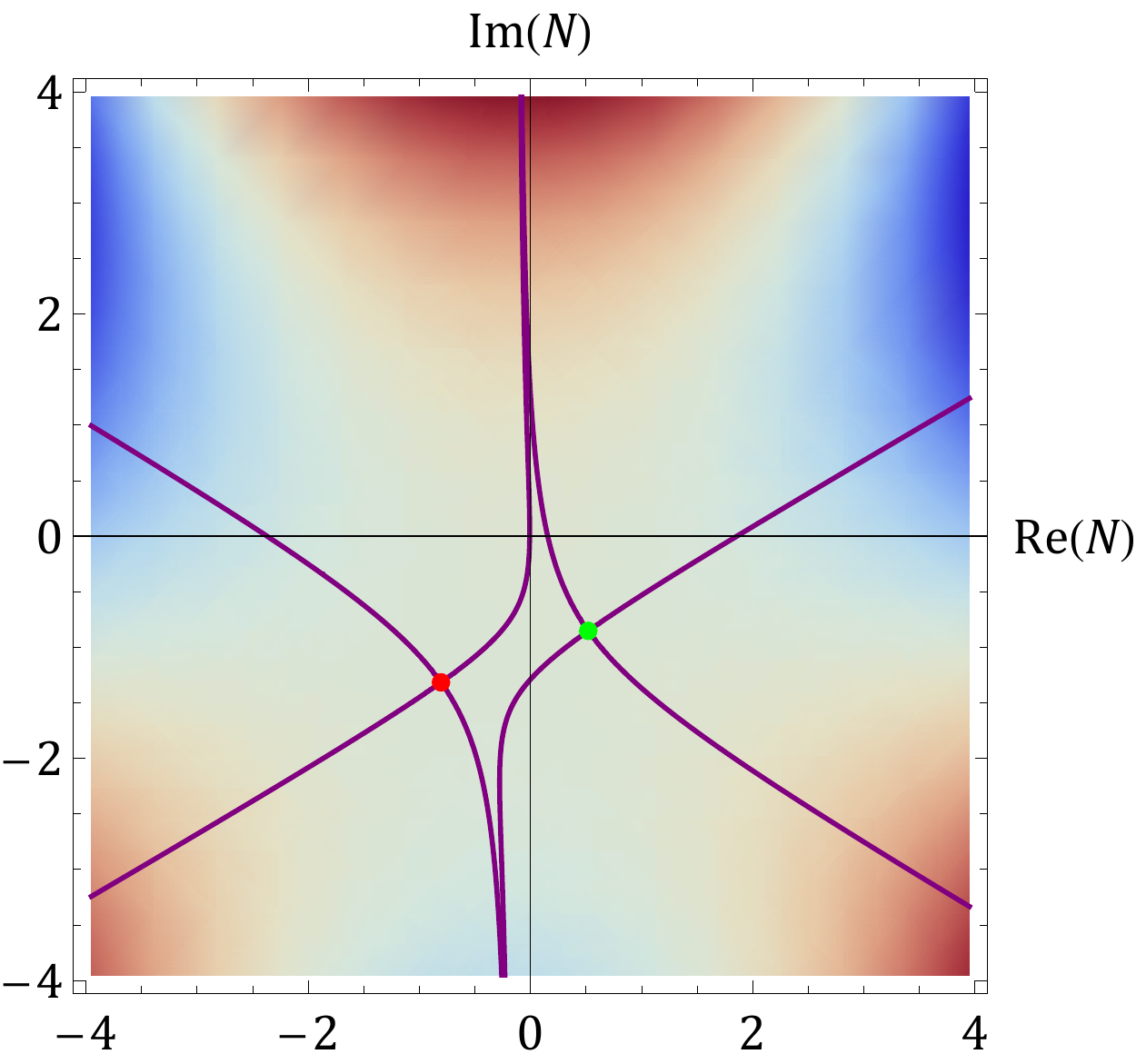}\caption{$N_+[\gamma_-]$ (green dot) and $N_-[\gamma_-]$ (red dot).}\label{fig:lapseint_complex1}
	\end{subfigure}\hspace{1cm}
	\begin{subfigure}{0.45\linewidth}
		\includegraphics[width=7cm]{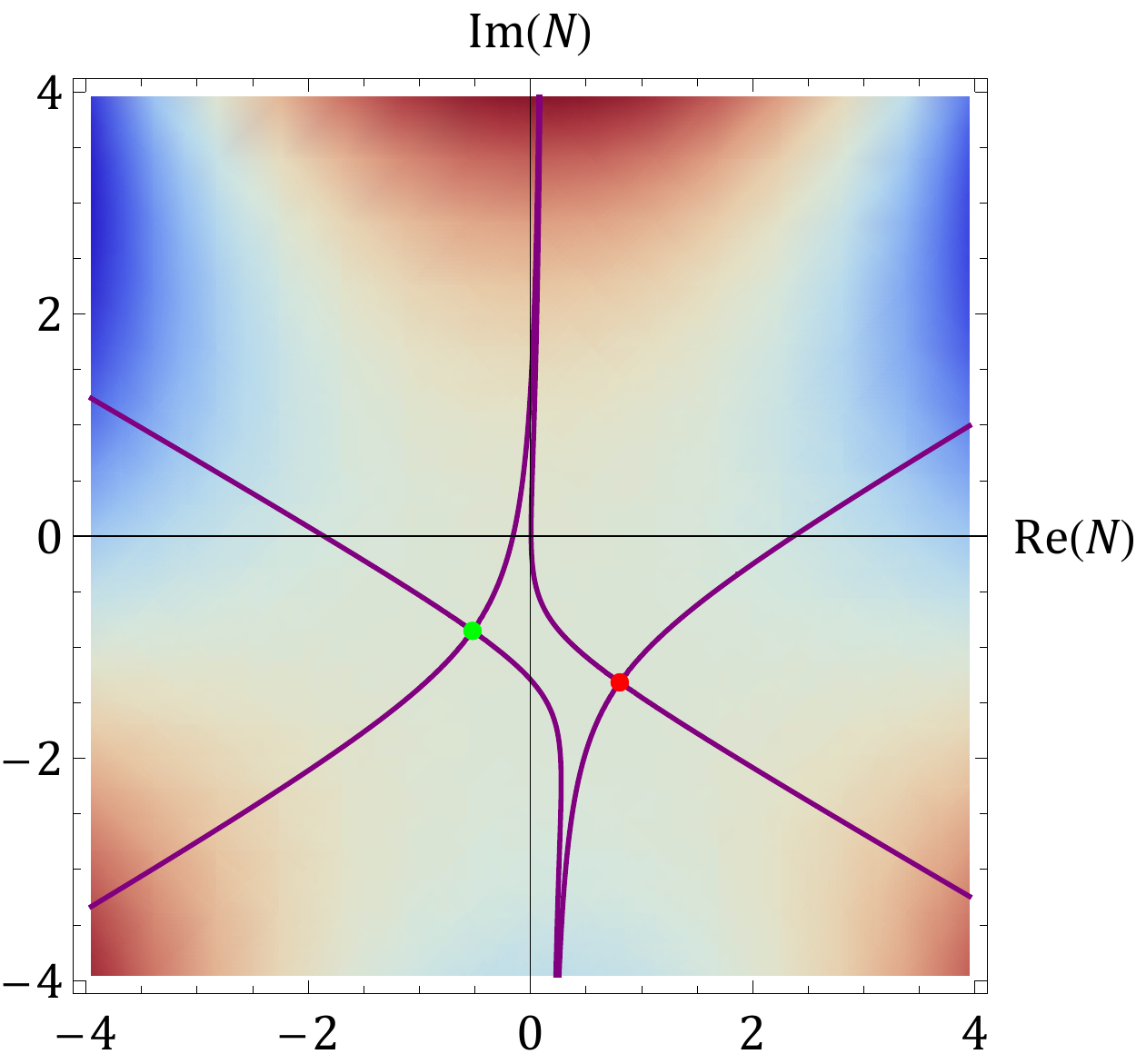}\caption{$N_-[\gamma_+]$ (green dot) and $N_+[\gamma_+]$ (red dot).}\label{fig:lapseint_complex2}
	\end{subfigure}
	\caption{\small Case $\alpha=1,\ \beta=0$, with $N$ complex: $\af=1$, $\phif=0.5$. Density plot of the weighting $\Re[iS_\text{saddle}^\text{on-shell}]$ and flow lines in the complex N plane. Steepest descent (ascent) contours are those lines emanating from the saddle points and reaching blue (red) regions. The path integral is defined on sums of steepest descent contours.}\label{fig:complexsaddle2}
\end{figure}  

Fig. \ref{fig:complexsaddle2} shows the steepest descent/ascent lines associated with the saddle points. We obtain one such figure for each value of $\gamma.$ In each graph there is thus one closed and one unclosed saddle point. Their thimbles can be summed to a contour that runs parallel to the real $N$ line -- this is the most natural contour of integration, and most closely related to a Lorentzian contour. (It does not correspond to an integration over exactly Lorentzian metrics, because the initial momentum conditions imply that near $\tau=0$ all metrics in the sum are necessarily complex.) Given that we are summing over the two relevant values of the initial conditions parameter $\gamma,$ the resulting no-boundary wave function, to leading order in $\hbar,$ is then given by
\begin{align}
\Psi[x_f,y_f] = e^{\frac{i}{\hbar}S(N_+[\gamma_-])}+ e^{\frac{i}{\hbar}S(N_-[\gamma_+])} +e^{\frac{i}{\hbar}S(N_+[\gamma_+])} +e^{\frac{i}{\hbar}S(N_-[\gamma_-])}\,.
\end{align}
The wave function is real, as the saddle points are pairwise complex conjugates of each other. 

It remains to determine which saddle point(s) actually dominate, i.e.~which saddles have the highest weighting. Recall that the saddle points are located at 
	\begin{equation}
		N_{\pm}=-\Pi_x\pm\sqrt{\xf+\Pi_x^2}\,.
	\end{equation}
We can thus write the (on-shell) saddle point action as
	\begin{align}
		S_\text{on-shell}&=\frac{N_\pm^3}{6}+\frac{N_\pm^2\Pi_x}{2}-\frac{N_\pm \xf}{2}+\frac{\yf\Pi_y-\xf\Pi_x}{2} \\
		&=\mp\frac{\sqrt{\xf+\Pi_x^2}^3}{3}+\frac{\Pi_x^3}{3}+\frac{\yf\Pi_y}{2}\,. \label{actionshell}
	\end{align}
	Let us now compute the values of $\Pi_x$ and $\Pi_y$ for the different saddle points.
	The two possible solutions for $\gamma$ are given from Eq.~\eqref{eq:gammasolalpha1} by
	\begin{equation}
		\gamma_\mp=\frac{\phif}{2}\mp\frac{i}{2}\arcsin\left(|\sinh(\phif)|\sqrt{\af^2\cosh[2](\phif)-1}\right)\,.\label{eq:solgamma}
	\end{equation}
	$\gamma_-$ is closing the geometry of the saddle point $N_+$, while $\gamma_+$ is closing the geometry of the saddle point $N_-$.
	We then obtain
	\begin{align}
		\Pi_{x\,\mp}&=i\cosh(2\gamma_\mp)		=\pm\sinh[2](\phif)\sqrt{\af^2\cosh[2](\phif)-1}+i\cosh[2](\phif)\sqrt{1-\af^2\sinh[2](\phif)}\,,\\
		\Pi_{y\,\mp}&=i\sinh(2\gamma_\mp)		=\pm\frac{\sinh(2\phif)}{2}\sqrt{\af^2\cosh[2](\phif)-1}+\frac{i\sinh(2\phif)}{2}\sqrt{1-\af^2\sinh[2](\phif)}\,.
	\end{align}
	Here we assume without loss of generality that $\phif>0$, and we recall that we are looking for solutions in the region of the phase space where $\af^2\cosh[2](\phif)>1$ and $\af^2\sinh[2](\phif)<1$. Thus note that the imaginary parts of $\Pi_\pm$ do not depend on the choice of saddle point, only the real parts differ by a sign. The weighting of the saddle point is determined by the imaginary part of the action, $e^{-\Im[S]}$. Hence from \eqref{actionshell} we see that only the first term will be important in comparing weightings. If we define
	 \begin{align}
		\sqrt{\xf+\Pi_{x\,\mp}^2}^3=\sqrt{A\pm iB}^3\equiv(\rho e^{\pm i\theta})^{3/2}=\rho^{3/2}e^{\pm3/2i\theta}\,,\label{eq:ABdef}
	\end{align}
	then we find
	\begin{align}
		&\Im\left[S_\text{on-shell}^\text{closed}[N_-[\gamma_+]]\right]-\Im\left[S_\text{on-shell}^\text{unclosed}[N_+[\gamma_+]]\right] \\
		=&\Im\left[S_\text{on-shell}^\text{closed}[N_+[\gamma_-]]\right]-\Im\left[S_\text{on-shell}^\text{unclosed}[N_-[\gamma_-]]\right] \\
		=&-\frac{2}{3}\rho^{3/2}\sin(\frac{3\theta}{2})\,.\label{eq:ImSclosed}
	\end{align}
	The saddle point that will contribute the most is the one with the smallest $\Im[S]$ value. This depends on the sign of the sine, and hence on the value of the angle $\theta$. With the definition \eqref{eq:ABdef} we have
	\begin{align}
		B&=2\sinh[2](\phif)\cosh[2](\phif)\sqrt{\af^2\cosh[2](\phif)-1}\sqrt{1-\af^2\sinh[2](\phif)}>0\,\Rightarrow\theta\in(0,\pi)\,;\\
		A&=\af^2\cosh(2\phif)+\sinh[4](\phif)(\af^2\cosh[2](\phif)-1)-\cosh[4](\phif)(1-\af^2\sinh[2](\phif))\,.
	\end{align}
	Here $B$ is always positive, while the sign of $A$ depends on the boundary conditions. The sine in Eq.~\eqref{eq:ImSclosed} will be positive as long as $\theta=\arccot(\frac{A}{B})<\frac{2\pi}{3}$, which implies $\cot(\theta)=\frac{A}{B}>-\frac{1}{\sqrt{3}}$, since the cotangent function is monotonically decreasing between $0$ and $\pi$.
	Therefore, when $A/B>-1/\sqrt{3}$, the sine of $3\theta/2$ is positive and the two closed saddle points have the biggest amplitude, while when $A/B<-1/\sqrt{3}$, the sine is negative and the two unclosed saddle points have the largest amplitude. We can plot the region in phase space $(\af,\phif)$ where $A/B>-1/\sqrt{3}$, and we find that it englobes the region where complex saddle point geometries are defined. Near the boundaries of this region,  $\af^2\cosh[2](\phif)=1$ or $\af^2\sinh[2](\phif)=1,$ the weightings approach each other, but in the interior the unclosed saddle points are exponentially suppressed compared to the closed saddle points. Thus we conclude that also in the case where a scalar field is included, the no-boundary wave function is dominated by compact, regular geometries.
	
	\begin{figure}[h!]
		\centering
		\includegraphics[width=10cm]{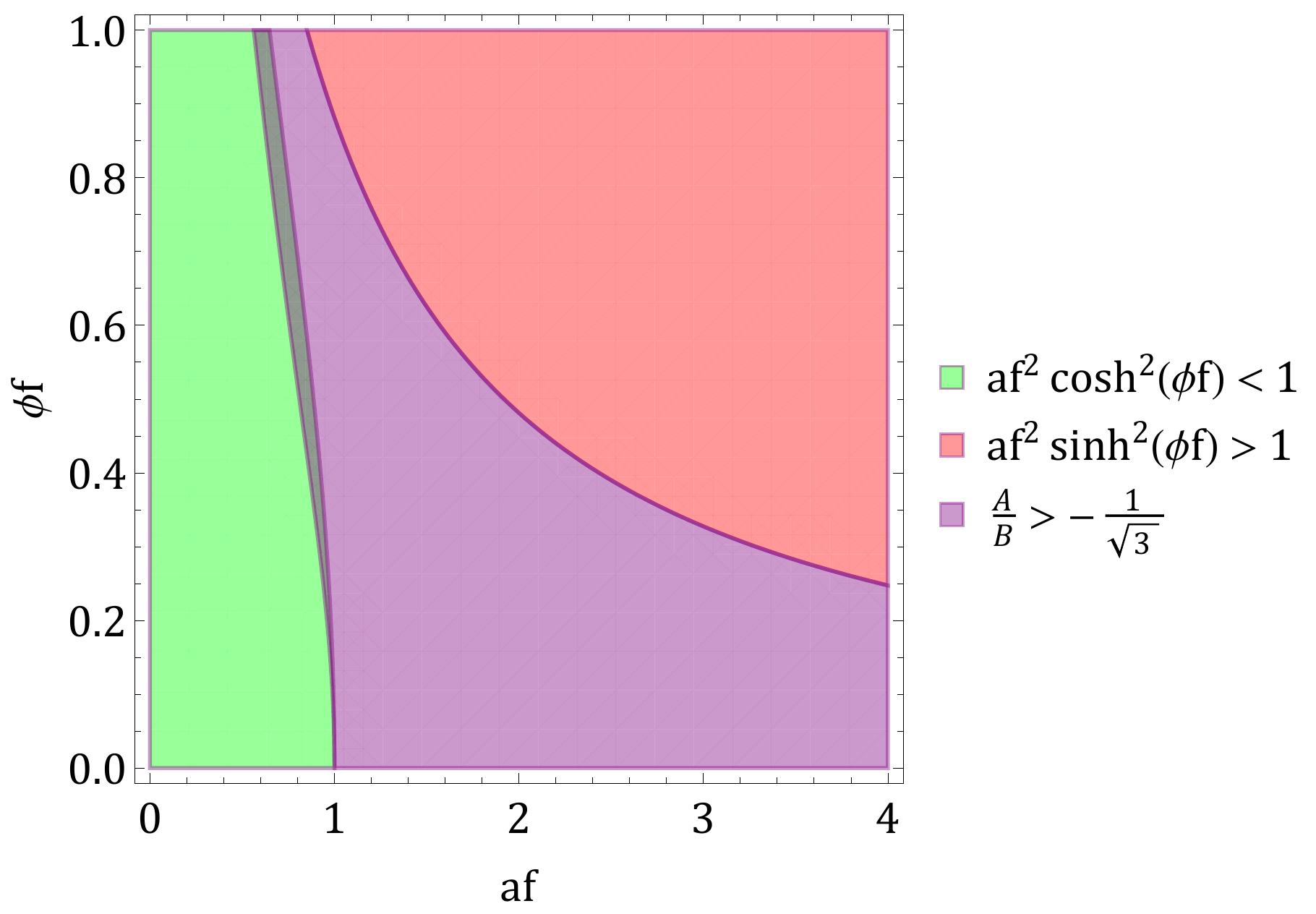}\caption{\small The region where the closed saddle points dominate over the unclosed one is englobing the region where complex saddle points are defined. Note that our analysis assumed that we were outside the green and red regions, so from this picture nothing can be inferred about these regions.}\label{fig:ABregions}
	\end{figure}
	
	Still, the interpretation of the wave function requires more refinements. As it stands, the wave function in a sum of four saddles, which are pairwise complex conjugates of each other. Thus the sum is dominated by not one, but two saddles of equal weight. Taken at face value, this would mean that there would be a strong interference between two universes. However, it was shown in \cite{Kiefer:1987ft,Halliwell:1989vw} that once perturbations are added, they lead to efficient decoherence of the two saddles as the universe grows. Thus, once $a_f \gg 1$ we may focus on a single (dominant) saddle point, say $N_+[\gamma_-]$. For this saddle point, the action may be expanded in the scalar field (which is required to be small for these saddle points to exist), with the result that
	\begin{align}
	S(N_+[\gamma_-]) &= -\frac{1}{3}(a_f^2-1)^{3/2} - \frac{1}{2}(a_f^2-1)^{1/2}a_f^2 \phi_f^2 
	 - \frac{i}{3}(1-\frac{3}{2}a_f^2 \phi_f^2) + {\cal{O}}(\phi_f^4)\,. \label{sadact}
	\end{align} 
	To quadratic order in the final value of the scalar field, the weighting is thus given by
\begin{align}
\abs{e^{\frac{i}{\hbar}S(N_+[\gamma_-])}} = e^{\frac{1}{3}(1-\frac{3}{2}a_f^2 \phi_f^2)}\,, \qquad a_f\phi_f \ll 1\,,
\end{align}
recovering the well known result that the no-boundary wave function gives the highest probability to histories that evolve low on the scalar potential. The on-shell action \eqref{sadact} further implies that at large scale factor we have to leading order
\begin{align}
\left|\frac{\partial \Im(S)}{\partial a_f}\middle/\frac{\partial \Re(S)}{\partial a_f}\right| \sim \frac{\phi_f^2}{a_f} \ll1 \,, \qquad \left|\frac{\partial \Im(S)}{\partial \phi_f}\middle/\frac{\partial \Re(S)}{\partial \phi_f}\right| \sim \frac{1}{a_f} \ll1 \,.
\end{align}
Thus, at large scale factor, the amplitude of the wave function varies very slowly compared to the phase. This demonstrates that this branch of the wave function becomes of WKB form, which implies that in this region of phase space a classical spacetime, with a classical background evolution, is predicted \cite{Hartle:2008ng,Lehners:2015sia}. Finally, we should note that the highest probability occurs for the limit where the scalar sits at the  minimum of the $\cosh(2\phi)$ potential. In this limit, $\gamma$ tends to zero and the unclosed saddles disappear. This limit in fact simply corresponds to the pure gravity case with a cosmological constant, except that now fluctuations of the scalar field are also included. These, however, provide a sub-dominant contribution to the wave function.

\vspace{0.5 cm}	
{\noindent\it{Region  II: imaginary saddle points.}}

In region II, where roughly speaking the universe is very small ($a_f^2 \cosh(2\phi_f)<1$), the solutions for $\gamma$ are purely real. Therefore the saddle point lapse values \eqref{eq:Nsaddlealpha1} will be purely imaginary, as well as the initial momenta $\Pi_x$ and $\Pi_y.$ Then the classical solutions $\bar{x}$ and $\bar{y}$ are purely real, see  \eqref{eq:solx}, \eqref{eq:soly}. If $\bar{a}=(\bar{x}^2-\bar{y}^2)^{1/4}$ is also real, we obtain a purely Euclidean geometry. Moreover the on-shell action evaluated on the saddle points  is imaginary from \eqref{eq:actiononshellN}. This means that the wave function of each saddle point geometry is a pure amplitude:
\begin{equation}
	\psi\propto e^\text{real number}\,;
\end{equation}
hence this doesn't lead to a WKB evolution. In this region of phase space, the universe is still in the nucleation phase, roughly the equivalent to the tunnelling phase of a particle under a barrier potential.

\begin{figure}
	\centering
	\begin{subfigure}{0.49\linewidth}
		\includegraphics[width=0.49\textwidth]{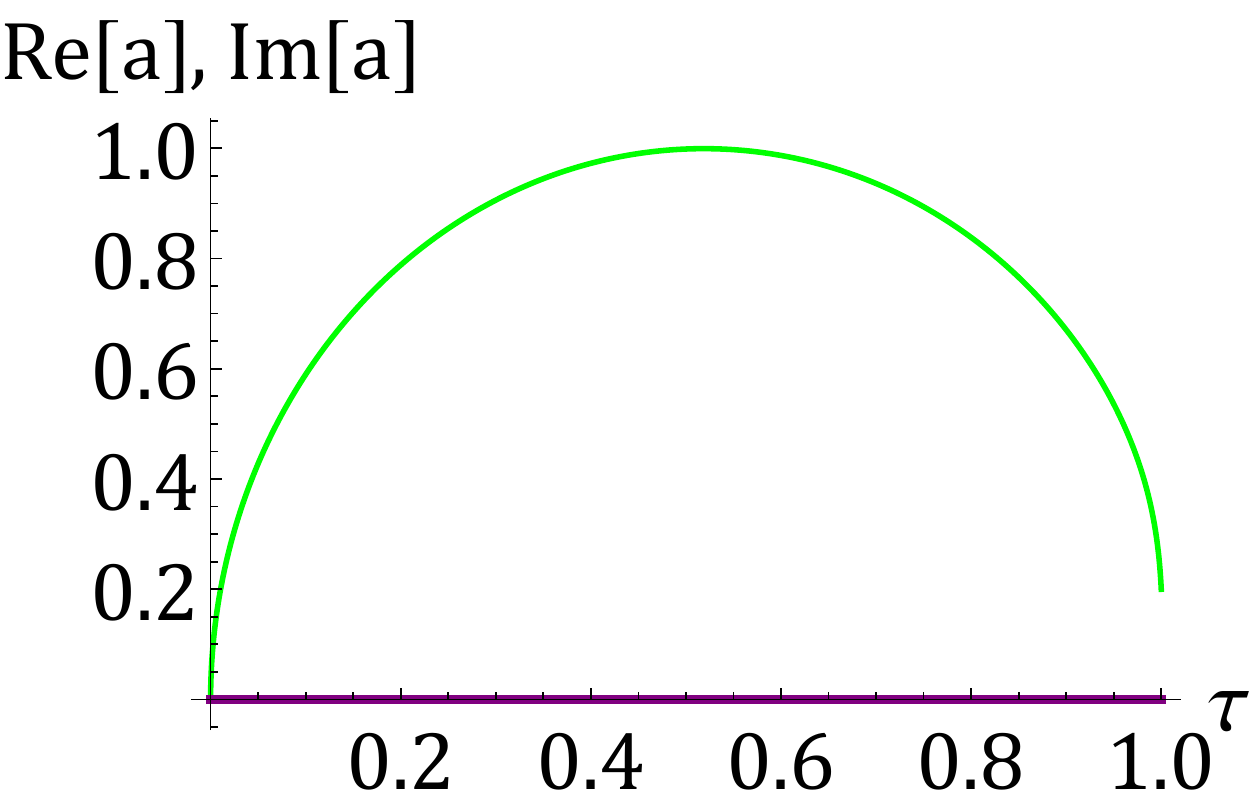}
		\includegraphics[width=0.49\textwidth]{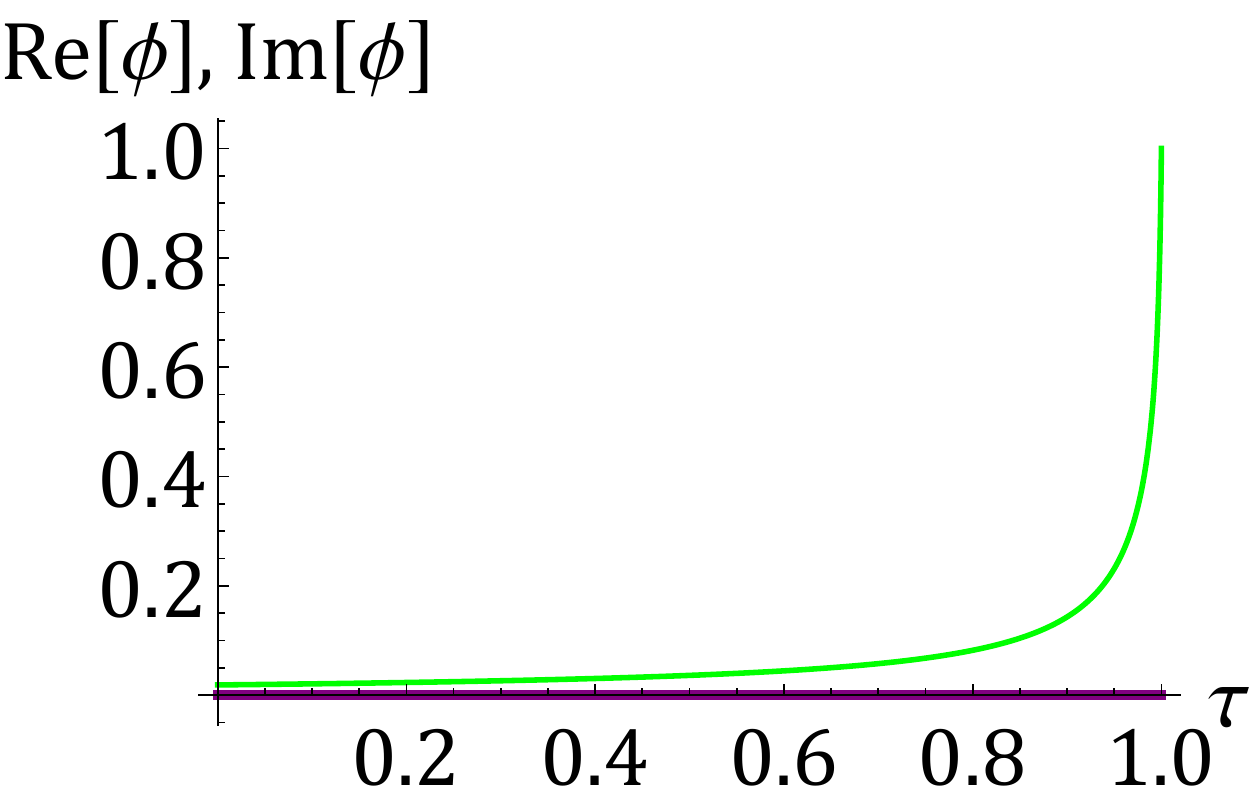}\caption{Closed Euclidean geometry $N_-[\gamma_-]$.}\label{fig:geom_a_imN1phi1}
	\end{subfigure}
	\begin{subfigure}{0.49\linewidth}
		\includegraphics[width=0.49\textwidth]{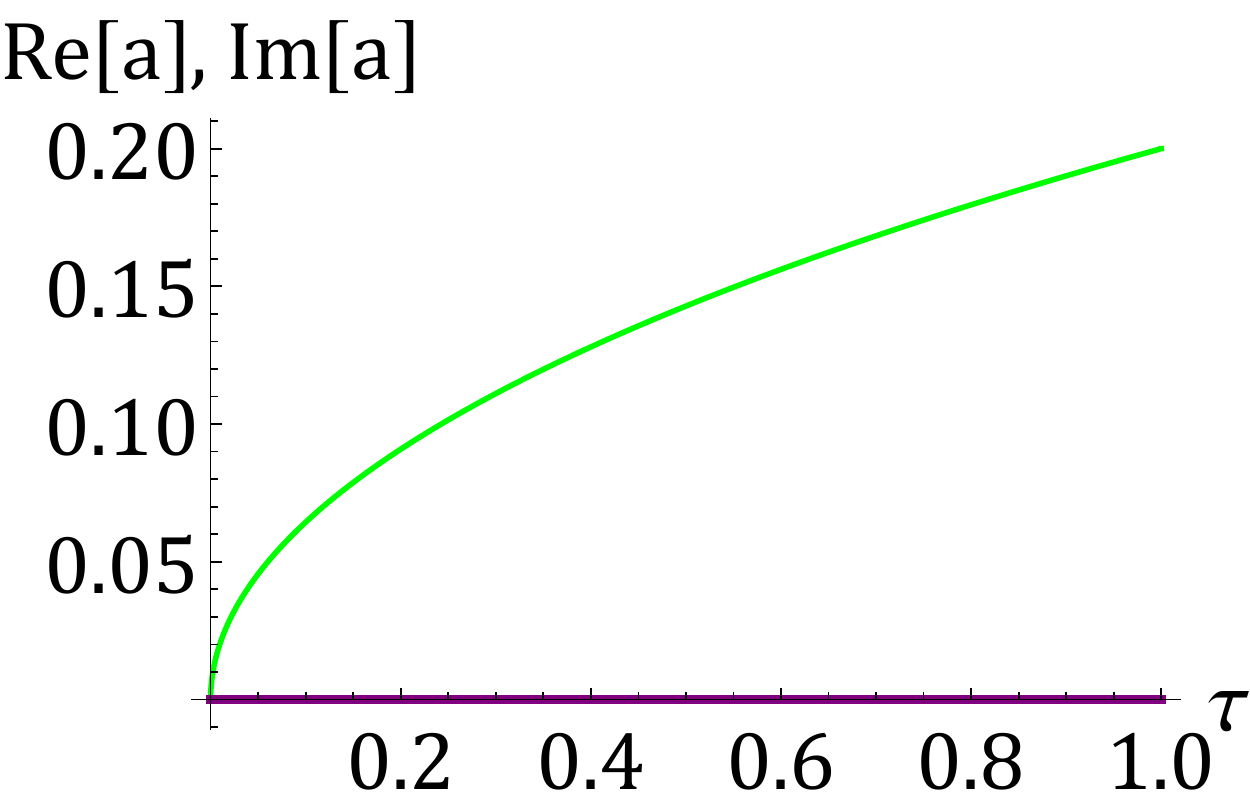}
		\includegraphics[width=0.49\textwidth]{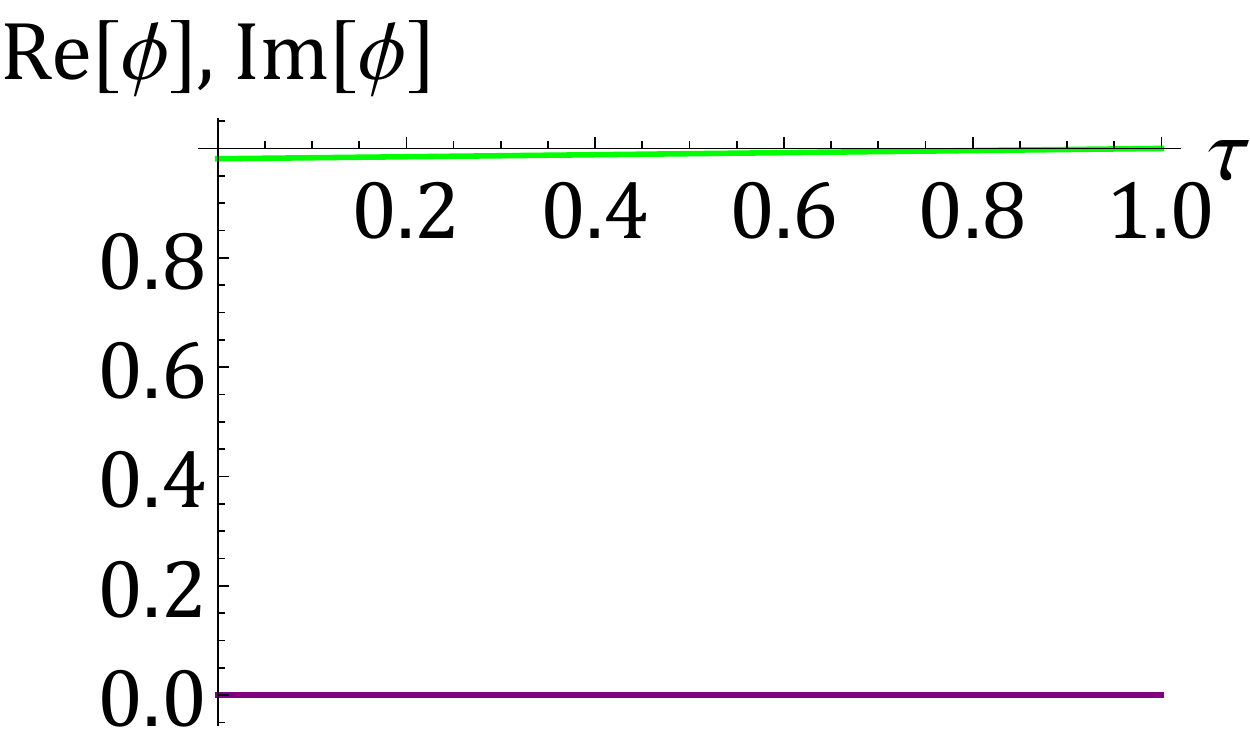}\caption{Closed Euclidean geometry $N_+[\gamma_+]$.}\label{fig:geom_a_imN2phi2}
	\end{subfigure}
	\begin{subfigure}{0.49\linewidth}
		\includegraphics[width=0.49\textwidth]{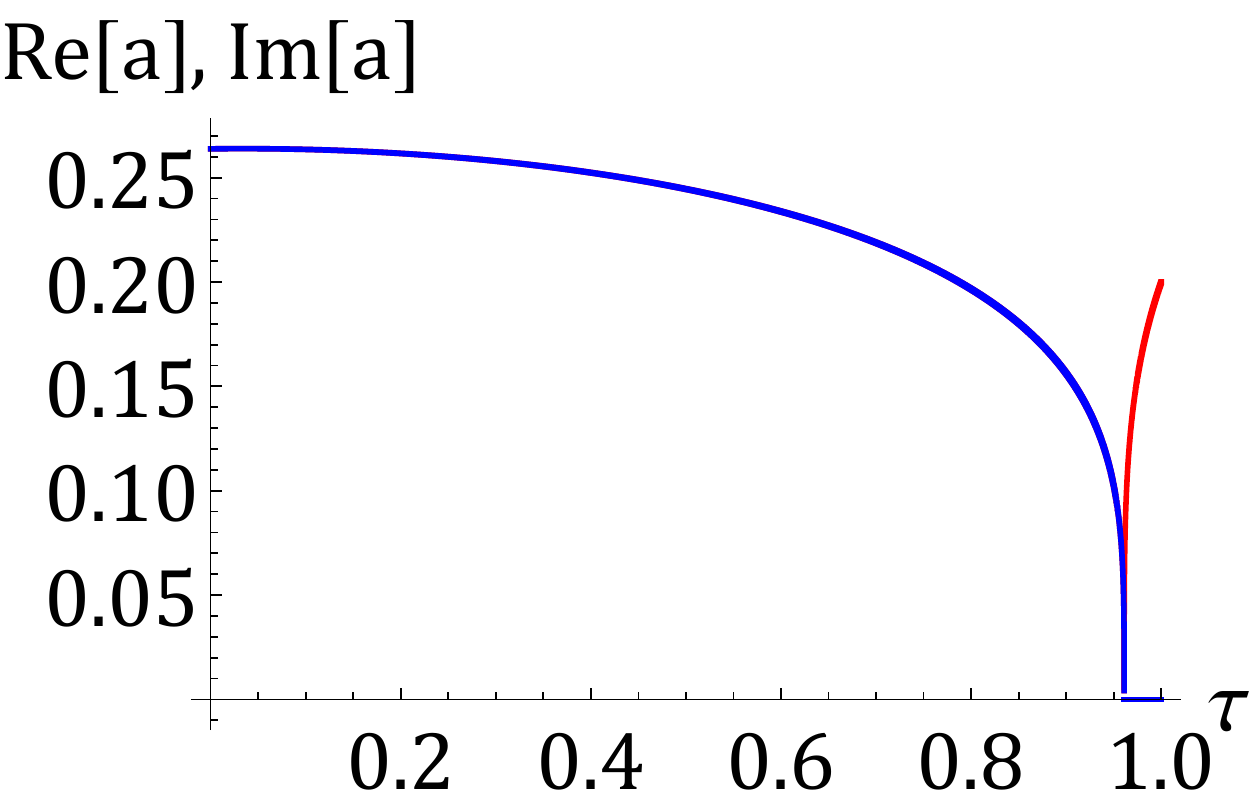}
		\includegraphics[width=0.49\textwidth]{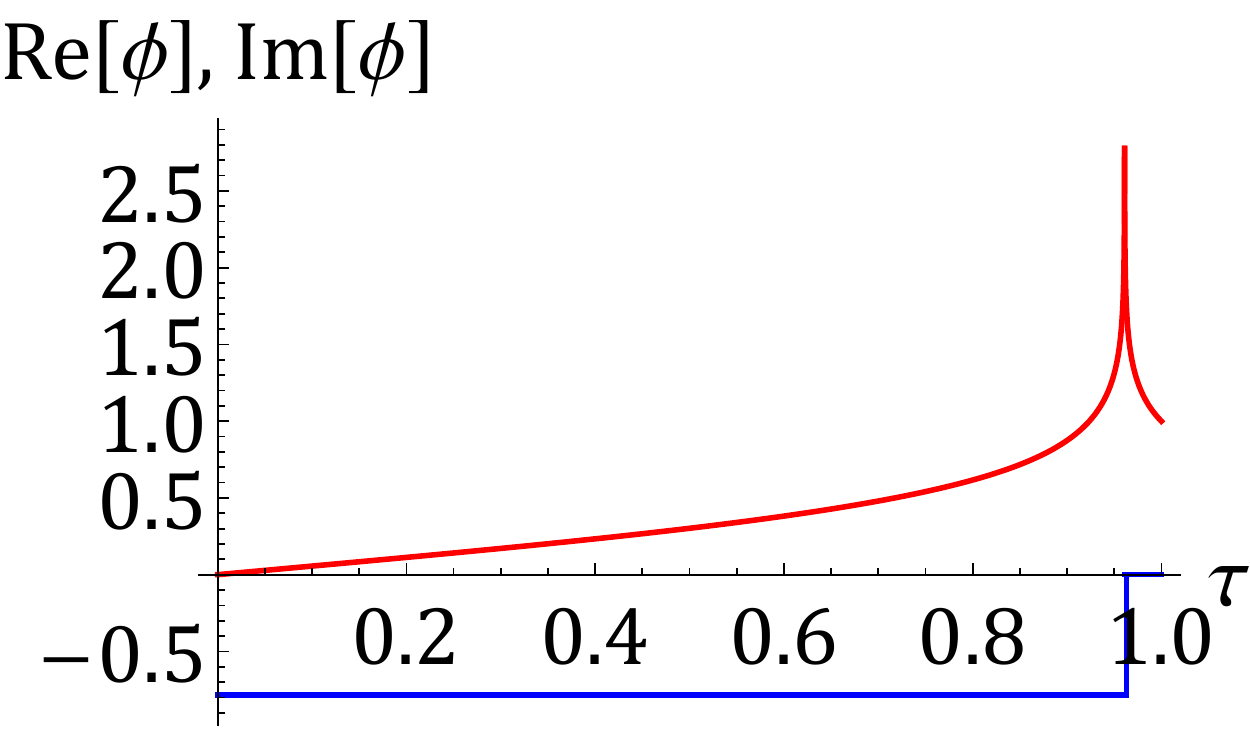}\caption{Unclosed Euclidean geometry $N_+[\gamma_-]$.}\label{fig:geom_a_imN2phi1}
	\end{subfigure}
	\begin{subfigure}{0.49\linewidth}
		\includegraphics[width=0.49\textwidth]{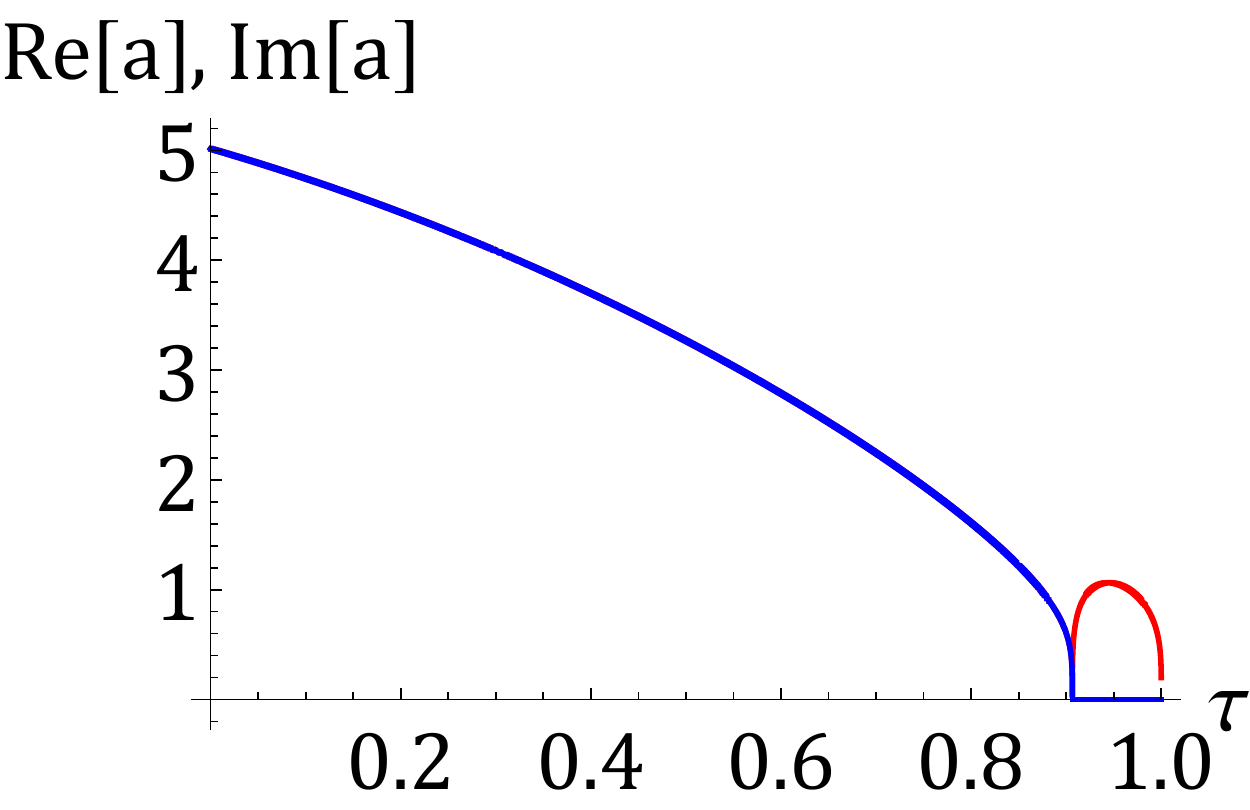}
		\includegraphics[width=0.49\textwidth]{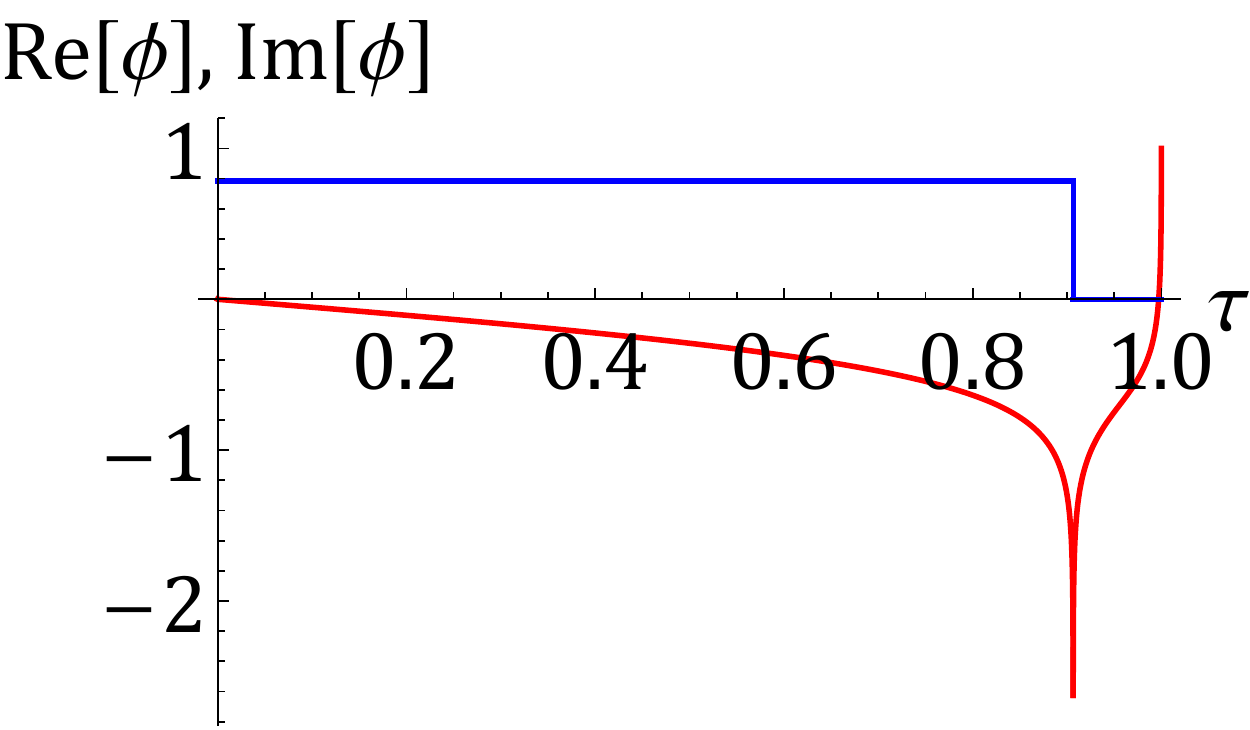}\caption{Unclosed Euclidean geometry $N_-[\gamma_+]$.}\label{fig:geom_a_imN1phi2}
	\end{subfigure}
	\caption{\small A numerical example of the field evolutions in region II. Here we have chosen $\alpha=1, \beta=0, \af=0.2$, $\phif=1$. For this example $\gamma_-=0.0189$, $\gamma_+=0.981 $, $N_-[\gamma_-]=-1.92 i$, $N_+[\gamma_-]=-0.0782 i$, $N_+[\gamma_+]=-0.0208 i$ and $N_-[\gamma_+]=-7.24 i$. We find two closed saddle point geometries -- $N_+[\gamma_+]$ and $N_-[\gamma_-]$ -- and two unclosed geometries -- $N_+[\gamma_-]$ and $N_-[\gamma_+]$. Real parts in green for closed saddle points, red for unclosed ones. Imaginary parts in purple for closed saddle points and blue for unclosed ones. Note that the imaginary part of the scalar typically jumps when the scale factor passes through zero, as the imaginary part of the scalar itself also changes there, given that it is obtained by taking a fourth root $a=(x^2-y^2)^{1/4}.$}\label{fig:imsaddlealpha1geom}
\end{figure}

We will now look at the saddle point solutions in a little more detail, see Fig.~\ref{fig:imsaddlealpha1geom} for a numerical example. For the closed saddle point solutions, $\bar{x}$ and $\bar{y}$ both start from $0$ when $\tau=0$. $\bar{x}$ describes a downward-facing parabola (because $N^2$ is negative), while $\bar{y}$ is a straight line. The gradient of $\bar{x}$ in $\tau=0$ is always bigger than that of $\bar{y}$:
\begin{equation}
	\abs{\frac{\dot{\bar{y}}}{\dot{\bar{x}}}}_0=\abs{\tanh(2\gamma)}<1\,;
\end{equation}
and finally at the end point of the trajectory $\tau=1$, the value of $\bar{x}$ is always above the value of $\bar{y}$ (since $\cosh(2\phi_f) > \sinh(2\phi_f)$). Hence $\bar{x}$ is always larger than $\bar{y}$, and $\bar{a}$ will be real for the whole trajectory. So, the geometry of the closed saddle points is purely Euclidean. For the unclosed saddle points, only $\bar{x}$ starts at $0$, so $\bar{x}^2-\bar{y}^2$ will start with a negative value at $\tau=0$. Because the end point value of $\bar{x}$ is always bigger than the end point of $\bar{y}$, this means that $\bar{x}^2-\bar{y}^2$ will be positive at $\tau=1$. Therefore by the mean value principle, there must be a $\tau_0\in(0,1)$ such that $\bar{a}(\tau_0)=0$: these unclosed geometries exhibit a singular bounce in the course of their evolution. These solutions are thus unphysical, because fluctuations will blow up at this singularity, rendering the action infinite. 

\begin{figure}
	\begin{subfigure}{0.45\linewidth}
		\includegraphics[width=7cm]{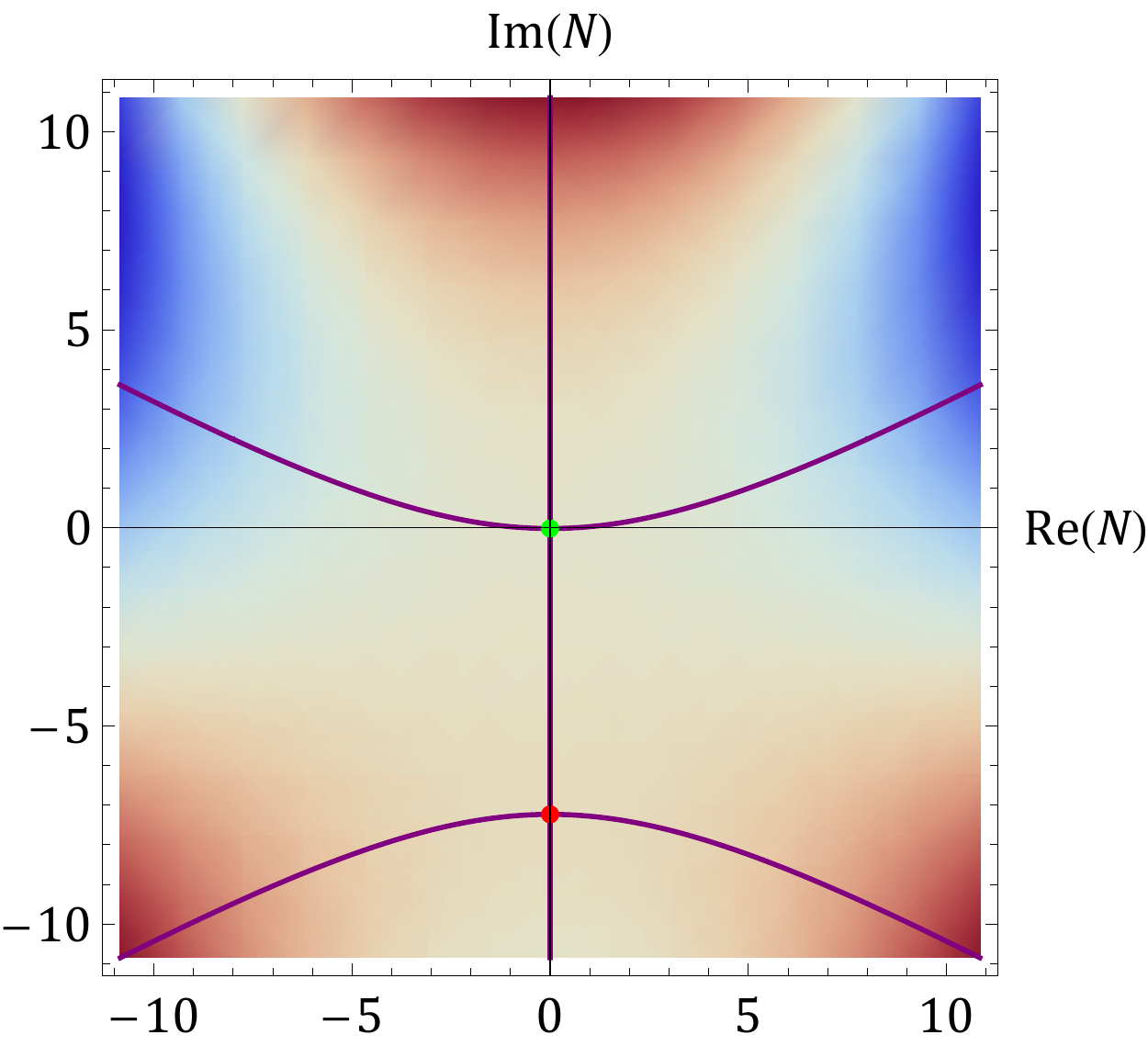}\caption{\small$N_-[\gamma_-]$ (green dot) and $N_+[\gamma_-]$ (red dot).}\label{fig:lapseint_im_gammam}
	\end{subfigure}\hspace{1cm}
	\begin{subfigure}{0.45\linewidth}
		\includegraphics[width=7cm]{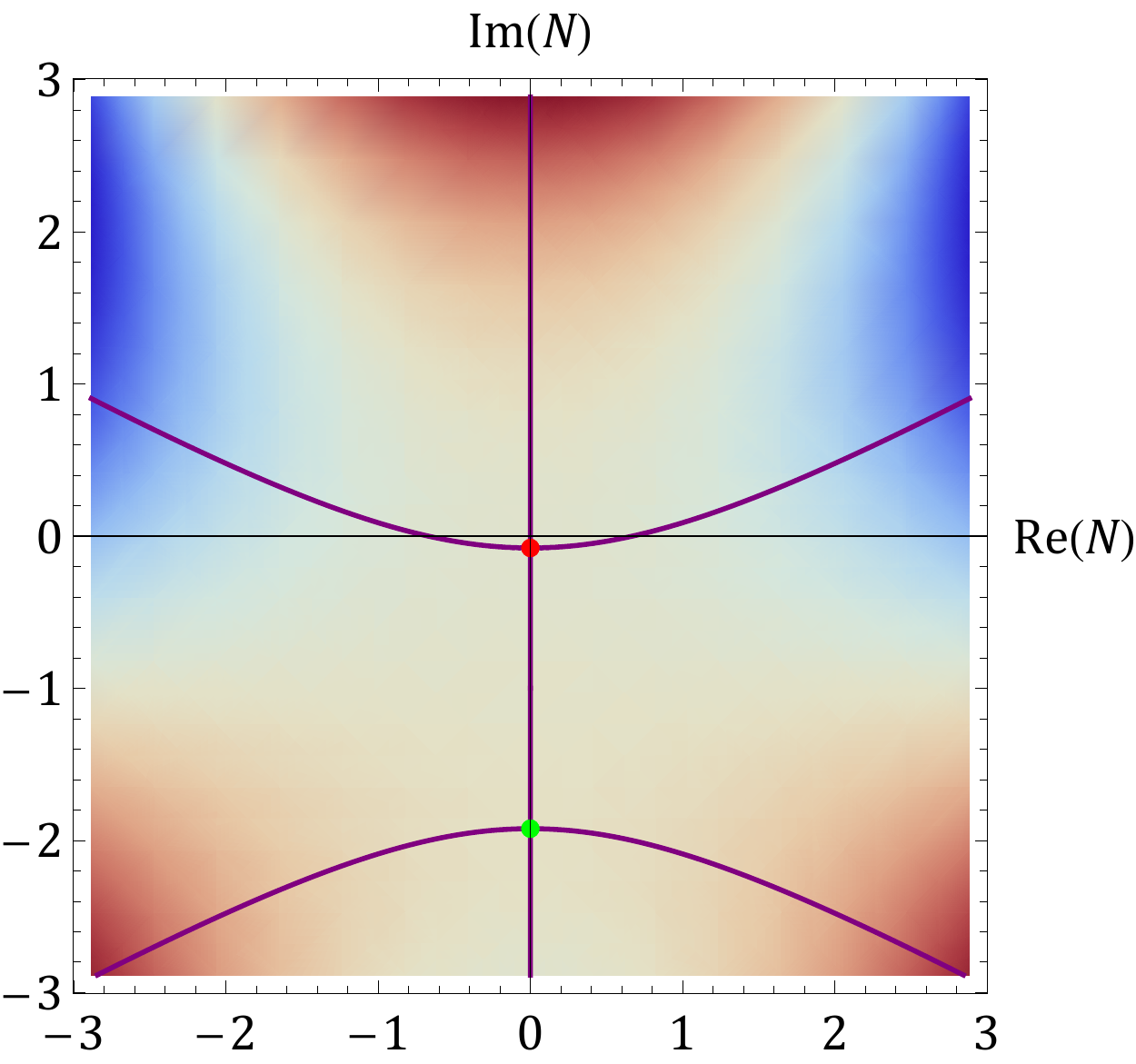}\caption{\small $N_+[\gamma_+]$ (green dot) and $N_-[\gamma_+]$ (red dot).}\label{fig:lapseint_im2}
	\end{subfigure}
	\caption{\small Case $\alpha=1,\ \beta=0$, $\af=0.2$, $\phif=1,$ with $N$ imaginary. Density plot of the weighting $\Re[iS_\text{saddle}^\text{on-shell}]$ and flow lines in the complex N plane.}\label{fig:imsaddlealpha1}
\end{figure}

To examine which saddle points contribute to the path integral we must study the thimbles in the complexified $N$ plane. A representative numerical example is shown in Fig.~\ref{fig:imsaddlealpha1}, with final values $\af=0.2$ and $\phif=1$. In the figure the closed saddle points are marked with green dots. The interesting feature is that their locations relative to the unclosed saddles are reversed for the two values of $\gamma.$ Hence, with the same contour of integration, we will necessarily get a contribution both from a closed and an unclosed saddle. If we stick to the same contour of integration as in region I, i.e.~a line parallel to the real $N$ axis, then for $\gamma_-$ the unclosed saddle will contribute, while for $\gamma_+$ the relevant saddle will be the closed one,
\begin{equation}
	\Psi[\xf,\yf] \approx e^{\frac{i}{\hbar} S_\text{on-shell}[N_+[\gamma_-]]} + \, e^{\frac{i}{\hbar} S_\text{on-shell}[N_+[\gamma_+]]}\,.
\end{equation}
As explained above, the unclosed saddles contain a singular bounce and thus, once perturbations are added, their action will in fact diverge. Thus such saddle points essentially remove themselves from the wave function. We should note that in \cite{Garay:1990re}, the point of view was taken that saddle points containing a singular bounce should be allowed, as one can deform the ``path'' taken in the complexified time plane as long as the end points remain unchanged, i.e.~as long as the evolution starts at $\tau=0$ and ends at $\int_0^1 N d\tau = N.$ Such a deformed evolution could then circumvent the singularity and render the field evolution regular. In view of Cauchy's theorem, nothing speaks against performing such deformations as long as the deformed path does not cross any singularities. Here, however, the original path actually does contain a singularity, at which all perturbations and matter configurations blow up -- cf. for instance the scalar field evolution shown in Fig.~\ref{fig:geom_a_imN2phi1}. Thus Cauchy's theorem cannot be applied and we are led to disregard saddle point geometries that contain spacetime singularities. Note that we take the constant lapse version of the saddle point geometry, rather than one of the deformed ones, as fundamental, given that the path integral is evaluated precisely in the gauge where $N$ is constant \cite{Teitelboim:1981ua}. We acknowledge that this point of view is debatable, and that this is an issue that deserves a deeper study in the future.

In light of the preceding discussion, with a single relevant saddle point remaining, the wave function may be approximated as
\begin{align}
\Psi & \approx e^{\frac{i}{\hbar}S(N_+[\gamma_+])}\,, \\ S(N_+[\gamma_+]) & = -\frac{i}{3}\left[ 1-(1-a_f^2)^{3/2}\right] + \frac{i}{2} \left[ 1-(1-a_f^2)^{1/2}\right]  a_f^2 \phi_f^2 + {\cal O}(\phi_f^4)\,.
\end{align}
The action is pure amplitude: it grows from zero at $a_f=0$ until the boundary of the domain $a_f^2 \cosh(2\phi_f)<1$ is reached. As the universe grows further, a phase will develop, as described previously for region I.

It is noteworthy that of the two closed saddle points, it is the one whose geometry approximately corresponds to less than a hemisphere that is relevant to the path integral, rather than the one whose geometry is larger than half of an hemisphere, cf. also Fig.~\ref{fig:imsaddlealpha1geom}. This is in agreement with previous findings in the pure gravity case \cite{Lehners:2021jmv} and implies that when the scale factor $a_f$ approaches zero, the dominant geometry is the vanishing geometry, rather than that of a full sphere, which would have signalled a dominant contribution from a non-trivial topology.

\vspace{0.5 cm}	
{\noindent\it{Region III: real saddle points.}}

In this region, $a_f^2 \sinh^2(2\phi)>1,$ the imaginary part of $\gamma$ implies that $\cosh(2\gamma)$ and $\sinh(2\gamma)$ are purely imaginary. This means that the values of $\Pi_x$ and $\Pi_y$ as well as the classical solutions $\bar{x}$ and $\bar{y}$ are always purely real. Hence the saddle point values of the lapse \eqref{eq:Nsaddlealpha1} are also purely real, and the on-shell action \eqref{eq:actiononshellN} evaluated on these saddle points as well.
Therefore whatever the relevant saddle points are, the wavefunction that we will get from the lapse integral will be of the form:
\begin{equation}
	\Psi[\xf,\yf]=\frac{8\hbar}{\pi}\sum e^{i\cdot\text{real number}}\,;
\end{equation}
which means that all the different saddle points have the same weighting, so this does not describe a true quantum state, but rather a collection of classical evolutions. However, we will now show that the classical solutions found in this region are not physical, because they exhibit a singular bounce in the course of their evolution. The $\bar{x}$ solution describes an upward facing parabola, while $\bar{y}$ is still a straight line.
For the closed saddle points, we start with $\bar{x}(0)=\bar{y}(0)=0$, and the initial gradients are
\begin{equation}
	\abs{\frac{\dot{\bar{y}}}{\dot{\bar{x}}}}_0=\abs{\coth(\Re[\gamma])}>1\,;
\end{equation}
therefore, since at the end point we still have $\bar{x}(1)>\bar{y}(1)$, there will be a $\tau_0$ such that the $x$ and $y$ curve cross, and we find $a(\tau_0)=0$. Closed geometries therefore contain a singular bounce, at which perturbations will blow up, and are therefore excluded as potential saddle points. For unclosed saddle points, we start from $x(0)^2-y(0)^2<0$ and we go to $x(1)^2-y(1)^2>0$, so again their must be a point $\tau_0\in(0,1)$ where the geometry vanishes, $a(\tau_0)=0$, and we find another singular bounce.

In this region, we therefore do not find any physically acceptable solutions, but this is not a contradiction because this region is by definition not continuously connected to the pure gravity case $\phif\to0$, so it just means that there exists a bound on the values of $\phif$ that lead to classical spacetime, and this bound is $\af^2\sinh[2](\phif)<1$.

%%%%%%%%%%%%%%%%%%%%%%%%%%%%%%%%%%%%%%%%%%%%%%%%%%%%%%%%%%%%

\subsection{$\alpha=-1\,,\ \beta=0$: $V(\phi)=-\cosh(2\phi)$.} \label{sec:ads}

The negative $\cosh(2\phi)$ potential is interesting to study, as potentials of this kind appear rather naturally in string compactifications (though the potential would then be expected to be bounded below, i.e.~it would turn upwards at large $|\phi|$). At the maximum of the potential, where the scalar field vanishes, we expect to recover the pure Anti-de Sitter case, studied in \cite{DiTucci:2020weq,Lehners:2021jmv}. In these works it was found that the AdS path integral is formally completely analogous to a no-boundary path integral, including the requirement for a regularity condition in the interior of the geometry. 

\begin{figure}
	\centering
	\begin{subfigure}{0.49\linewidth}
		\includegraphics[width=0.49\textwidth]{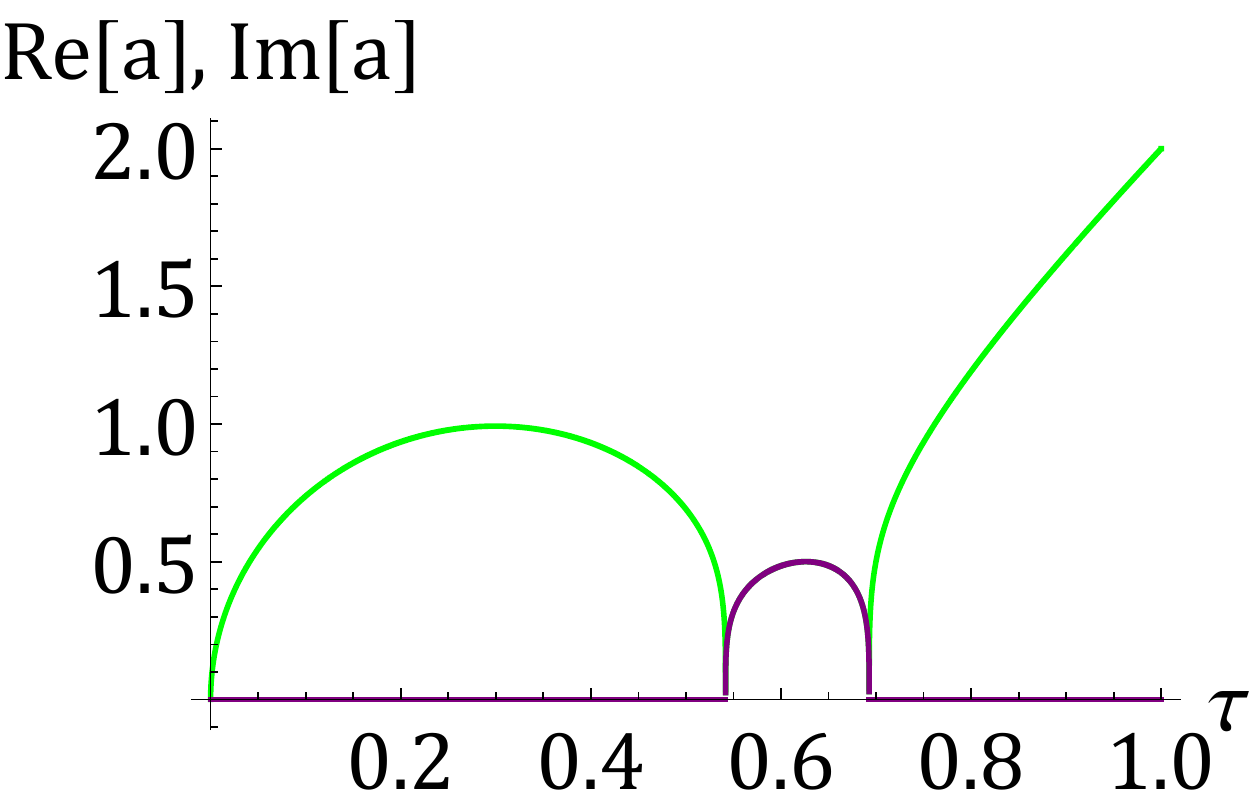}
		\includegraphics[width=0.49\textwidth]{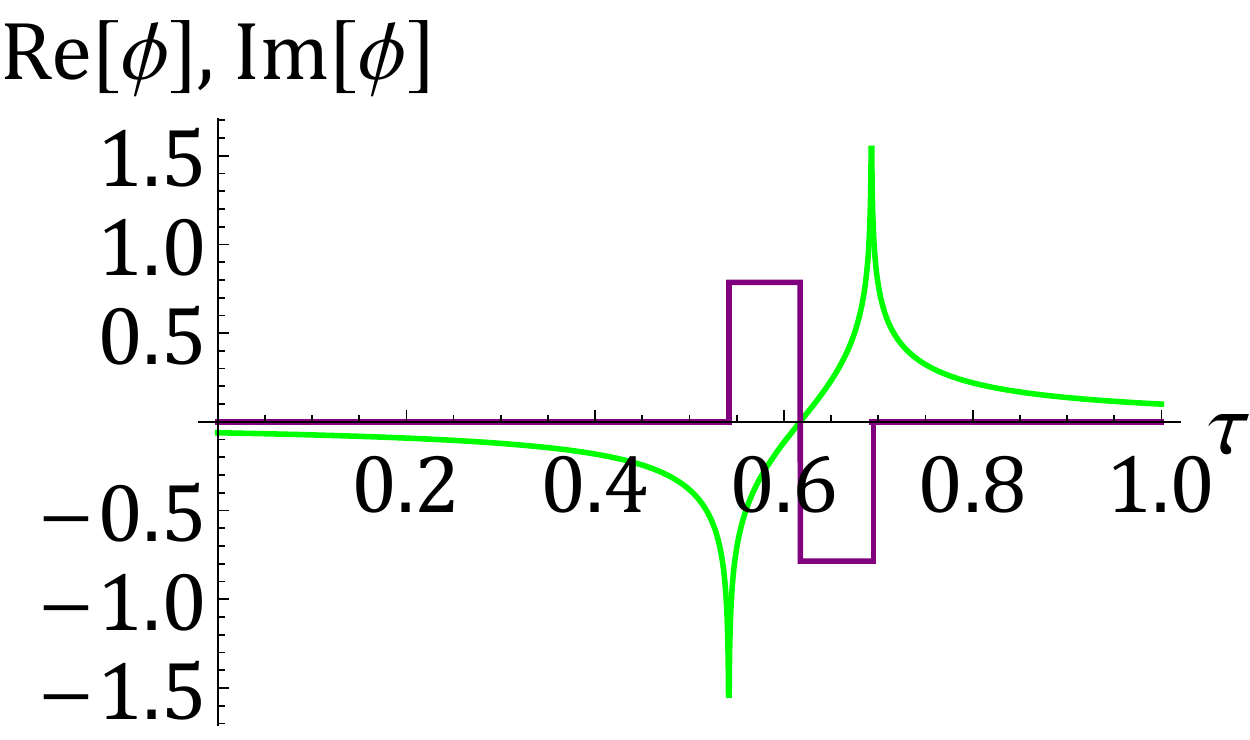}\caption{Closed Euclidean geometry $N_+[\gamma^\text{I}_-]$.}\label{fig:alpham1_geom_a_imN1phi1}
	\end{subfigure}
	\begin{subfigure}{0.49\linewidth}
		\includegraphics[width=0.49\textwidth]{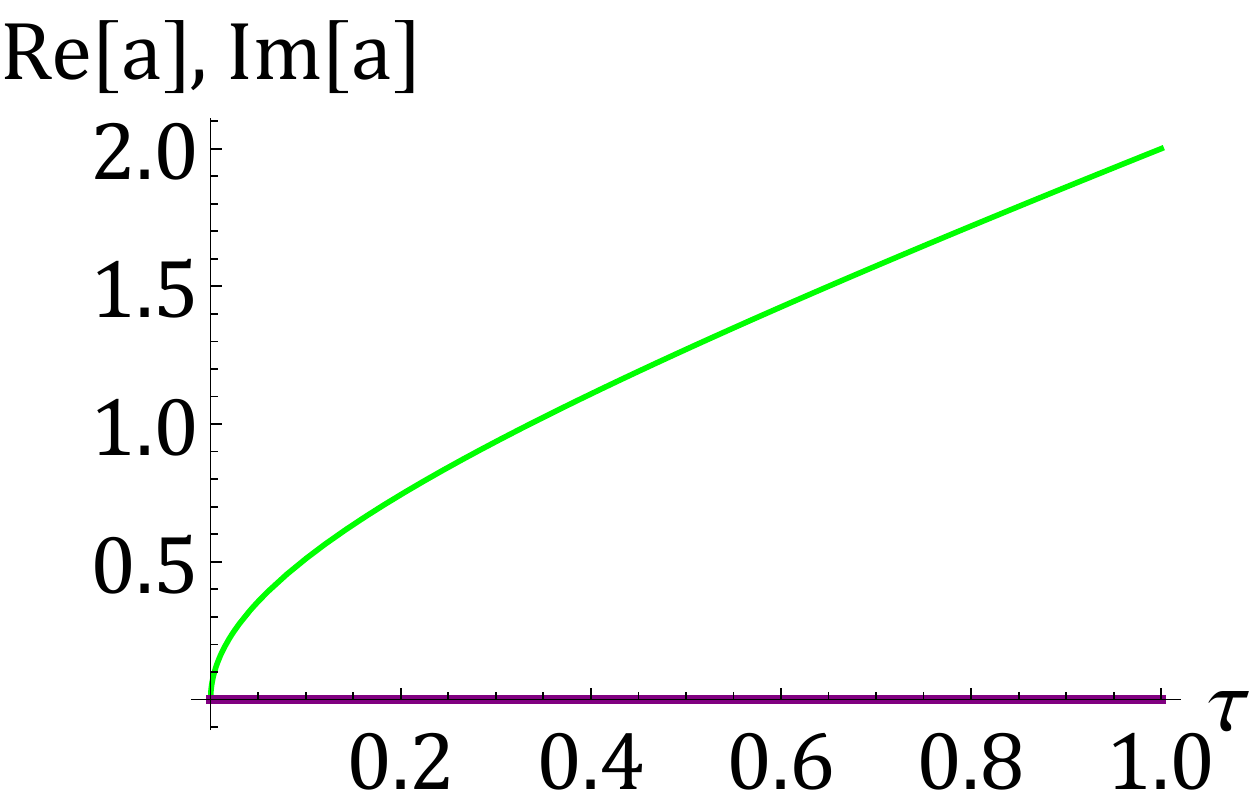}
		\includegraphics[width=0.49\textwidth]{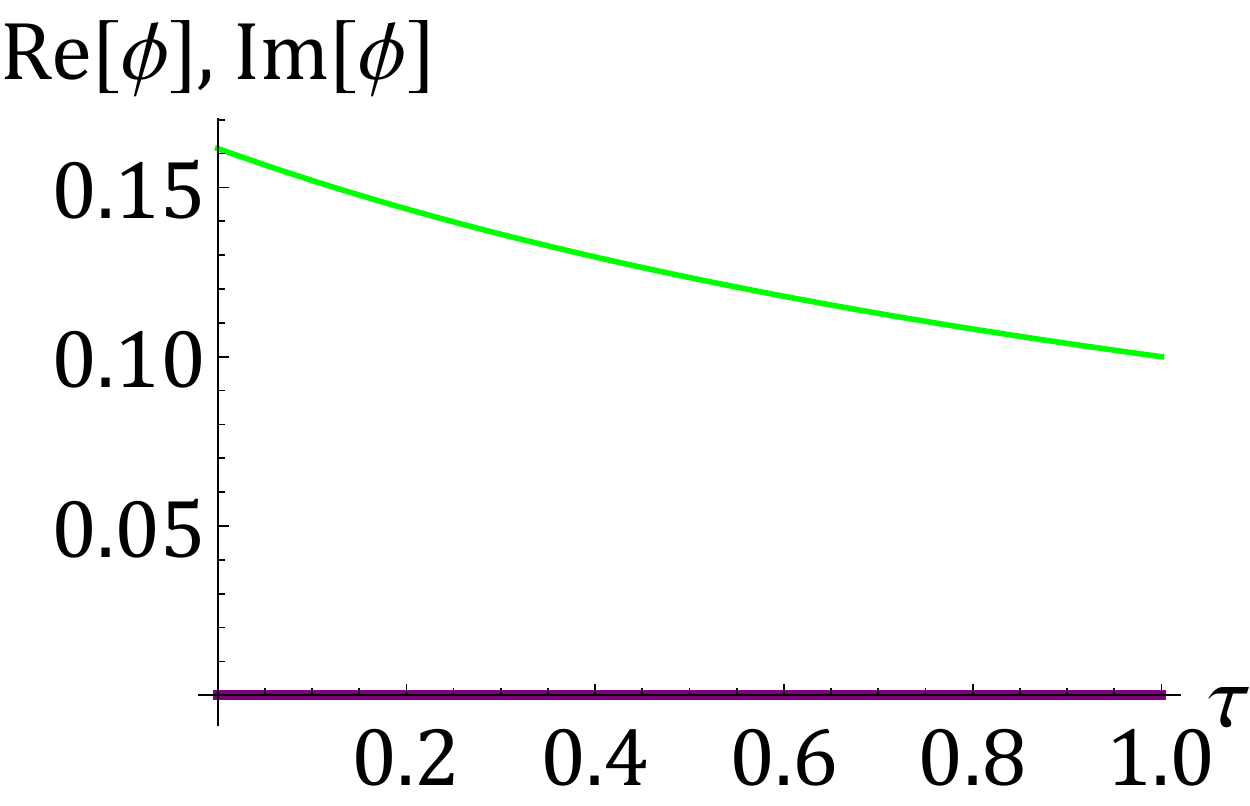}\caption{Closed Euclidean geometry $N_-[\gamma^\text{II}_+]$.}\label{fig:alpham1_geom_a_imN2phi2}
	\end{subfigure}
	\begin{subfigure}{0.49\linewidth}
		\includegraphics[width=0.49\textwidth]{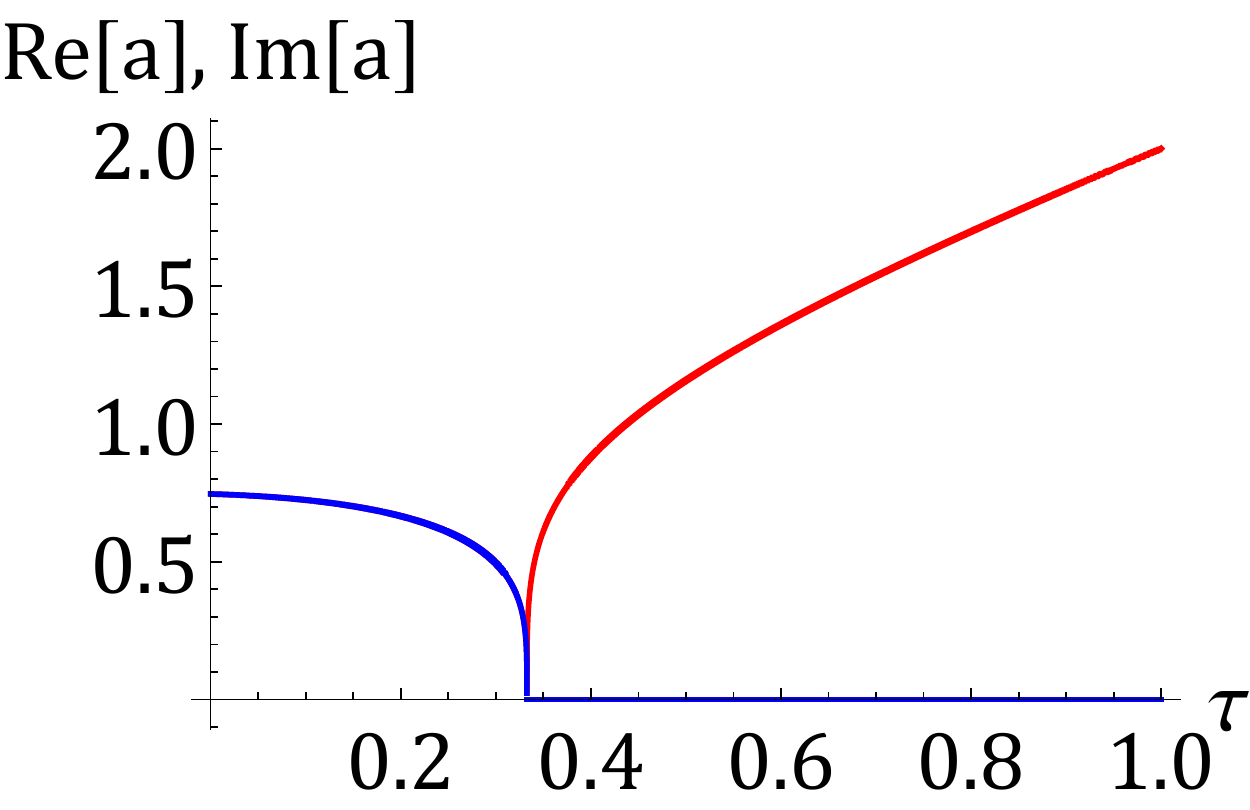}
		\includegraphics[width=0.49\textwidth]{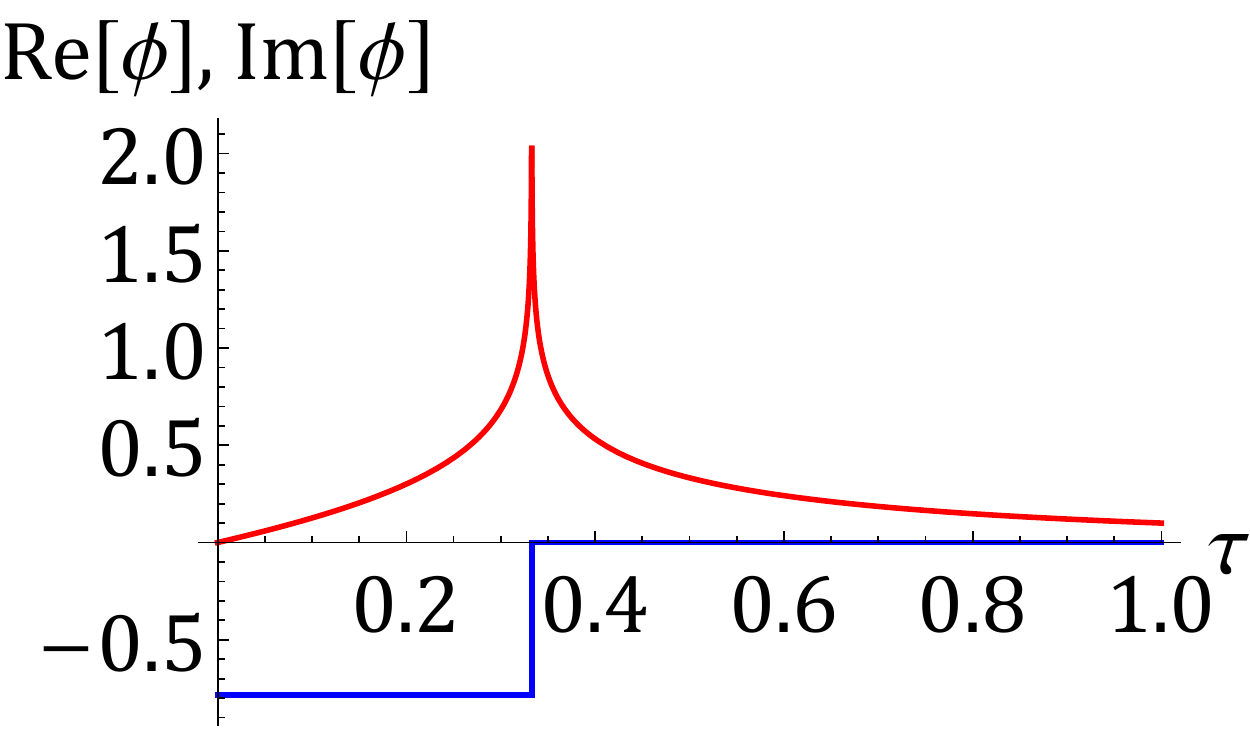}\caption{Unclosed Euclidean geometry $N_-[\gamma^\text{I}_-]$.}\label{fig:alpham1_geom_a_imN2phi1}
	\end{subfigure}
	\begin{subfigure}{0.49\linewidth}
		\includegraphics[width=0.49\textwidth]{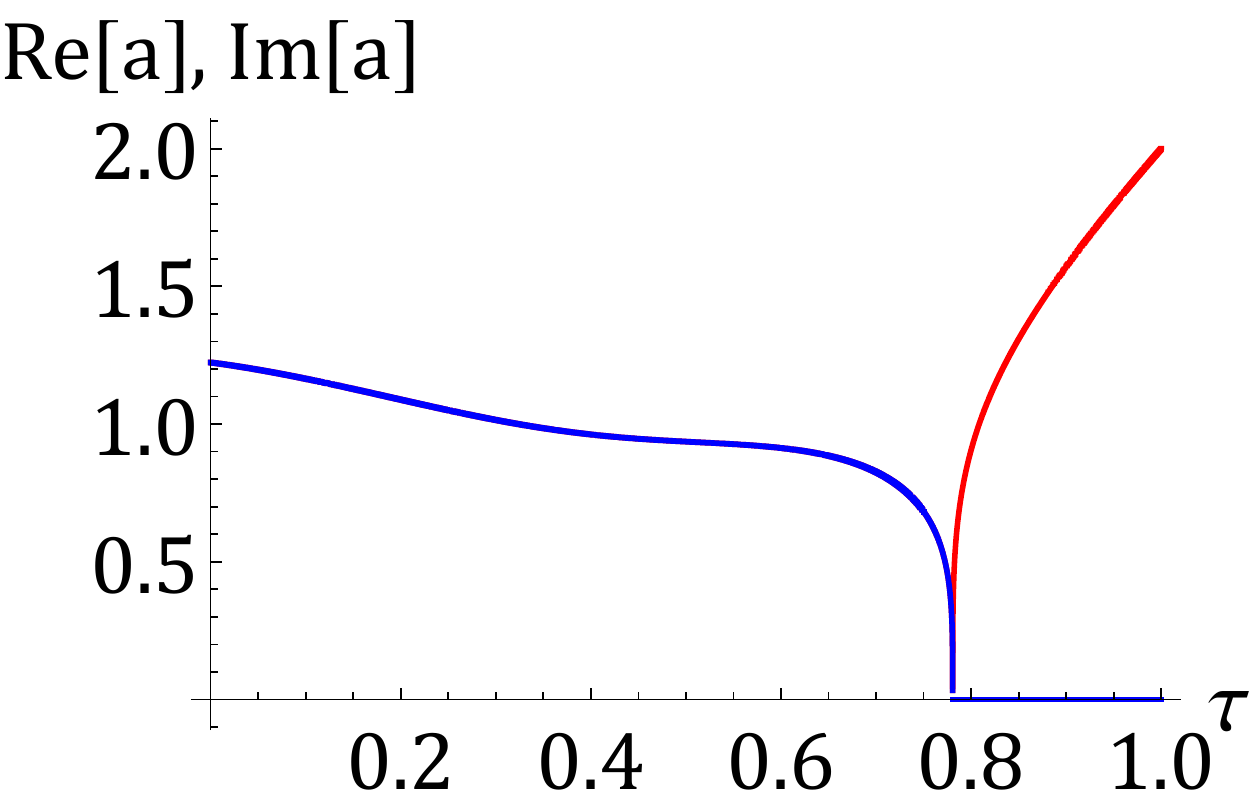}
		\includegraphics[width=0.49\textwidth]{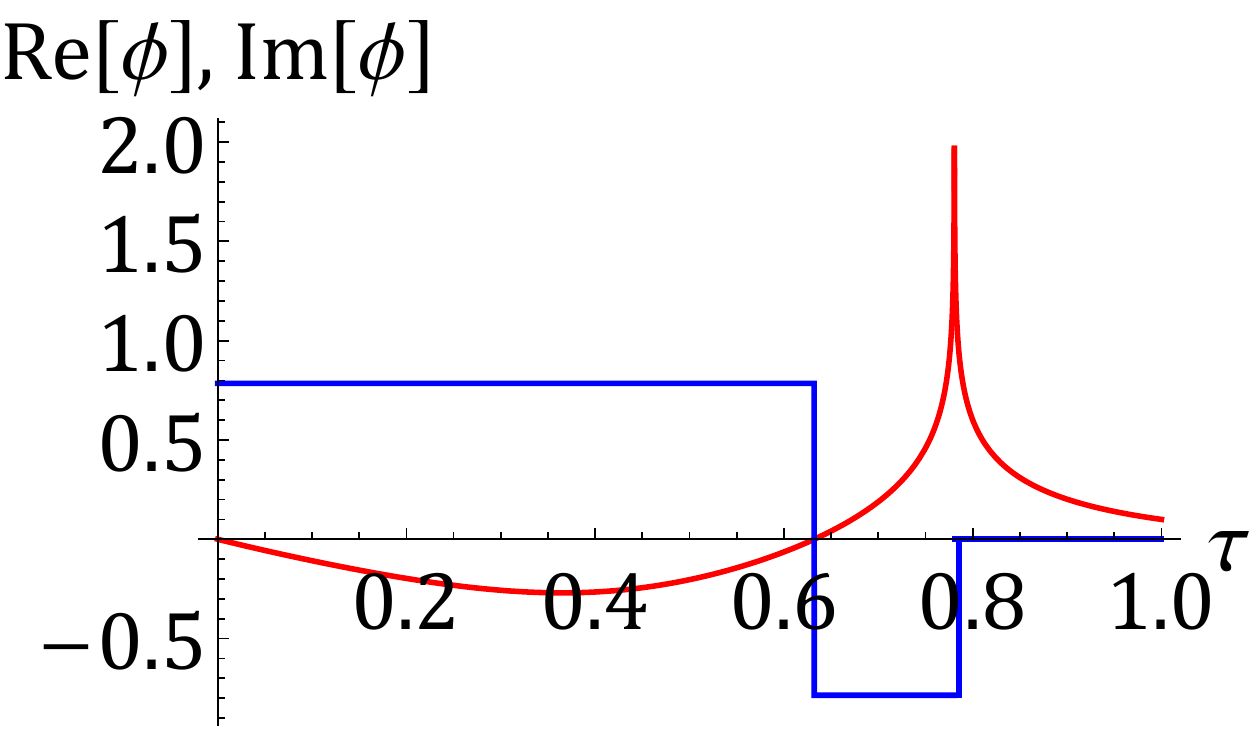}\caption{Unclosed Euclidean geometry $N_+[\gamma^\text{II}_+]$.}\label{fig:alpham1_geom_a_imN1phi2}
	\end{subfigure}
	\caption{\small Case $\alpha=-1,\ \beta=0$, $\af=2$, $\phif=0.1,$ with $N$ imaginary. The relevant numerical values are$ \gamma^\text{I}_-=-0.0615\,, \, N_+[\gamma^\text{I}_-]=3.26 i\,,\, N_-[\gamma^\text{I}_-]=-1.25 i\,; \, \gamma^\text{II}_+=0.162\,,\, N_-[\gamma^\text{II}_+]=-1.23 i\,,\, N_+[\gamma^\text{II}_+]=3.33 i\,.$ Geometry associated to the different saddle points. Same conventions as in Fig.~\ref{fig:complexsaddle1}.}\label{fig:imsaddlealpham1geom}. 
\end{figure}

Here, with a scalar included, we now have saddle points at
\begin{equation}
	N^\text{saddle}_{\pm}=i\cosh(2\gamma)\pm\sqrt{-\cosh[2](2\gamma)-\af^2\cosh(2\phif)}\,.\label{eq:saddleexpalpham1}
\end{equation} 
The $\gamma$ values that lead to closed geometries are given by (see appendix \ref{appendix:gamma})
\begin{equation}
	\left\lbrace
	\begin{aligned}
		&\gamma^{\text{I}}_{\mp}=\mp\frac{1}{2}\cosh[-1](\sqrt{1+\frac{\sinh[2](2\phif)}{4}\left(2+\af^2\cosh(2\phif)-\sqrt{\left(2+\af^2\cosh(2\phif)\right)^2-\af^4}\right)})\,;\\
		&\gamma^{\text{II}}_{\mp}=\mp\frac{1}{2}\cosh[-1](\sqrt{1+\frac{\sinh[2](2\phif)}{4}\left(2+\af^2\cosh(2\phif)+\sqrt{\left(2+\af^2\cosh(2\phif)\right)^2-\af^4}\right)})\,.
	\end{aligned}
	\right.\label{eq:gammasolalpham1}
\end{equation}
The solutions $\gamma^{\text{I}}_-$ and $\gamma^{\text{II}}_+$ are defined for $\phif>0$, while $\gamma^{\text{I}}_+$ and $\gamma^{\text{II}}_-$ are defined for $\phif<0$. The solutions I close the saddle point $N_+$, and the solutions II close the saddle point $N_-$.
At each phase space point, i.e.~for each choice of final boundary conditions $(a_f,\phi_f),$ we get a total of four saddle point solutions, two closed ($N_+[\gamma^{\text{I}}]$ and $N_-[\gamma^{\text{II}}]$), and two unclosed ($N_+[\gamma^{\text{II}}]$ and $N_-[\gamma^{\text{I}}]$). It is straightforward to see that all $\gamma$ values are real, and thus the saddle points  \eqref{eq:saddleexpalpham1} are pure imaginary. Hence we may deduce from \eqref{eq:actiononshellN} that the on-shell action will also be purely imaginary: $e^{iS}$ is only an amplitude and does not contain any phase. Thus no classical spacetime will be implied. Examples of specific solutions are shown in Fig.~\ref{fig:imsaddlealpham1geom}. The two saddle points associated with $\gamma^\text{I}$, as well as the unclosed saddle associated to $\gamma^\text{II}$ have a singular bounce within their evolution, they are therefore suppressed by the blowing-up fluctuations. Using similar techniques as for the $\alpha=1$ case, one can show this is always the case. Therefore the only relevant solution is the closed saddle point geometry $N_-[\gamma_+^\text{II}]$. Note that for this saddle point, the scalar field is purely real valued, and in fact runs up the potential as the scale factor grows. This is characteristic of approximately Anti-de Sitter space, where even potential maxima with negative second derivatives are stable, as long as they satisfy the Breitenlohner-Freedman bound \cite{Breitenlohner:1982bm}, as our potential does (if one undoes the re-scaling \eqref{eq:rescaling}, then $V=-\cosh(\sqrt{\frac{2}{3}}\Phi)$ so that the effective mass is $m^2=-\frac{2}{3}>m^2_{BF}=-\frac{9}{4}$).

\begin{figure}
	\begin{subfigure}{0.45\linewidth}
		\includegraphics[width=7cm]{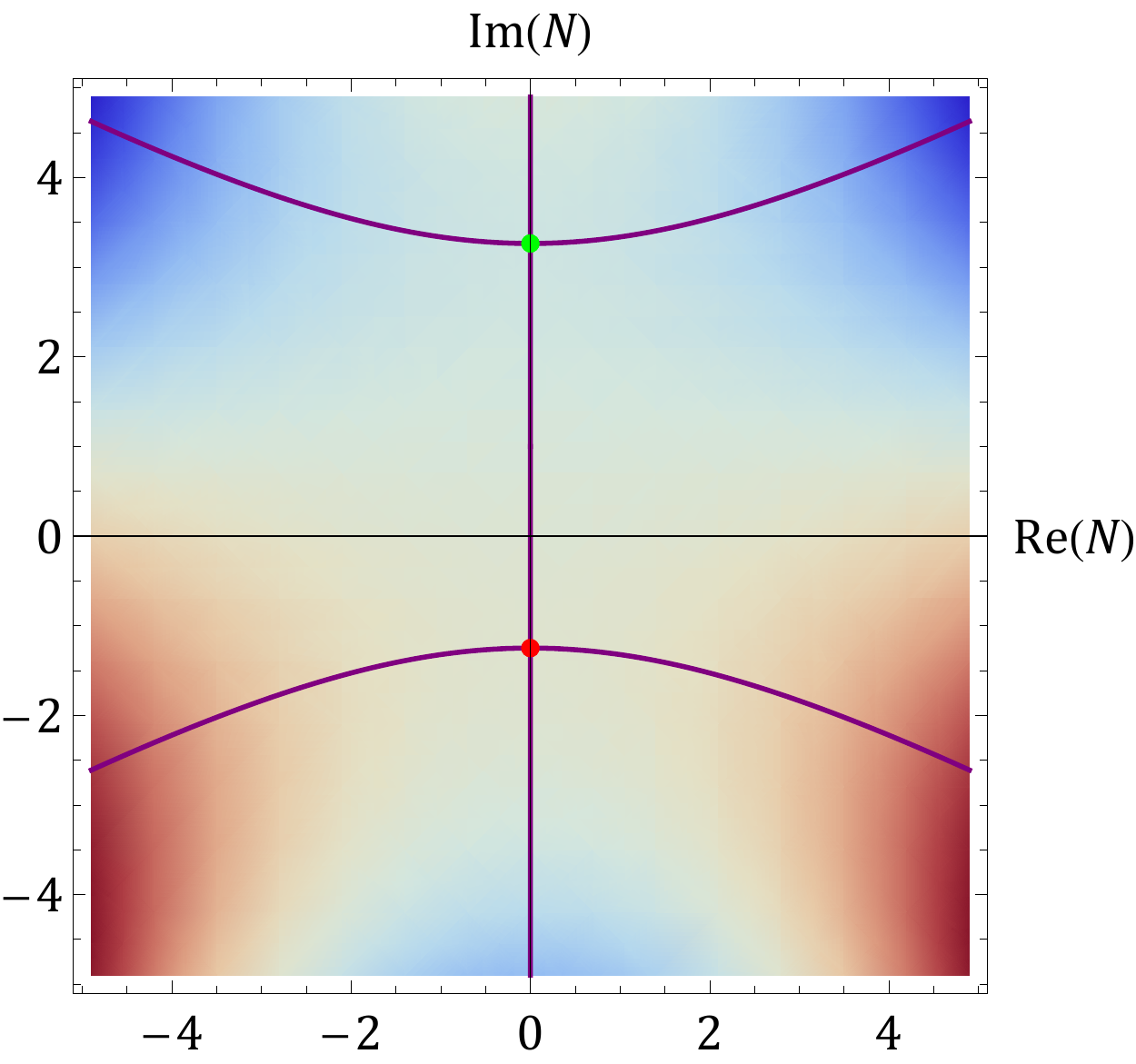}\caption{\small$N_+[\gamma^\text{I}_-]$ (green dot) and $N_-[\gamma^\text{I}_-]$ (red dot).}\label{fig:alpham1_lapseint_im_gammam}
	\end{subfigure}\hspace{1cm}
	\begin{subfigure}{0.45\linewidth}
		\includegraphics[width=7cm]{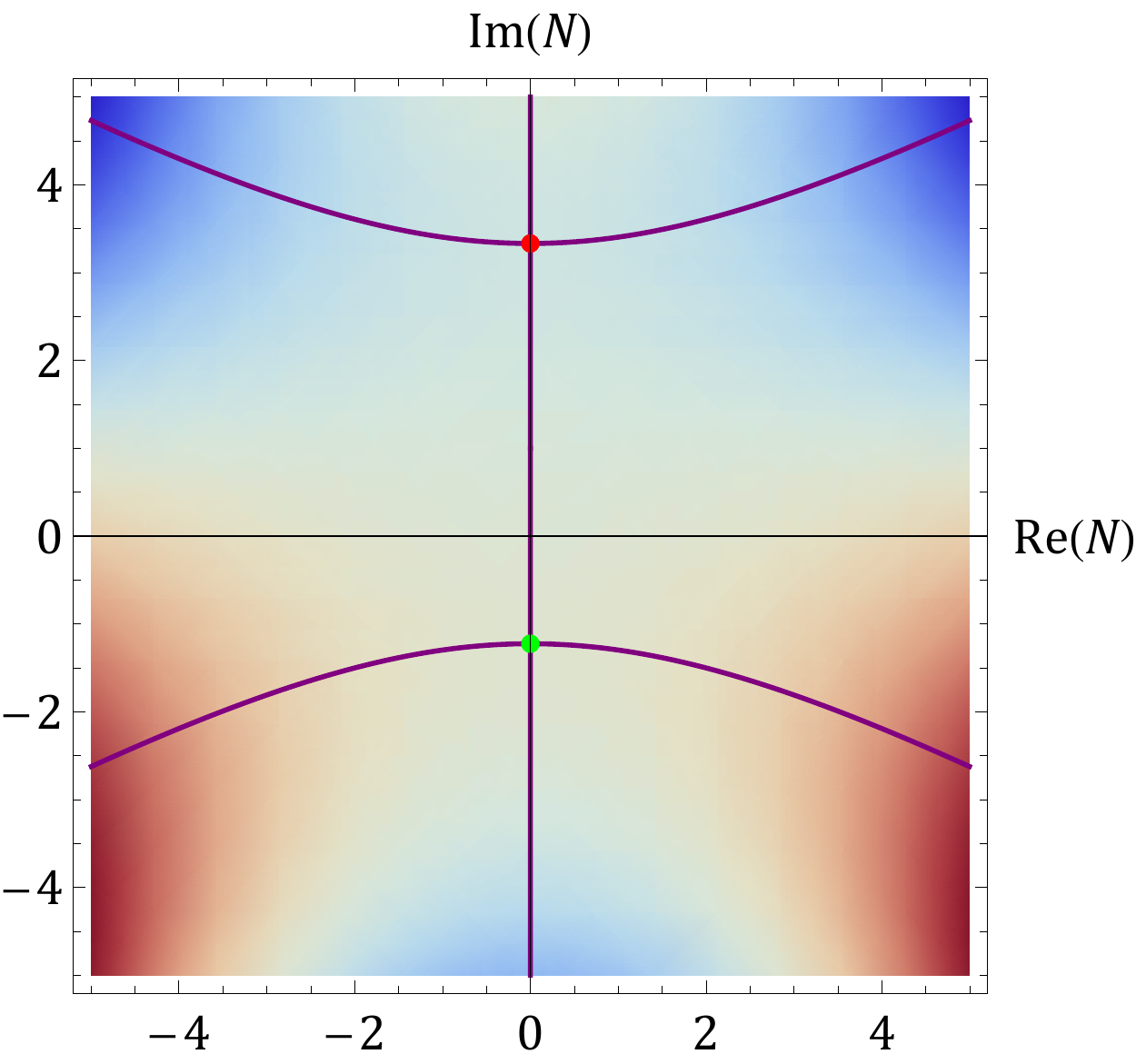}\caption{\small $N_-[\gamma^\text{II}_+]$ (green dot) and $N_+[\gamma^\text{II}_+]$ (red dot).}\label{fig:alpham1_lapseint_im2}
	\end{subfigure}
	\caption{\small Case $\alpha=-1,\ \beta=0$, $\af=2$, $\phif=1$. Density plot of the weighting $\Re[iS_\text{saddle}^\text{on-shell}]$ and flow lines in the complex N plane. Same conventions as in Fig.~\ref{fig:complexsaddle2}.} \label{fig:imsaddlealpham1}
\end{figure}

The structure of the flow lines is again most easily grasped with the help of a numerical example, see Fig.~\ref{fig:imsaddlealpham1}. The closed saddle point is selected by Picard-Lefschetz theory only if the defining contour of integration contains the lower part of the imaginary lapse axis. In fact, given that we expect the wave function to be real valued (especially in view of a potential CFT dual \cite{Maldacena:1997re}), one should define the path integral as a sum of two contours: the first from negative imaginary infinity up to the saddle point in the upper half plane and then on to either the left or right, following the steepest descent contour; plus a contribution from a contour that is reflected across the imaginary lapse axis. These two contours will add up two complex conjugate contributions, just as was described for the pure gravity case in \cite{DiTucci:2020weq,Lehners:2021jmv}. The wave function may then be estimated as
\begin{align}
\Psi & \approx e^{\frac{i}{\hbar}S(N_-[\gamma_+^\text{II}])}\,, \\ S(N_-[\gamma_+^\text{II}]) & =  -\frac{i}{3}\left[ (a_f^2+1)^{3/2} - 1\right] - \frac{i}{2} \left[(a_f^2 + 1)^{1/2} - 1\right]  a_f^2 \phi_f^2 + {\cal O}(\phi_f^4)\,.
\end{align}
For completeness, let us mention that in applications to AdS/CFT one would let the scale factor $a_f \to \infty$ and add the appropriate counter terms to remove the volume divergence in the wave function. 

We may conclude that the prescription to define the path integral with a momentum condition in the interior, rather than a Dirichlet condition, also works once a scalar field is included. 

%%%%%%%%%%%%%%%%%%%%%%%%%%%%%%%%%%%%%%%%%%%%%%%%%%%%%%%%%%%
%%%%%%%%%%%%%%%%%%%%%%%%%%%%%%%%%%%%%%%%%%%%%%%%%%%%%%%%%%%

	\subsection{$\alpha=\beta$: $V(\phi)=\alpha\exp(2\phi)$.}
	
\begin{figure}[h!]
	\includegraphics[width=8cm]{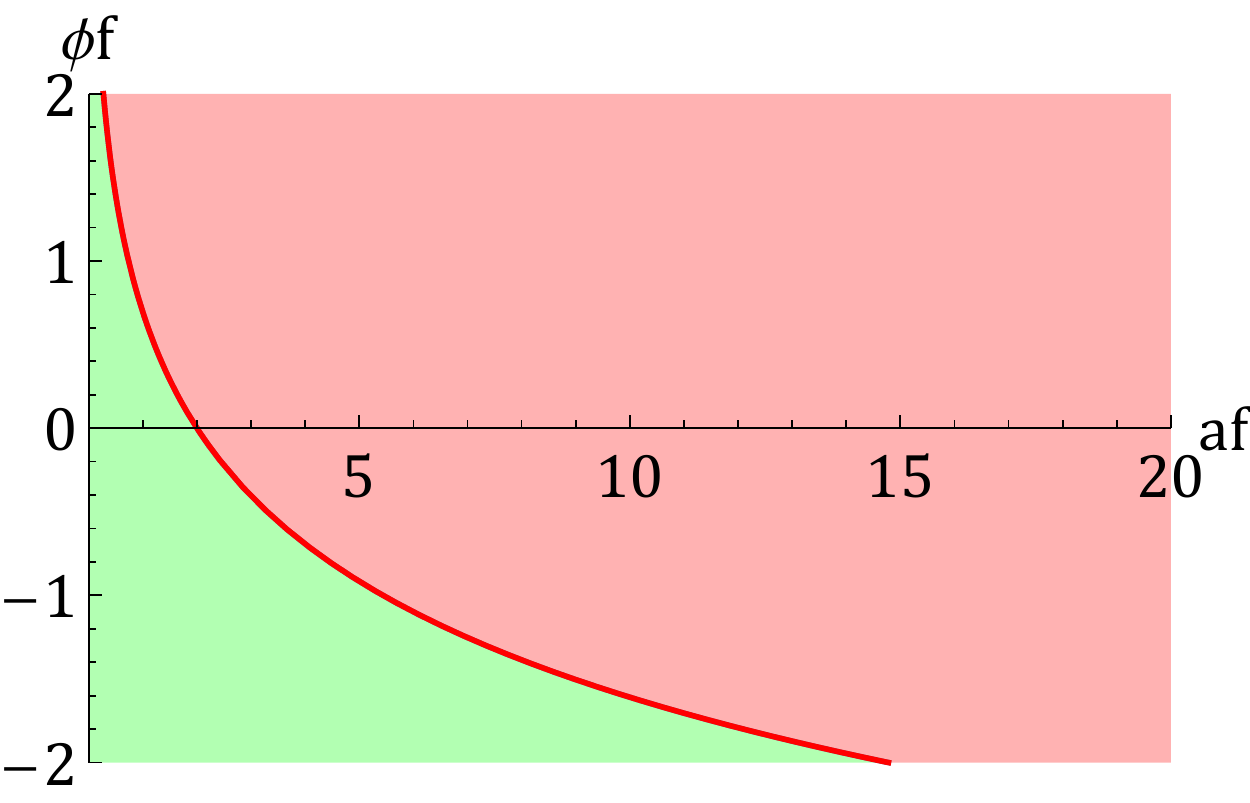}{\caption{\small Phase space regions delimited by the condition $\af^2\frac{\alpha}{2}e^{2\phif}=1$ (red curve), with $\alpha=1/2$. In the red region, the saddle point value for the lapse is purely real. In the green region, the saddle point value for the lapse is purely imaginary.}\label{fig:phasespacealphaeqbeta}}
	\end{figure}

	When $\alpha=\beta,$ the potential becomes even simpler, and is given by a single exponential $V(\phi)\propto e^{2\phi}.$ Classically, this potential allows for inflationary scaling solutions, with constant equation of state \cite{Garay:1990re}. However, because the potential is nowhere flat,  it is not possible to achieve a de Sitter limit. The classical scaling symmetry, which consists of a re-scaling of the metric accompanied by a shift in the scalar field, explains why the analysis is simplified for this potential. For instance, we are left with a unique expression for the saddle point, cf. \eqref{eq:saddle},
		\begin{equation}
		N_\text{saddle}=-i\frac{\af^2}{2}\exp(2\phif-2\gamma)\,.\label{eq:saddlealphaeqbeta}
	\end{equation} 
	We find that the values of $\gamma$ closing the saddle point geometry in this case are given by:
	\begin{equation}
		\left\lbrace
		\begin{aligned}
			&\gamma^\text{I}_{(n)}=\frac{1}{4}\ln(\frac{1-\af^2\frac{\alpha}{2}e^{2\phif}}{e^{-4\phif}})+in\cdot\frac{\pi}{2}\quad\text{for}\ n\in\lbrace0,1\rbrace\,;\quad\\
			&\gamma^\text{II}_{(n)}=\frac{1}{4}\ln(\frac{\af^2\frac{\alpha}{2}e^{2\phif}-1}{e^{-4\phif}})+i\left(\frac{\pi}{4}+n\cdot \frac{\pi}{2}\right)\quad\text{for}\ n\in\lbrace0,1\rbrace\,;\\
		\end{aligned}
		\right.
	\end{equation}
	and their regions of validity, represented in Fig.~\ref{fig:phasespacealphaeqbeta} for $\alpha=1/2$, are
	\begin{equation}
		\text{region I:}\quad\af^2\frac{\alpha}{2}e^{2\phif}<1\,;\qquad\text{region II:}\quad\af^2\frac{\alpha}{2}e^{2\phif}>1\,.
	\end{equation}
	Note that in this case there is only one saddle point for each value of $\gamma,$ and it is closed.  This is a special case of Picard's little theorem, see appendix \ref{appendix:picardthm}). Let us study the geometries of these solutions. 
	
	In region I the solution for $\gamma$ implies that $\cosh(2\gamma)$ and $\sinh(2\gamma)$ are both real, so that the momenta $\Pi_x$ and $\Pi_y$ are purely imaginary. The saddle point values for the lapse \eqref{eq:saddlealphaeqbeta} are thus also purely imaginary, as are the on-shell actions evaluated on these saddle point values. The wave function associated to these geometries is a pure amplitude, without phase. Thus no classical spacetime is predicted in this region.
	%%%%%%%%%%%%%%%%%%%%%%%%%%%%%%%%%%%%%%%%%%%%%%%%%%%%%%%%%%%%%%%%%%%%%%%%%%%%%%%%%%%%%%%%%%%%%%%%%%%%%%%%%%%%%%%

In region II, the solution for $\gamma$ instead leads to $\cosh(2\gamma)$ and $\sinh(2\gamma)$ being pure imaginary, so that the momenta $\Pi_x$ and $\Pi_y$ are real. The saddle point values, and the on-shell actions, are consequently real. In principle, we would then expect the wave function to represent a collection of classical solutions, all with equal weighting. However, as was also already noted in \cite{Garay:1990re}, the geometries all contain a singular bounce. One may see this as follows: the background solutions $\bar{x}$ and $\bar{y}$ both start at zero and are real valued, with $\bar{x}$ being a parabola with a local minimum, while $\bar{y}$ is a parabola with a local maximum. Given that at the final time $\bar{x}(1)> \bar{y}(1),$ there necessarily exists an intermediate time at which $|\bar{x}|=|\bar{y}|,$ where the scale factor vanishes. At this time all matter fields and perturbations blow up and cause a divergence in the local curvature as well as the action. As explained in section \ref{subsubsec:alpha1beta0}, we exclude such solutions and in fact find that for the pure exponential potential, no regular no-boundary solutions exist which would predict a classical evolution of the fields.

		%%%%%%%%%%%%%%%%%%%%%%%%%%%%%%%%%%%%%%%%%%%%%%%%%%%%%%%%%%%%%%%%%%%%%%%%%%%%%%%%%%%%%%%%%%%%%%%%%%%%%%%%%%%%%%	%%%%%%%%%%%%%%%%%%%%%%%%%%%%%%%%%%%%%%%%%%%%%%%%%%%%%%%%%%%%%%%%%%%%%%%%%%%%%%%%%%%%%%%%%%%%%%%%%%%%%%%%%%%%%%

	\subsection{$\alpha=0\,,\ \beta=1$: $V(\phi)=\sinh(2\phi)$.}
	
	Finally, we will look at the case $\alpha=0, \beta=1,$ i.e.~at the pure $\sinh(2\phi)$ potential. At large $\phi_f\gg1,$ we expect to recover the results of the $\cosh(2\phi)$ potential, where we did not find instantons that would predict a classical spacetime. However, at small $\phi$ we expect significant differences, as, just as for the exponential potential, there is no minimum where the scalar field can sit. Then at negative $\phi,$ the potential is unbounded below and quite steep, but not steep enough to allow for ekpyrotic solutions. Thus there is no region in this potential which obviously allows for useful/realistic instanton solutions. This expectation will be confirmed by the detailed analysis that follows.
	
	The saddle point values \eqref{eq:saddle} of the lapse here read
	\begin{equation}
		N^{\textrm{saddle}}_{\pm}=i\sinh(2\gamma)\pm\sqrt{-\sinh[2](2\gamma)-\af^2\sinh(2\phif)}\,.\label{eq:saddlealpha0}
	\end{equation}
	Again using similar techniques than in appendix \ref{appendix:gamma}, we find the following values of $\gamma$ that can potentially close these saddle point geometries:
	\begin{equation}
	\left\lbrace
	\begin{aligned}
		\gamma_{(n)}^{\text{I}}&=-\frac{1}{2}\cosh[-1](\frac{\cosh(2\phi)}{2}\sqrt{2-\af^2\sinh(2\phif)+\sqrt{\af^4+\left(2-\af^2\sinh(2\phif)\right)^2}})\\
		&\quad+i\frac{n\cdot\pi}{2}\ \text{with}\ n\in\lbrace0,1\rbrace\,;\\
		\gamma^{\text{II}}_{(n)}&=\frac{1}{2}\cosh^{-1}\bigg(\sqrt{1-\frac{\cosh[2](2\phif)}{4}\Big(2-\af^2\sinh(2\phif)-\sqrt{\af^4+(2-\af^2\sinh(2\phif))^2}\Big)}\bigg)\\
		&\quad+i\left(\frac{\pi}{4}+\frac{n\cdot\pi}{2}\right),\ n\in\lbrace0,1\rbrace\,.
	\end{aligned}
	\right.
	\end{equation}
	These expressions are both defined on the whole of phase space.

	\begin{figure}
	\centering
	\begin{subfigure}{0.49\linewidth}
		\includegraphics[width=0.49\textwidth]{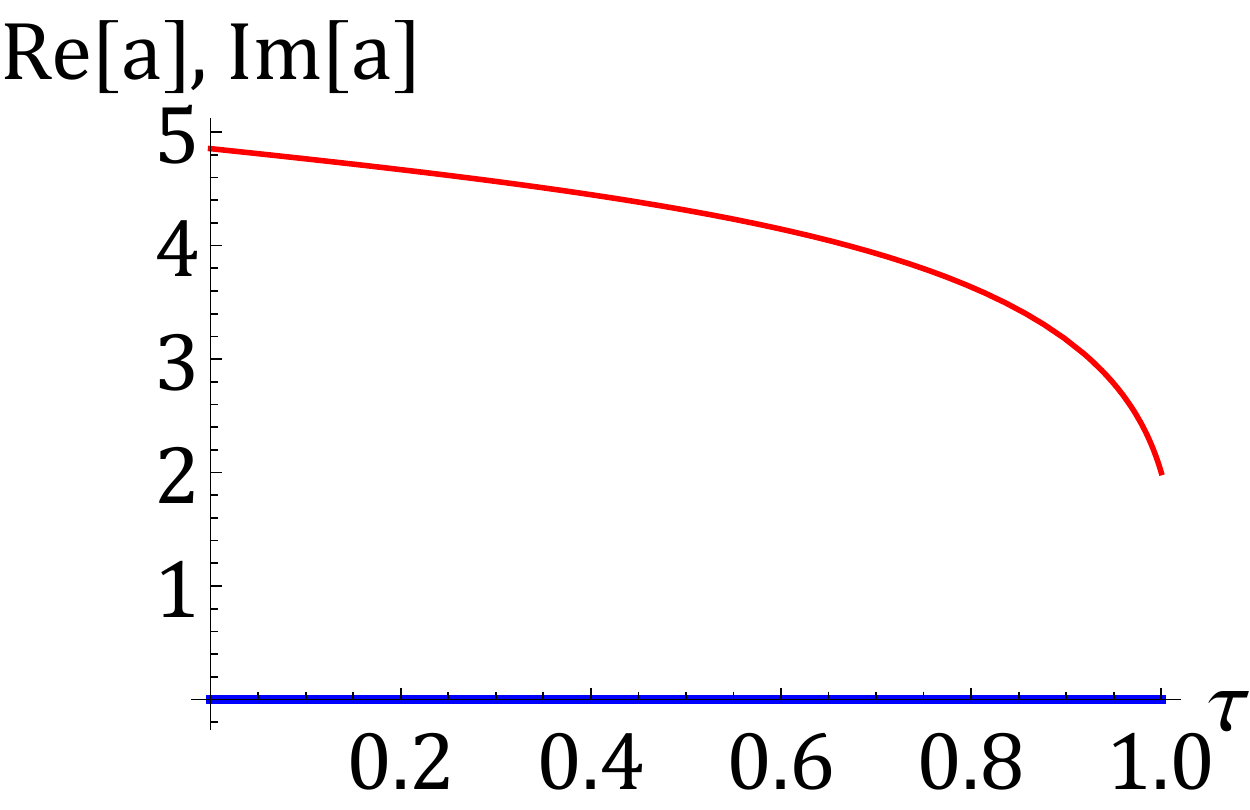}
		\includegraphics[width=0.49\textwidth]{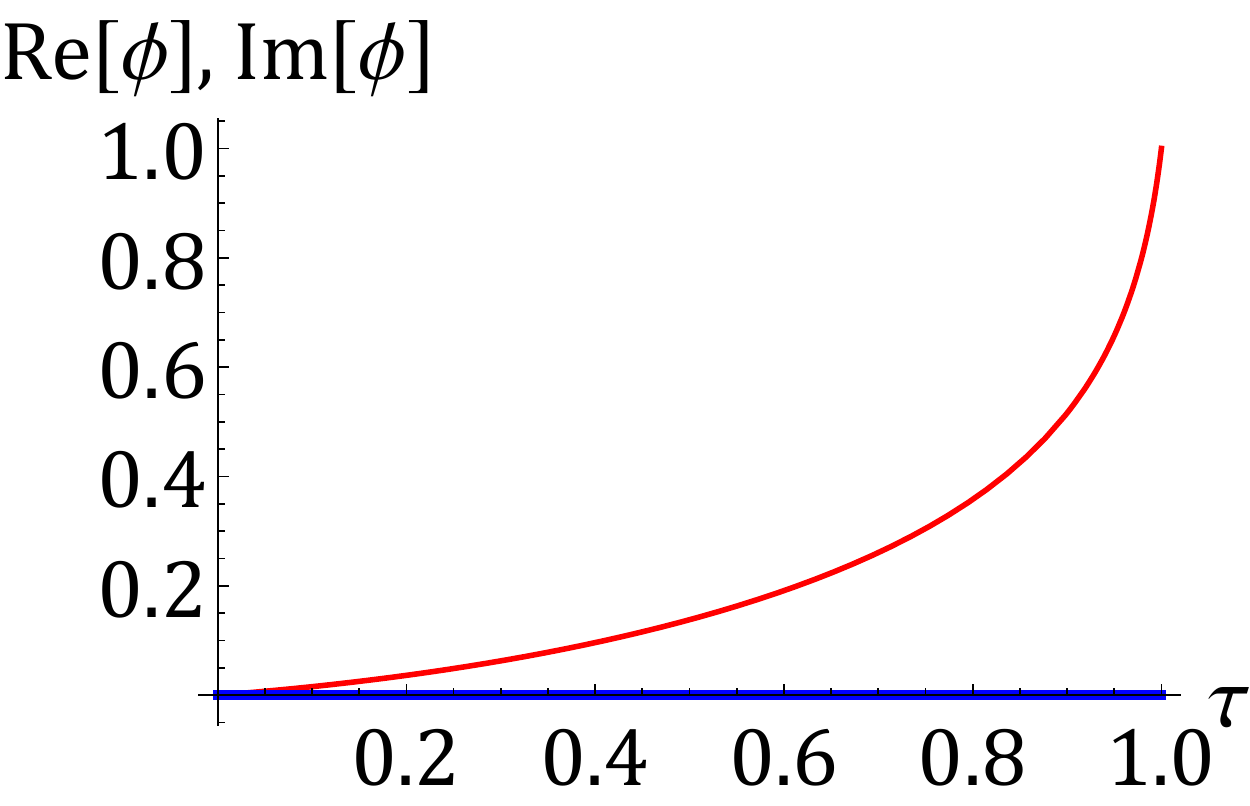}\caption{Unclosed Euclidean geometry $N_+[\gamma^\text{I}_{(0)}]$.}\label{fig:alpha0_geom_a_imN1phi1}
	\end{subfigure}
	\begin{subfigure}{0.49\linewidth}
		\includegraphics[width=0.49\textwidth]{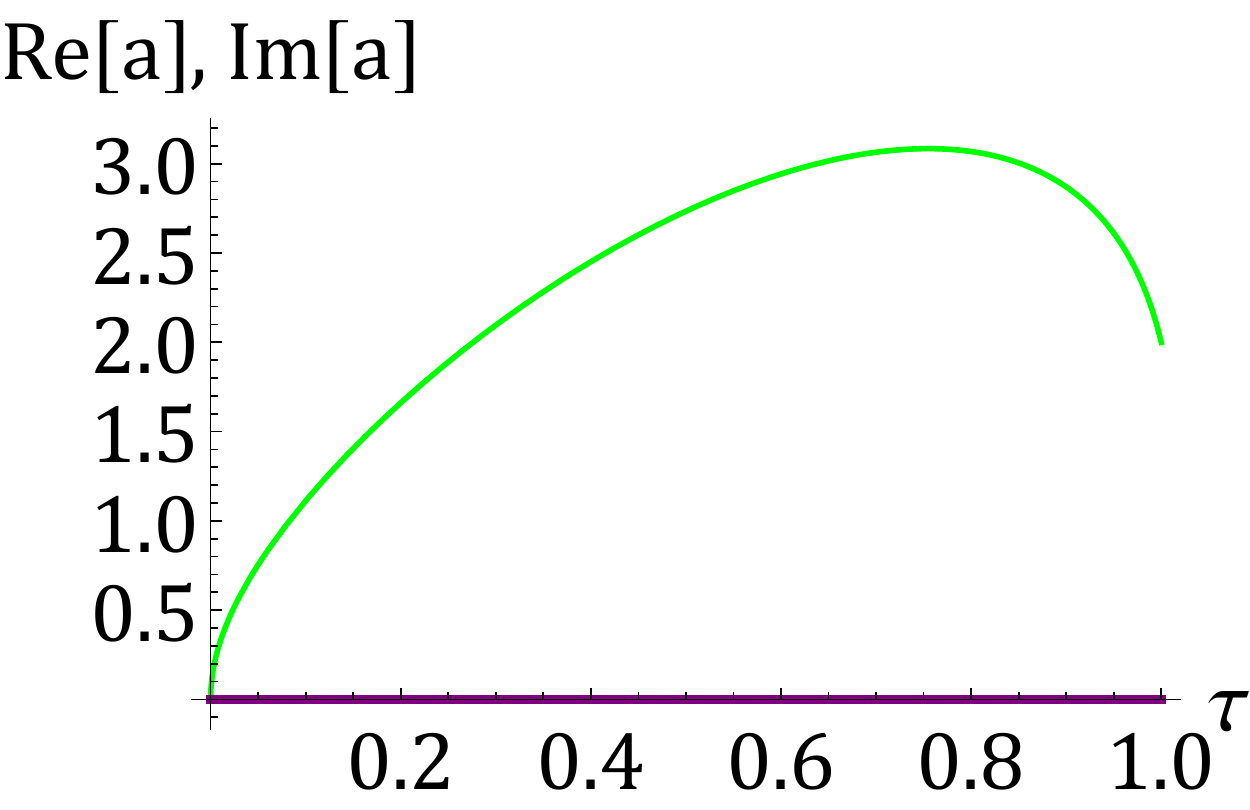}
		\includegraphics[width=0.49\textwidth]{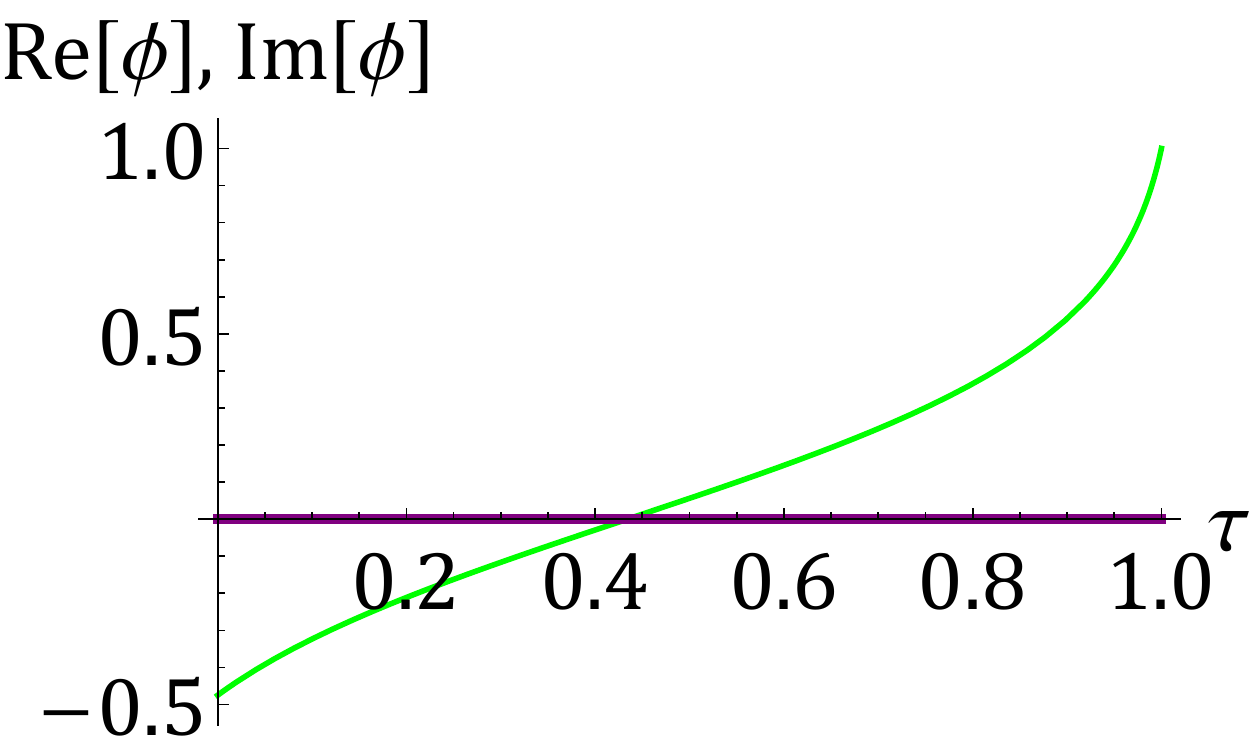}\caption{Closed Euclidean geometry $N_-[\gamma^\text{I}_{(0)}]$.}\label{fig:alpha0_geom_a_imN2phi1}
	\end{subfigure}
	\caption{\small Case $\alpha=0,\ \beta=1$,  $\af=2$, $\phif=1,$ with $N$ imaginary. The relevant numerical values are $\gamma^\text{I}_{(0)}=-0.475\,, \, N_+[\gamma^\text{I}_{(0)}]=2.87 i\,,\, N_-[\gamma^\text{I}_{(0)}]=-5.06 i\,.$ Geometry associated to the different saddle points. Same conventions as in Fig.~\ref{fig:complexsaddle1}.}\label{fig:sinh1} 
\end{figure}

\begin{figure}
		\includegraphics[width=7cm]{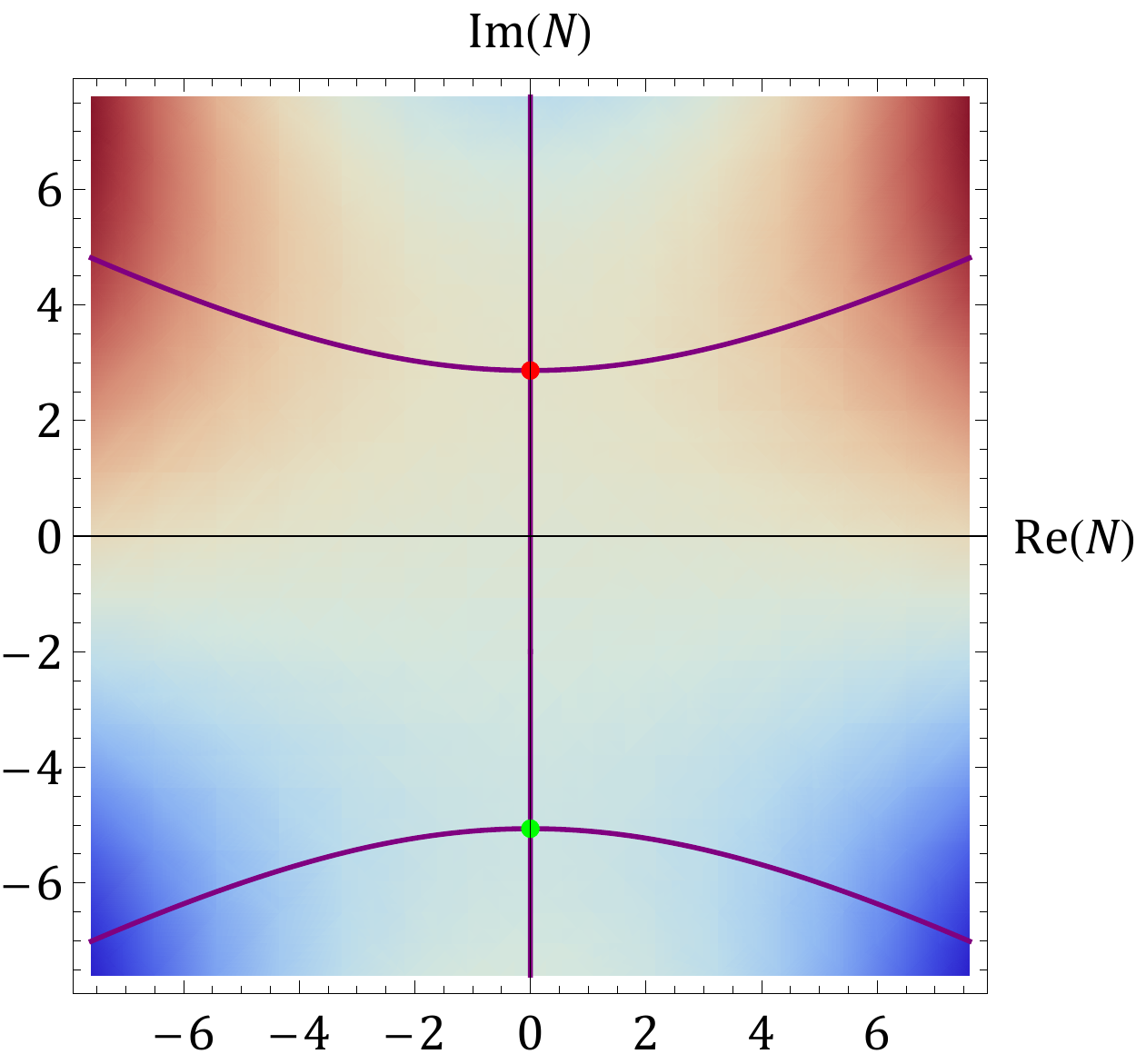}
	\caption{\small Case $\alpha=0,\ \beta=1$, $\af=2$, $\phif=1$. Density plot of the weighting $\Re[iS_\text{saddle}^\text{on-shell}]$ and flow lines in the complex N plane. Same conventions as in Fig.~\ref{fig:complexsaddle2}. $N_-[\gamma^\text{I}_{(0)}]$ (green dot) and $N_+[\gamma^\text{I}_{(0)}]$ (red dot).} \label{fig:sinh2}
\end{figure}

	The solutions $\gamma^\text{I}$ have $\Im[\gamma^\text{I}]=n\cdot\pi/2$, therefore $\cosh(2\gamma^\text{I})=(-1)^n\cosh(2\Re[\gamma^{\text{I}}])\in\mathbb{R}$ and $\sinh(2\gamma^\text{I})=(-1)^n\sinh(2\Re[\gamma^\text{I}])\in\mathbb{R}$. Hence the momenta $\Pi_x$ and $\Pi_y$ are purely imaginary and the saddle point values of the lapse also. As before, we impose that $Im(\Pi_x)>0$ and this eliminates the values $\gamma^I_{(1)}.$ The geometries are Euclidean, and the remaining ones are shown in Fig.~\ref{fig:sinh1}. We once again obtain a closed and an unclosed geometry. The associated flow lines are shown in Fig.~\ref{fig:sinh2}. One needs to define the integral on a contour that runs parallel to the real $N$ line, but this time one needs to shift it into the lower half plane for convergence (this is because $\alpha=0$, cf. Eq. \eqref{eq:actiononshellN}). The relevant saddle point is then the closed saddle at $N_-[\gamma^I_{(0)}],$ and the wave function can be approximated as
	\begin{align}
	\Psi & \approx e^{\frac{i}{\hbar}S(N_-[\gamma^I_{(0)}])}\,, \\ S(N_-[\gamma^I_{(0)}]) & = -\frac{i}{2}\left[ 3 a_f^2 (\sqrt{a_f^4+4}+2)^{1/2} -(\sqrt{a_f^4+4}-2)^{3/2}\right] \nonumber \\ & +\frac{i}{2}(\sqrt{a_f^4+4}-2)^{1/2}a_f^2\phi_f+ {\cal O}(\phi_f^2)\,.
	\end{align}
	The wave function is pure amplitude. This could describe the nucleation of the universe at small scale factor, but there is no follow-up saddle point that could describe the subsequent evolution of the universe. Hence we conclude that these saddle points do not predict the emergence of classical spacetime. In fact, the wave function in this case resembles more the quasi-AdS case of section \ref{sec:ads}. From this point of view, it is understandable that the scalar field starts at negative values (for the closed saddle), but then, to finally reach positive values of the potential, the universe recollapses, which allows the scalar to run up to positive values, see again Fig.~\ref{fig:sinh1}. It would be interesting to see whether the present solutions have an application in finite bulk extensions of AdS/CFT \cite{McGough:2016lol}.

	For the solutions $\gamma^\text{II}$, we have $\Im[\gamma^\text{I}]=\pi/4+n\cdot\pi/2$, so $\cosh(2\gamma^\text{I})=i(-1)^n\sinh(2\Re[\gamma^{\text{I}}])\in i\mathbb{R}$ and $\sinh(2\gamma^\text{I})=i(-1)^n\cosh(2\Re[\gamma^\text{I}])\in i\mathbb{R}$. The momenta $\Pi_x$ and $\Pi_y$, as well as the saddle point values of the lapse are purely real. However, using analogous arguments to those used in previous paragraphs, one may straightforwardly show that all these saddle point geometries contain a singular bounce. For the same reasons as those mentioned previously, namely that such geometries lead to divergences of the curvature and the action once perturbations or additional matter fields are included, we deem such solutions to be unphysical and discard them. Thus we are unfortunately left without any complex saddle points, so that for this potential we conclude that the no-boundary wave function does not predict the emergence of classical evolution. In some sense, this is a very good example showing the predictive power of the no-boundary proposal: only in suitably flat regions of the potential does the no-boundary wave function exhibit classical (inflationary) spacetime evolution (while in sufficiently steep and negative regions of the potential the same could be achieved via ekpyrotic no-boundary instantons \cite{Battarra:2014xoa,Battarra:2014kga}). This is thus a good example showing that the classicality of the early universe should be seen as an important clue as to its origin.

	%%%%%%%%%%%%%%%%%%%%%%%%%%%%%%%%%%%%%%%%%%%%%%%%%%%%%%%%%%%%%
	%%%%%%%%%%%%%%%%%%%%%%%%%%%%%%%%%%%%%%%%%%%%%%%%%%%%%%%%%%%%%

\section{Discussion} \label{sec:discussion}

In this paper we have been able to show, in the context of a minisuperspace model, that the no-boundary proposal can be defined as a sum over regular metrics and scalar field configurations. The regularity condition has the advantage that in the sum over metrics, one is considering metrics on which perturbations are damped, rather than enhanced. This is a great advantage compared to defining the no-boundary path integral as a sum over compact geometries, since in the latter case one is forced to include unstable configurations too. The price to pay is that a sum over regular metrics does not guarantee that the geometries are also closed. Off-shell the geometries are certainly not closed in general, and in fact the best one can hope for is that the saddle points are closed. As we have demonstrated, even this does not occur in full generality: alongside saddle points that are both regular and compact, we found saddle points that are unclosed. Moreover, we argued that such unclosed saddle points cannot be avoided. However, they were always subdominant, so that the semi-classical no-boundary wave function ended up being dominated by a closed, regular geometry with a regular scalar field configuration, as originally intended. 

Our construction highlights a puzzle inherent in the no-boundary proposal. This puzzle is not new, but it has come into renewed focus here. The puzzle is that imposing regularity alone is insufficient to fully fix the initial conditions. It only fixes them up to a complex number $\gamma$ which, as we have discussed, can often be interpreted as the initial value of the scalar field. (Thus, the same problem arises when attempting to define the no-boundary integral over compact metrics.) This number must be fixed such that the path integral may contain saddle points that are closed (for generic values of $\gamma$ the saddle points will all be unclosed). The puzzle is that the required values of $\gamma$ depend on the final conditions, i.e.~as the universe grows and the scalar field evolves, $\gamma$ necessarily has to change. When the wave function is of WKB form, $\gamma$ changes less and less as the universe grows \cite{Hartle:2008ng}, but nevertheless it is not constant. This is as non-local/non-causal as it gets: at every moment in the history of the universe, it suggests that the universe must ``recalculate'' its entire wave function, from the very beginning of space and time. The conclusion we can draw from this observation is that either quantum gravity really is this drastically non-local, or there is something wrong with the prescription that we have outlined.

We can conceive of one possible way in which the prescription above could be refined, such that the puzzle disappears: this is to require the scalar field to always take real values. A motivation for such a condition stems from the recent discussions of allowable metrics, where the criterion of allowability is derived from demanding convergence of path integrals over real scalars and p-form fields, see the works of Kontsevich-Segal \cite{Kontsevich:2021dmb} and Witten \cite{Witten:2021nzp}, based on earlier work by Louko-Sorkin \cite{Louko:1995jw} (consequences for minisuperspace models started being explored in \cite{Lehners:2021mah,Rajeev:2021xit}). For positive potentials, the only place where $\gamma$ turned out to be real and where the saddle point geometries are regular and lead to a classical spacetime, was at the minimum of the $\cosh(2\phi)$ potential. More generally, the scalar field value is both real and constant at extrema of the scalar potential. Thus, at such locations of the potential, the initial conditions for the no-boundary wave function remain unchanged over time. For the negative potential studied in section \ref{sec:ads}, we also found real, yet evolving, values of the scalar field. However, the corresponding geometries were Euclidean, and in Euclidean space there is no notion of causality, so that the puzzle disappears in any event. 

What would a restriction to extrema of the scalar potential imply for observations? First, we should note that the no-boundary wave function scales roughly as $e^{1/(\hbar V(\phi))},$ implying that low values of the potential come out as preferred. Thus, in a theory with many scalar fields (as expected in string theoretic models, for instance), the most likely configuration would be for all scalars to reside in the lowest minima of the potential. This is in fact a desirable feature, as it can help explain why all coupling constants in our universe are found not to vary over time. However, with such a configuration, the universe is also left empty. If one scalar instead starts out at a local maximum of the potential, then there is the possibility for this scalar to drive an inflationary phase during which density perturbations may also be generated, and structure may form in the universe. Due to quantum fluctuations, the scalar field would then also eventually roll off the potential maximum, allowing inflation to end. Of course, to make this precise and see if this scenario is truly viable, one would have to understand the precise structure of the scalar potential, and in particular one would have to find out if there exists a local maximum in which sufficiently large density perturbations can be generated, alongside potential minima that are sufficiently stable. This is a substantial unsolved open problem. However, what this discussion illustrates is that, given a suitable dynamical theory, the no-boundary proposal remains an appealing, consistent (currently the only?) candidate for a theory of initial conditions.

	%%%%%%%%%%%%%%%%%%%%%%%%%%%%%%%%%%%%%%%%%%%%%%%%%%%%%%%%%%%%%
	%%%%%%%%%%%%%%%%%%%%%%%%%%%%%%%%%%%%%%%%%%%%%%%%%%%%%%%%%%%%%

\acknowledgments

CJ and JLL gratefully acknowledge the support of the European Research Council in the form of the ERC Consolidator Grant CoG 772295 ``Qosmology''.

	%%%%%%%%%%%%%%%%%%%%%%%%%%%%%%%%%%%%%%%%%%%%%%%%%%%%%%%%%%%%%
	%%%%%%%%%%%%%%%%%%%%%%%%%%%%%%%%%%%%%%%%%%%%%%%%%%%%%%%%%%%%%

	%%% APPENDIX %%%	
	\appendix

	\section{Fluctuation integrals} \label{sec:fluct}
	
	The Wheeler-DeWitt equation in ``position'' space is obtained from the constraint \eqref{constraint}
	\begin{equation}
		\left[\hbar^2\left(\partial_x^2-\partial_y^2\right)-\frac{1}{4}\left(1-\alpha x-\beta y\right)\right]\Psi=0\,.
	\end{equation}
	Assuming that the integrations of $x$ and $y$ have already been performed implies that the wave function can be written as an ordinary integral over the lapse, but with a currently unknown measure factor $m(N),$
	\begin{equation}
		\Psi[\xf,\yf]=\int\dd N\,m(N)\,e^{iS_0^{\text{on-shell}}[\xf,\yf,N]/\hbar}\,.
	\end{equation}
	We abbreviate the on-shell (in $x$ and $y$) action \eqref{eq:actiononshellN} simply by $S$, and $\xf$, $\yf$ by $x$, $y.$ Then
	\begin{equation}
		\hbar^2(\partial_x^2-\partial_y^2)\Psi=\int\dd N\,m(N)\hbar^2\left(\frac{iS_{,xx}}{\hbar}-\frac{(S_{,x})^2}{\hbar^2}-\frac{iS_{,yy}}{\hbar}+\frac{(S_{,y})^2}{\hbar^2}\right)e^{iS/\hbar}\,.\label{eq:psideriv}
	\end{equation}
	Starting from  \eqref{eq:actiononshellN}, the explicit expressions are given by
	\begin{align}
		&S_{,x}=-\frac{\alpha N}{2}-\frac{\Pi_x}{2}\,;\quad S_{,xx}=0\,;\quad S_{,y}=-\frac{\beta N}{2}-\frac{\Pi_y}{2}\,;\quad S_{,yy}=0\,;\\
		&\Rightarrow(S_{,y})^2-(S_{,x})^2=\frac{1}{4}\left(N^2(\alpha^2-\beta^2)-2N(\alpha\Pi_x+\beta\Pi_y)+\Pi_y^2-\Pi_x^2\right)\,;\\
		&S_{,N}=\frac{N^2(\alpha^2-\beta^2)}{2}+N(\alpha\Pi_x+\beta\Pi_y)-\frac{(\alpha x+\beta y)}{2}\,;\\
		&\Rightarrow\ \textrm{if}\ \Pi_y^2-\Pi_x^2=1,\ \ (S_{,y})^2-(S_{,x})^2=-\frac{S_{,N}}{2}+\frac{(1-\alpha x-\beta y)}{4}\,.
	\end{align}
	Inserting into the WdW equation we obtain
	\begin{equation}
		0=\int\dd N m(N)\vdot\left[-\frac{S_{,N}}{2}\right]e^{iS/\hbar}\ \Leftrightarrow\ 0=\int\dd N \left(-\frac{i\hbar}{2}\right)m_{,N}e^{iS/\hbar}\ \Leftrightarrow\ m=\textrm{const.}
	\end{equation}
which verifies that, with mixed Neumann-Dirichlet boundary conditions, the fluctuation integrals lead to no additional $N$ dependence in the integrand of the lapse integral.

	%%%%%%%%%%%%%%%%%%%%%%%%%%%%%%%%%%%%%%%%%%%%%%%%%%%%%%%%%%%%%%%%%%%%%%%%%%%%%%%%%%%%%%%%%%%%%%%%%%%%%%%%%%%%%%%
%%%%%%%%%%%%%%%%%%%%%%%%%%%%%%%%%%%%%%%%%%%%%%%%%%%%%%%%%%%%%%

	\section{Solving  for $\gamma$} \label{appendix:gamma}
	
	In this appendix, we will present the details on how to solve for the initial conditions parameter $\gamma,$ under the condition that a closed saddle point geometry should exist. We will restrict the analysis to the $\cosh(2\phi)$ potential ($\alpha=\pm1, \beta=0$). For the other potentials, the analysis is analogous.

	%\subsection{$\alpha=1$ and $\beta=0$}

\vspace{0.5 cm}	
{\noindent\it{$\alpha=+1$ and $\beta=0$}}

	We need to solve the following equation for $\gamma\in\mathbb{C}$:
		\begin{equation}
		\af^2\sinh[2](2\phif)+4\sinh(2\gamma)\sinh(2\gamma-2\phif)=0\,.
		\end{equation}
	The real and imaginary parts of this equation must both hold separately. The imaginary part yields:
	\begin{equation}
		2\af^2\sin(4\Im[\gamma])\sinh(4\Re[\gamma]-2\phif)=0\,;
	\end{equation}
	so we need either $\Im[\gamma]=k\cdot\pi/4$ for $k\in\mathbb{Z}$, or $\Re[\gamma]=\phif/2$. We now analyse the real part of equation \eqref{eq:closed_saddle} for these different cases.
	\begin{enumerate}
		\item $\Re[\gamma]=\phif/2$: then the real part of \eqref{eq:closed_saddle} becomes
		\begin{equation}
			\sin[2](2\Im[\gamma])=(\af^2\cosh[2](\phif)-1)\sinh[2](\phif)\,.\label{eq:solrealpartcomplexN}
		\end{equation}
		Solutions to this equation only exist if the right hand side is between $0$ and $1$, which translates into
		\begin{equation}
			\af^2\cosh[2](\phif)>1\quad \text{and}\quad \af^2\sinh[2](\phif)<1\,.\label{eq:conditioncomplexalpha1beta0}
		\end{equation}
		When these conditions are satisfied, equation \eqref{eq:solrealpartcomplexN} possesses four families of solutions
		\begin{equation}
			\left\lbrace
			\begin{aligned}
				&\gamma^{1,2(k)}=\frac{\phif}{2}\mp\frac{i}{2}\left(\arcsin(\abs{\sinh(\phif)}\sqrt{\af^2\cosh[2](\phif)-1})+2\pi\cdot k\right);\\
				&\gamma^{3,4(k)}=\frac{\phif}{2}\mp\frac{i}{2}\left(\pi-\arcsin(\abs{\sinh(\phif)}\sqrt{\af^2\cosh[2](\phif)-1})+2\pi\cdot k\right).
			\end{aligned}
			\right.
			\,\forall k\in\mathbb{Z}\,.\label{eq:solphi0complex}
		\end{equation}
		We still need to make sure that we obtain only saddle points that admit stable fluctuations, i.e.~which imply the correct orientation for the Wick rotation. This means that we must get rid of spurious solutions for which $a\dot{a}|_0\to-iN$ (rather than $a\dot{a}|_0\to+iN$) when $\phi\to0$, i.e.~we only keep solutions where
		\begin{equation}
			\Pi_x=\left.\frac{\dot{x}}{2N}\right|_0=\left.\frac{a\dot{a}\cosh(2\phi)+a^2\dot{\phi}\sinh(2\phi)}{N}\right|_0\xrightarrow{\phi\to0}\left.\frac{a\dot{a}}{N}\right|_0=+i\,.
		\end{equation}
		Writing $\Pi_x$ in terms of $\gamma$, we obtain
		\begin{equation}
			\Pi_x=i\cosh(2\gamma)=i\Big(\cosh(\phif)\cos(2\Im[\gamma])+i\sinh(\phif)\sin(2\Im[\gamma])\Big)\,.\label{eq:Pixexpression}
		\end{equation}
		Expanding this expression at zeroth order in $\phif$, the second term vanishes and we find, by plugging in the solutions \eqref{eq:solphi0complex} for $\gamma$:
		\begin{equation}
			\left\lbrace
			\begin{aligned}
				&\Pi_x^{1,2(k)}=i\cos\left(\mp\arcsin(\abs{\sinh(\phif)}\sqrt{\af^2\cosh[2](\phif)-1})+2\pi\cdot k\right)\xrightarrow{\phif\to0}+i\,;\\
				&\Pi_x^{3,4(k)}=i\cos\left(\mp\pi\pm\arcsin(\abs{\sinh(\phif)}\sqrt{\af^2\cosh[2](\phif)-1})+2\pi\cdot k\right)\xrightarrow{\phif\to0}-i\,.
			\end{aligned}\right.
		\end{equation}
		We therefore exclude the last two families of solutions, $\gamma^{3,4(k)}$, and keep only the first two, $\gamma^{1,2 (k)}$. Since $\gamma$ only enters expressions through its hyperbolic sine or cosine, the value of $k$ is unimportant and we can simply set it to $0$. In the end we only have two values for $\gamma$:
		\begin{equation}
			2\gamma_{\mp}=\phif\mp i\arcsin(\abs{\sinh{\phif}}\sqrt{\af^2\cosh[2](\phif)-1})\,.\label{eq:gammaNcomplex}
		\end{equation}
		The saddle point for the lapse computed from these $\gamma$ values will be complex.
		\item $\Im[\gamma]=k\cdot\pi/2$, $k\in\mathbb{Z}$: $\sin(2\Im[\gamma])=0$ and $\cos(2\Im[\gamma])=(-1)^k$. All expressions then only depend on whether $k$ is odd or even, so we can restrict to $k=0,1$. The real part of the equation \eqref{eq:closed_saddle} yields:
		\begin{equation}
			\af^2\sinh[2](2\phif)=4\sinh(2\Re[\gamma])\sinh(2\phif-2\Re[\gamma])\,.\label{eq:conditionimaginary}
		\end{equation}
		This admits a solution only if
		\begin{align}
			&\af^2\sinh[2](2\phif)<4\abs{\max_{\Re[\gamma]}\Big(\sinh(2\Re[\gamma])\sinh(2\phif-2\Re[\gamma])\Big)}=4\sinh[2](\phif)\,;\nonumber\\
			&\Leftrightarrow\  \af^2\cosh[2](\phif)<1\,.\label{eq:condimaginaryalpha1beta0}
		\end{align}
		When this condition is satisfied, we find two solutions for each $k$ value:
		\begin{align}
			\gamma_{\mp}^{(k)}=\frac{1}{2}&\cosh^{-1}\bigg(\frac{1}{2}\Big[4+\sinh[2](2\phif)(2-\af^2\cosh(2\phif))\\
			&\,\mp\sinh[2](2\phif)\sqrt{(2-\af^2\cosh(2\phif))^2-\af^4}\Big]^{1/2}\bigg)+ik\cdot\frac{\pi}{2}\,;\nonumber
		\end{align}
		which implies that
		\begin{align}
			\Pi_{x\,\mp}=i\cosh(2\gamma_\mp)=&\frac{i(-1)^k}{2}\bigg[4+\sinh[2](2\phif)(2-\af^2\cosh(2\phif))\\
			&\,\mp\sinh[2](2\phif)\sqrt{(2-\af^2\cosh(2\phif))^2-\af^4}\bigg]^{1/2}\in i\mathbb{R}\,,\nonumber
		\end{align}
		\begin{align}
			\Pi_{y\,\mp}=i\sinh(2\gamma_{\mp})&=\frac{i(-1)^k}{2}\abs{\sinh(2\phif)}\bigg[(2-\af^2\cosh(2\phif))\mp\sqrt{(2-\af^2\cosh(2\phif))^2-\af^4}\bigg]^{1/2} \nonumber  \\ & \in i\mathbb{R}\,.
		\end{align} We look at the limit of $\Pi_x$ when $\phif\to0$ to exclude spurious solutions, and we find
		\begin{equation}
			\Pi_{x\,\mp}\to i(-1)^k\,;
		\end{equation}
		thus we only keep $k=0$ and we reject $k=1$. Finally the $\gamma$ values are
		\begin{align}
			\gamma_{\mp}=\frac{1}{2}\cosh^{-1}\bigg(\sqrt{1+\frac{\sinh[2](2\phif)}{4}\Big((2-\af^2\cosh(2\phif))\mp\sqrt{(2-\af^2\cosh(2\phif))^2-\af^4}\Big)}\bigg)\,.\nonumber
		\end{align}
		Because these $\gamma$ values are real, the saddle points for the lapse will be purely imaginary.
		\item $\Im[\gamma]=\pi/4+k\cdot\pi/2$: $\sin(2\Im[\gamma])=(-1)^k$ and $\cos(2\Im[\gamma])=0$.	All expressions will only depend on whether $k$ is odd or even, so we can restrict to $k=0,1$ only. The real part of equation \eqref{eq:closed_saddle} becomes
		\begin{equation}
			\af^2\sinh[2](2\phif)-4\cosh(2\Re[\gamma])\cosh(2\Re[\gamma]-2\phif)=0\,,
		\end{equation}
		which admits a solution only if 
		\begin{align}
			&\af^2\sinh[2](2\phif)>4\min_{\Re[\gamma]}(\cosh(2\Re[\gamma])\cosh(2\Re[\gamma]-2\phif))=4\cosh[2](\phif)\,,\nonumber\\
			&\Leftrightarrow\ \af^2\sinh[2](\phif)>1\,.\label{eq:condrealalpha1beta0}
		\end{align}
		This condition being fulfilled, we find the four following solutions:
		\begin{align}
			\gamma_\mp^{(k)}=&\mp\frac{1}{2}\cosh[-1](\frac{1}{2}\sinh(2\phif)\sqrt{\af^2\cosh(2\phif)-2\mp\sqrt{\left(\af^2\cosh[2](2\phif)-2\right)^2-\af^4}})\nonumber\\
			&+i\left(\frac{\pi}{4}+\frac{k\cdot\pi}{2}\right)\,,\quad k\in\lbrace0,1\rbrace\,.\label{eq:gammavalue1Nreal}
		\end{align}
		In this case it is not possible to take the limit $\phif\to0$, as this would contradict the condition \eqref{eq:condrealalpha1beta0}. The saddle points for the lapse will be purely real.
	\end{enumerate}

%%%%%%%%%%%%%%%%%%%%%%%%%%%%%%%%%%%%%%%%%%%%%%%%%%%%%%%%%%%%%%

\vspace{0.5 cm}	
{\noindent\it{$\alpha=-1$ and $\beta=0$}}
	%\subsection{$\alpha=-1$ and $\beta=0$}
	
The saddle point expressions are:
\begin{equation}
	N^\text{saddle}_{\pm}=i\cosh(2\gamma)\pm\sqrt{-\cosh[2](2\gamma)-\af^2\cosh(2\phif)}\,.\label{eq:saddlealpha-1}
\end{equation}
The constraint equation automatically yields $x|_0=0$. Requiring the initial geometry to close hence implies 
\begin{equation}
	y|_0=\yf-2N\Pi_y=0\,,\quad\Leftrightarrow\quad N_\text{closed}=\frac{\af^2\sinh(2\phif)}{2i\sinh(2\gamma)}\,.
\end{equation}
We thus get the following equation for a closed saddle point geometry, that we will solve for $\gamma$:
\begin{equation}
	\af^2\sinh[2](2\phif)-4\sinh(2\gamma)\sinh(2\gamma-2\phif)=0\,.\label{eq:alphaminus1closedsaddle}
\end{equation}
Taking the imaginary part of this expression, we find
\begin{align}
	2\af^2\sin(4\Im[\gamma])\sinh(4\Re[\gamma]-2\phif)=0\,.
\end{align}
This is solved by $\Im[\gamma]=k\cdot\pi/4$ for $k\,\in\,\mathbb{Z}$, or $\Re[\gamma]=\phif/2$. For each of these cases we examine the real part of equation \eqref{eq:alphaminus1closedsaddle}:
\begin{enumerate}
	\item $\Re[\gamma]=\phif/2$: then we find 
	\begin{align}
		4\sin[2](2\Im[\gamma])=-\af^2\sinh[2](2\phif)-4\sinh[2](\phif)\,.
	\end{align}
	This equation does not admit any solution for $\af>0$ and $\phif\in\mathbb{R}$.
	\item $\Im[\gamma]=k\cdot\pi/2$, $k\,\in\,\mathbb{Z}$: then $\cos(2\Im[\gamma])=(-1)^k$ and $\sin(2\Im[\gamma])=0$. Again we only keep $k=\lbrace0,1\rbrace$. The real part of the equation for $\gamma$ is
	\begin{align}
		\af^2\sinh[2](2\phif)=4\sinh(2\Re[\gamma])\sinh(2\Re[\gamma]-2\phif)\,;\label{eq:alpha-1}
	\end{align}
	which yields the following constraint:
	\begin{equation}
		\af^2\sinh[2](2\phif)>4\min_{\Re[\gamma]}(\sinh(2\Re[\gamma])\sinh(2\Re[\gamma]-2\phif))\,.
	\end{equation}
	This minimum lies at $\Re[\gamma]=\phif/2$ and hence implies $\af^2\sinh[2](2\phif)>-4\sinh[2](\phif)$. This condition is always satisfied, thus the equation \eqref{eq:alpha-1} admits solutions for all values $\af>0$ and $\phif\in\mathbb{R}$, which are:
	\begin{equation}
		\left\lbrace
		\begin{aligned}
			&\Re[\gamma^{\text{I}}_{\mp}]=\mp\cosh[-1](\sqrt{1+\frac{\sinh[2](2\phif)}{4}\left(2+\af^2\cosh(2\phif)-\sqrt{\left(2+\af^2\cosh(2\phif)\right)^2-\af^4}\right)})\,;\\
			&\Re[\gamma^{\text{II}}_{\mp}]=\mp\cosh[-1](\sqrt{1+\frac{\sinh[2](2\phif)}{4}\left(2+\af^2\cosh(2\phif)+\sqrt{\left(2+\af^2\cosh(2\phif)\right)^2-\af^4}\right)})\,.
		\end{aligned}
		\right.
	\end{equation}
	We can further constrain the value of $k$ by looking at the limit of $\Pi_x$ when $\phif\to 0$. In that case, \eqref{eq:alpha-1} directly implies $\sinh[2](2\Re[\gamma])\to0$, so $\gamma\to0$. Then
	\begin{equation}
		\Pi_x=i\cosh(2\gamma)= i(-1)^k\cosh(2\Re[\gamma])\xrightarrow[\gamma\to0]{} i(-1)^k\,;
	\end{equation}
	so we recover the right limit $\Pi_x\to+i$ only for $k=0$, so $\Im[\gamma]=0$.
	\item $\Im[\gamma]=\pi/4+k\cdot\pi/2$, $k\,\in\,\mathbb{Z}$: here we have $\cos(2\Im[\gamma])=0$ and $\sin(2\Im[\gamma])=(-1)^k$, so
	\begin{align}
		&\af^4\sinh[2](2\phif)=-4\af^2\cosh(2\Re[\gamma])\cosh(2\Re[\gamma]-2\phif)\nonumber\,;\\
		\Rightarrow\ &
		\af^2\sinh[2](2\phif)<4\max_{\Re[\gamma]}(-\cosh(2\Re[\gamma])\cosh(2\Re[\gamma]-2\phif))\,.
	\end{align}
	This maximum lies at $\Re[\gamma]=\phif/2$ and thus $\af^2\sinh[2](2\phif)<-4\cosh[2](\phif)$. This can never be satisfied for real values of $\af$ and $\phif.$
	\end{enumerate}

	%%%%%%%%%%%%%%%%%%%%%%%%%%%%%%%%%%%%%%%%%%%%%%%%%%%%%%%%%%%%%%%%%%%%%%%%%%%%%%%%%%%%%%%%%%%%%%%%%%%%%%%%%%%%%%%%

	\section{Picard's little theorem}\label{appendix:picardthm}
	
	\textit{"If a function $f : \mathbb{C}\to\mathbb{C}$ is entire and non-constant, then the set of values that $f(z)$ assumes is either the whole complex plane or the plane minus a single point."}
	An entire function is a holomorphic function defined on the whole complex plane, and a function $f:\mathbb{C}\to\mathbb{C}$ is holomorphic if and only if the application $\mathbb{C}\to\mathbb{C}:z\to f^{\prime}(z)$ is continuous on $\mathbb{C}$, and $f$ is $\mathbb{C}$-differentiable, i.e., the corresponding function $F:\mathbb{R}^2\to\mathbb{R}^2:(\Re[z],\Im[z])\to(\Re[f(z)],\Im[f(z)])$ is differentiable at all points $(\Re[z],\Im[z])=(x,y)$, and for $f(x+iy)=u(x,y)+iv(x,y)$, the Cauchy-Riemann relations are satisfied:
	\begin{equation}
		\left\lbrace
		\begin{aligned}
			&\frac{\partial u}{\partial x}(x,y)=\frac{\partial v}{\partial y}(x,y)\,,\\
			&\frac{\partial u}{\partial y}(x,y)=-\frac{\partial v}{\partial x}(x,y)\,.
		\end{aligned}
		\right.
	\end{equation}
	Based on this definition, one can show that the function 
	\begin{equation}
		V:\mathbb{C}\to\mathbb{C}:z\to\alpha\cosh(z)+\beta\sinh(z)\,,
	\end{equation}(i.e., our potential) is holomorphic. 
	Generically, for a potential that is an entire function, this theorem means that the constraint equation at $\tau=0,$ $a|_0^2V(\phi|_0)=0,$ is not only satisfied for closed geometries where $a|_0=0$, but also for situations where $V(\phi|_0)=0$, which generically exist. There is one well-known exception given by the exponential potential, $V(\phi)=e^{\phi}$, which only vanishes at infinity when $\Re[\phi]<0$. This is precisely the case we study when $\alpha=\beta$. 
	That case put aside, this explains why for every $\gamma$, we find two regular saddle points; one that closes and for which $\phi|_0=\gamma$; and one that remains unclosed, for which $\phi|_0\neq\gamma$, but for which $\phi|_0$ will be such that $V(\phi|_0)=0$.
	
	Moreover this is not specific to this potential, but is a generic feature of all entire functions. Therefore the only assumption we are making is that this potential of the scalar field can be extended to the complex plane. This is however a strong assumption, that can be questioned, see e.g.~the discussion section \ref{sec:discussion}.

\bibliographystyle{utphys}
\bibliography{NoBoundaryScalarField}

\end{document}